\documentclass[12pt,sort&compress]{elsarticle}

\usepackage{afterpage}
\usepackage{enumitem}
\usepackage{mystyle}
\usepackage{algpseudocode}
\usepackage{algorithm}
\usepackage{amsmath}
\usepackage{amssymb}
\usepackage{subfiles}
\graphicspath{{Images/}{../Images/}}
\usepackage{subfigmat} 
\usepackage{relsize}
\usepackage{pst-all}
\usepackage[normalem]{ulem}
\usepackage{cancel}
\usepackage{caption}
\usepackage{cleveref}

\newcommand\figref[1]{Figure \ref{fig:#1}} 
\newcommand\tabref[1]{Table \ref{tab:#1}} 
\newcommand\eref[1]{Eq. (\ref{eq:#1})} 
\newcommand\p[1]{\partial{#1}}

\begin{document}

\begin{frontmatter}

\title{Projection-based reduced order modeling and data-driven artificial viscosity closures for incompressible fluid flows}

\author[CMU]{Aviral Prakash \corref{cor1}}
\ead{aviralp@andrew.cmu.edu}
\author[CMU]{Yongjie Jessica Zhang}
\cortext[cor1]{Corresponding author}

\address[CMU]{Department of Mechanical Engineering, Carnegie Mellon University, Pittsburgh, PA 15213, USA}
\begin{abstract}
    
    Projection-based reduced order models rely on offline-online model decomposition, where the data-based energetic spatial basis is used in the expensive offline stage to obtain equations of reduced states that evolve in time during the inexpensive online stage. The online stage requires a solution method for the dynamic evolution of the coupled system of pressure and velocity states for incompressible fluid flows. The first contribution of this article is to demonstrate the applicability of the incremental pressure correction scheme for the dynamic evolution of pressure and velocity states. The evolution of a large number of these reduced states in the online stage can be expensive. In contrast, the accuracy significantly decreases if only a few reduced states are considered while not accounting for the interactions between unresolved and resolved states. The second contribution of this article is to compare three closure model forms based on global, modal and tensor artificial viscosity approximation to account for these interactions. The unknown model parameters are determined using two calibration techniques: least squares minimization of error in energy approximation and closure term approximation. This article demonstrates that an appropriate selection of solution methods and data-driven artificial viscosity closure models is essential for consistently accurate dynamics forecasting of incompressible fluid flows.

\end{abstract}


\end{frontmatter}

\section{Introduction}

Over the past decade, improvements in computational hardware and advancements in physical simulations have pushed the envelope of complexity of scientific applications that can be modeled with adequate accuracy. High-fidelity simulations of many challenging applications demand extensive computational resources and long computational time. The expensive nature of these simulations prohibits their use in the context of multi-query applications, real-time simulation and dynamics forecasting. In such situations, reduced order models (ROMs) are an attractive alternative as they can potentially simulate engineering systems at a lower computational overhead without a significant loss in accuracy. ROMs often rely on offline-online task decomposition, where the computationally expensive component is ideally performed in the offline stage and typically carried out only once. In contrast, the inexpensive online stage is performed when the model is deployed in the actual application. 

Projection-based ROMs \cite{Sirovich1987, Aubry1988} are the most popular among model reduction approaches. These models rely on high-fidelity data to obtain spatial basis, using modal decomposition methods such as proper orthogonal decomposition (POD) \cite{Lumley1967, Lumley1981} or reduced basis \cite{Hesthaven2016}, and optimally representing the data in terms of reduced states. Using this basis representation in the relevant physical equation and projecting the solution to a lower-dimensional subspace, a set of equations for the evolution of reduced states is obtained. The reduced states are classified as resolved states, which are the most energetic states whose dynamics are predicted, and unresolved states, which are less energetic states whose dynamics are neglected. These equations can be solved in the online stage of ROMs to get the dynamic evolution of the solution. The cost of the online stage can scale highly with the number of resolved states; for example, in the case of Navier-stokes equations, the cost can scale as $O(r^3)$ where $r$ is the number of resolved states. A lower number of resolved states must be considered for such applications. This selection comes at the cost of more unresolved states, which negatively influences the dynamics of the resolved states. Several approaches have been considered in ROM literature to account for the interactions between resolved and unresolved states either directly \cite{Aubry1988, Sirisup2004, Wang2012} or indirectly \cite{Bergmann2009, Carlberg2011}. Closure models are direct approaches that add additional terms to the dynamic evolution of the resolved states to account for their interaction with unresolved states. A comprehensive summary of several approaches under this category is given in \cite{Ahmed2021}. Common indirect approaches include stabilization methods \cite{Bergmann2009, Carlberg2011, Lee2017, Parish2020, Parish2023} that typically use Petrov-Galerkin projection, which implicitly adds terms to the dynamic evolution of reduced states to account for the interaction of unresolved states. Some of these Petrov-Galerkin projection methods can also be considered a variational multiscale method \citep{Bergmann2009, Iliescu2014} and categorized as closure models. These Petrov-Galerkin projection ROMs often represent the dynamics of resolved states based on the optimality conditions \cite{Lee2017, Parish2020}. 

In this article, we focus on the closure models to account for the influence of unresolved states on the resolved states. These models can be broadly classified into physics-driven and data-driven models. Physics-driven models \cite{Aubry1988, Rempfer1994, Couplet2003} typically use physical insight to add terms to equations of the resolved states to account for their interaction with unresolved states. For example, artificial/eddy viscosity model forms \cite{Aubry1988, Couplet2003, Borggard2011, Wang2012} are based on the assumption that unresolved states are primarily responsible for dissipating energy from the system. Augmented versions of these models based on modal artificial viscosity enable energy dissipation characteristics to vary for individual resolved states \cite{Rempfer1994, Couplet2003, Sirisup2004} and provide more accurate dynamics of resolved states \cite{Osth2014}. Alternate models based on the theory developed in the context of large eddy simulation (LES) \citep{Xie2017}, variational multiscale method \citep{Iliescu2014, Reyes2020} and regularization \citep{Xie2018b} have also been developed. The model parameters in these models are typically obtained using several calibration techniques like manual tuning, statistical averages \citep{Couplet2003, Osth2014}, data-assimilation \citep{Cordier2013, Protas2015, Ahmed2020} and least squares minimization \citep{Couplet2003}. With the growing interest in modern machine learning algorithms, data-driven models for closure modeling in ROMs \cite{Xie2018, Ahmed2021} and LES \citep{Duraisamy2019, Prakash2022, Prakash2023a, Prakash2023c} are becoming popular. These data-driven models rely on raw data to determine the approximate operators or correction to the operators in the resulting dynamic system. Often, these models assume a model form, such as models with a linear term \citep{Koc2022}, models with both linear and quadratic terms \cite{Xie2018, Mohebujjaman2019} and neural network model forms \cite{San2018}. As the development and deployment of ROMs are typically in a data-rich environment, data-driven closure models can directly leverage the data used for constructing the energetic spatial basis. These models are generally ill-posed and require regularization to work well \cite{Cordier2010, Xie2018}. The selection of an appropriate regularization scheme and parameters depends on the application scenario and model selection, thereby adding an expensive model-tuning process to obtain efficient ROMs. 

In addition to the closure problem, ROMs of incompressible fluid flows require unique solution methods for coupled continuity and momentum equations in the incompressible Navier-Stokes equations. Most ROMs for incompressible flows neglect the role of pressure gradient in the momentum equations by assuming the discretely divergence free nature of the velocity data. However, in most situations, this assumption is not satisfied. For example, data from commonly used numerical methods, like finite element methods with Lagrange elements, is also not discretely divergence free. Furthermore, obtaining discretely divergence free velocities may not be possible with common experimental measurement techniques. ROMs of shear layers are sensitive to the approximation of pressure gradient; thereby, neglecting this term can reduce ROM accuracy \cite{Noack2005}. Accurate dynamics prediction of pressure states is also essential for several applications such as fluid-structure interaction \citep{Caiazzo2014}. Therefore, ROMs that avoid such assumptions and accurately solve coupled equations of velocity and pressure reduced states are essential. In these coupled equations, the role of the pressure states is to enforce the incompressibility condition. Improper solution methods can result in spurious pressure modes or pressure instabilities and several studies have focused on mitigating them \citep{Caiazzo2014, Ballarin2015, Decaria2020, Robino2020}.

Based on the discussion above, an ideal ROM for incompressible flows needs to have two key features: 1) an accurate and stable solution method for the dynamic evolution of velocity and pressure states and 2) an adequate closure model to account for interactions between unresolved and resolved states while reducing the costly human-in-loop involvement of parameter tuning, such as parameter tuning while adding regularization or manually tuning parameters for physics-driven models. Inspired by these challenges, the main contributions of this article are two-fold:
\begin{itemize}
    \item The incremental pressure correction scheme to solve coupled equations of velocity and pressure states while providing a second-order accurate temporal evolution. 
    \item New artificial viscosity closure models with unknown parameters determined using two data-driven calibration techniques: least squares minimization of error in energy approximation and error in closure term approximation. 
\end{itemize}
The first contribution is to demonstrate the applicability of incremental pressure correction schemes through numerical experiments. This solution method has been commonly used in FOM \citep{Guermond2006}; however, to the extent of the authors knowledge, this solution method has not been used previously in the context of ROMs. This solution method gives us velocity and pressure states without requiring the pressure and velocity states to satisfy \textit{inf-sup} condition, thereby avoiding the need for a more expensive supremizer enrichment method \cite{Ballarin2015}. The second contribution is to demonstrate the applicability of several artificial viscosity closure models for resolving the closure problem in ROMs. In addition to the two physics-driven model forms discussed above, which are global \cite{Aubry1988} and modal \cite{Rempfer1994, Couplet2003} artificial viscosity model forms, we also propose a tensor artificial viscosity model form. This model form is inspired by the concept of tensor eddy viscosity subgrid models for LES \cite{Prakash2023b, Abba2023}. The unknown parameters for these three model forms are learned using the least squares minimization technique, typically used for data-driven closure models. In this article, we compare six closure models obtained by combining each of these three physics-driven model forms with two data-driven calibration techniques for determining unknown model parameters. These closure models lie between physics-driven and data-driven models and we refer to them as data-driven artificial viscosity closure models. Unlike the data-driven closure models that approximate the quadratic operator \citep{Xie2018, Mohebujjaman2019}, these closure models did not require additional regularization. Together with the proposed solution method for incompressible flows, we validate and compare the proposed six closure models for forecasting the flow dynamics over a 2-D cylinder at three Reynolds numbers.

The outline of this article is given below. Section 2 discusses the incompressible Navier-Stokes equations. Section 3 discusses POD for dimensionality reduction. Section 4 mentions the construction of projection-based ROMs with a POD basis while highlighting the closure problem and several models proposed in this work to overcome it. This section also discusses the incremental pressure correction scheme for solving coupled equations for resolved velocity and pressure states. Section 5 validates the solution method for ROMs and several closure models for an incompressible fluid flow case: flow over a 2-D cylinder. Section 6 concludes this article by summarizing its main contributions and highlighting possible directions for future research. 

\section{Incompressible Navier-Stokes Equations}
\label{sec:NSE}
Incompressible Navier-Stokes equations are commonly solved to obtain discrete approximations, denoted by superscript $h$, of velocity, $\bm{u}^h : [0, T] \times \Omega \to \mathbb{R}$ and pressure, $p^h : [0, T] \times \Omega \to \mathbb{R}$, in an incompressible flow in a domain $\Omega \subset \mathbb{R}^d$ of dimensionality $d$ at time $T \in \mathbb{R}^+$. The approximate velocity and pressure lie in respective function spaces, $\bm{u}^h \in \mathcal{V}^h$ and $p^h \in \mathcal{Q}^h$, which are appropriately chosen to represent the flow physics. These equations consist of the continuity equation
\begin{equation}
    \nabla \cdot \bm{u}^h = 0,
    \label{eq:NSE_cont}
\end{equation}
\noindent which ensures the conservation of mass, and the momentum equation,
\begin{equation}
    \frac{\partial \bm{u}^h}{\p t} + \bm{u}^h \cdot \nabla \bm{u}^h = -\nabla p^h + \nu \nabla^2 \bm{u}^h,
    \label{eq:NSE_mom}    
\end{equation}
\noindent which ensures the conservation of flow momentum. From here onwards, we will remove superscript $h$ and use $\bm{u}$ and $p$ to refer to discrete approximations of velocity and pressure, respectively. The solution to incompressible Navier-Stokes equations exhibits a saddle-point structure, which is accompanied by the role of the pressure to enforce the incompressibility constraint. For this scenario, the equation that governs the dynamic evolution of pressure is unavailable, unlike compressible Navier-Stokes equations, where such an equation can be obtained. The saddle point nature of these equations often restricts selecting appropriate function spaces for velocity and pressure. For such saddle point problems, the Ladyzhenskaya–Babuška–Brezzi (LBB) or \textit{inf-sup} condition \cite{Babuska1973, Brezzi1974} is a sufficient condition to ensure the uniqueness of the solution. 


\section{Proper Orthogonal Decomposition}
\label{sec:POD}

An effective ROM relies on constructing the basis and determining reduced states that can effectively represent the flow dynamics. POD is one of the most common methods for dimensionality reduction and obtaining an effective spatial basis for the dynamic system \cite{Lumley1967, Lumley1981}. Velocity and pressure fields are decomposed using POD as,
\begin{equation}
    \bm{u} (\bm{x},t) \approx \bm{u}_m (\bm{x}) + \sum_{i=1}^n a_i (t) \bm{\phi}_i (\bm{x}) \qquad \text{and} \quad p (\bm{x},t) \approx p_m (\bm{x}) + \sum_{i=1}^n b_i (t) \psi_i (\bm{x}),
    \label{eq:PODvelandpres}
\end{equation}
\noindent where $n$ is the number of modes/states, $a_i : [0, T] \to \mathbb{R}$ and $b_i : [0, T] \to \mathbb{R}$ are the temporally evolving reduced states of velocity and pressure respectively, $\bm{\phi}_i: \Omega \to \mathbb{R}$ and $\psi_i: \Omega \to \mathbb{R}$ are spatial modes of velocity and pressure respectively and $\bm{u}_m$ and $p_m$ are the reference velocity and pressure respectively which are taken to be respective mean values in this article. When finite element methods are used to solve PDEs, the spatial modes are unknown and determined by assembling a large system of equations and solving them. Meanwhile, ROMs use high-fidelity simulation or experimental data as spatial modes to reduce the dimensionality of the problem and study the temporal evolution of these reduced states. POD basis optimally represents the energy of incompressible flows: $(\bm{u},\bm{u})/2$. Through numerical experiments, we observed that the coupled approach, where velocity and pressure POD basis are computed together \cite{Bergmann2009}, did not result in a monotonic decrease in energy error with increased modes. Therefore, this article follows a decoupled approach where velocity and pressure POD basis are computed separately \cite{Caiazzo2014}. 

For generating velocity and pressure POD basis, the velocity data, $\bm{X}_u (\bm{x})$, and pressure data, $\bm{X}_p (\bm{x})$, at each spatial coordinate location, $\bm{x} \in \mathbb{R}^d$, are represented as,
\begin{equation}
    \bm{X}_u (\bm{x}) := [\bm{u}^{1} (\bm{x}) \; \bm{u}^{2} (\bm{x})\; \cdot \cdot \cdot \; \bm{u}^{n} (\bm{x})] \qquad \text{and} \qquad    \bm{X}_p (\bm{x}) := [p^{1} (\bm{x})\; p^{2} (\bm{x})\; \cdot \cdot \cdot \; p^{n} (\bm{x})], 
\end{equation}
\noindent where $\bm{u}^i (\bm{x}) = \bm{u}(\bm{x},t_i) - \bm{u}_m$ and $p^i (\bm{x}) = p (\bm{x}, t_i) - p_m$ for $t_i \in \{t_1, t_2, \cdot \cdot \cdot, \; t_n \}$. The selection of the POD velocity modes is based on minimizing the approximation error,
\begin{equation}
    \bm{\Phi} = \underset{\hat{\bm{\Phi}} = [\bm{\phi}_1,\bm{\phi}_2,...,\bm{\phi}_{n}]}{\text{arg min}} \sum_{i=1}^{n} \Big\vert \Big\vert \bm{u}^{i} (\bm{x}) - \sum_{i=1}^n a_i (t_i) \bm{\phi}_i (\bm{x}) \Big\vert \Big\vert^2_{L^2}, 
\end{equation}
such that
\begin{equation}
    (\bm{\phi}_i, \bm{\phi}_j ) = \delta_{ij},\;  \text{for} \quad i,j = 1, 2, \cdot \cdot \cdot ,\; n \;, 
\end{equation}
\noindent where $\vert \vert \bm{v}(\bm{x}) \vert \vert_{L^2}^2 = \int_{\Omega} \bm{v}(\bm{x}) \cdot \bm{v}(\bm{x})  d \bm{x}$ is the norm introduced by the inner product $( \cdot, \cdot)$ and $\delta_{ij}$ is the Kronecker delta. Other weighed norms and inner products have been considered in the past \cite{Rowley2004, Kalashnikova2010} for better stability properties. However, these are not considered in this article. The velocity POD modes are obtained using the relation,
\begin{equation}
    \bm{\Phi} = \bm{X}_u \bm{\hat{\Phi}} \sqrt{\bm{\Lambda}_u^{-1}}.
\end{equation}
\noindent The components of $\bm{\hat{\Phi}} (= [ \bm{\hat{\phi}}_1 \; \bm{\hat{\phi}}_2 \; \cdot \cdot \cdot \; \bm{\hat{\phi}}_n ])$ and components of $\bm{\Lambda}_u$, which is a diagonal matrix ($\bm{\Lambda}_u = \text{diag}(\lambda^u_1, \lambda^u_2, \cdot \cdot \cdot, \lambda^u_n)$), are obtained by solving the eigenproblem
\begin{equation}
    \bm{R}^u \bm{\hat{\phi}}_i = \lambda^u_i \bm{\hat{\phi}}_i,
\end{equation}
\noindent where the $ij^{th}$ component of the velocity correlation matrix, $\bm{R}^u$, is given as
\begin{equation}
    R^u_{ij} =  \int_{\Omega} \bm{u}^i (\bm{x}) \cdot \bm{u}^j (\bm{x}) d \bm{x}.
\end{equation}

Similarly, the selection of POD pressure modes is based on minimizing the approximation error, 
\begin{equation}
    \Psi = \underset{\hat{\bm{\Psi}} = [\bm{\psi}_1,\bm{\psi}_2,...,\bm{\psi}_{n}]}{\text{arg min}} \sum_{i=1}^{n} \Big\vert \Big\vert p^{i} (\bm{x}) - \sum_{i=1}^n b_i (t_i) \psi_i (\bm{x}) \Big\vert \Big\vert^2_{L^2} ,
\end{equation}
such that
\begin{equation}
    (\psi_i, \psi_j ) = \delta_{ij}, \;\text{for} \quad i,j = 1, 2, \cdot \cdot \cdot ,\; n \;.
\end{equation}
\noindent The pressure POD modes are obtained using the relation,
\begin{equation}
    \bm{\Psi} = \bm{X}_p \bm{\hat{\Psi}} \sqrt{\bm{\Lambda}_p^{-1}},
\end{equation}
\noindent where components of $\bm{\hat{\Psi}} (= [ \hat{\psi}_1 \; \hat{\psi}_2 \; \cdot \cdot \cdot \;\hat{\psi}_n ])$ and components of $\bm{\Lambda_p}$, which is a diagonal matrix ($\bm{\Lambda_p} = \text{diag}(\lambda^p_1, \lambda^p_2, \cdot \cdot \cdot, \lambda^p_n)$), are obtained by solving the eigenproblem,
\begin{equation}
    \bm{R}^p \hat{\psi}_i = \lambda^p_i \hat{\psi}_i. 
\end{equation}
\noindent The $ij^{th}$ component of the pressure correlation matrix, $\bm{R}^p$, is determined using,
\begin{equation}
    R^p_{ij} =  \int_{\Omega} p^i (\bm{x}) \cdot p^j (\bm{x}) d \bm{x}.
\end{equation}
The orthonormality of velocity and pressure modes ensures that the reduced states at time $t_j$ are obtained using,
\begin{equation}
    a_i (t_j) = ( \bm{\phi}_i, \bm{u} (t_j) ) \quad \text{and} \quad b_i (t_j) = ( \psi_i, p (t_j) ).
\end{equation}
\noindent The temporal evolution of reduced states governs the dynamics of the solution. Depending on the availability of computational resources, it may be computationally unaffordable to determine the dynamics of all $n$ reduced states. Therefore, these reduced states are further decomposed as resolved (denoted by subscript $c$) and unresolved states (denoted by subscript $f$). Keeping this in mind, POD for velocity can be rewritten as,
\begin{equation}
    \bm{u} (\bm{x},t) \approx \bm{u}_c (\bm{x},t) + \bm{u}_f (\bm{x},t),
    \label{eq:PODvelThree}
\end{equation}
\noindent where
\begin{equation}
    \bm{u}_c (\bm{x},t) = \bm{u}_m (\bm{x}) + \sum_{i=1}^r a_i (t) \bm{\phi}_i (\bm{x}) \quad \text{and} \quad \bm{u}_f (\bm{x},t) = \sum_{i=r+1}^n a_i (t) \bm{\phi}_i (\bm{x}).
\end{equation}
\noindent The reduced states $a_i$ for $i = 1, 2, \cdot \cdot \cdot, \; r$ are the resolved velocity states and $a_i$ for $i = r+1, r+2, \cdot \cdot \cdot, \; n$ are the unresolved velocity states. Similarly, POD for pressure is rewritten as,
\begin{equation}
    p (\bm{x},t) \approx p_c (\bm{x},t) + p_f (\bm{x},t),
    \label{eq:PODpresThree}
\end{equation}
\noindent where
\begin{equation}
    p_c (\bm{x},t) = p_m (\bm{x}) + \sum_{i=1}^r b_i (t) \psi_i (\bm{x}) \quad \text{and} \quad p_f (\bm{x},t) = \sum_{i=r+1}^n b_i (t) \psi_i (\bm{x}).
\end{equation}
\noindent The reduced states $b_i$ for $i = 1, 2, \cdot \cdot \cdot, \; r$ are the resolved pressure states and $b_i$ for $i = r+1, r+2, \cdot \cdot \cdot, \; n$ are the unresolved pressure states. Even though different number of velocity and pressure states can be used, we use the same number of states in this article.

\section{Reduced order modeling}

As the introduction suggests, ROMs are critical for multi-query applications, such as optimization and uncertainty quantification, real-time control, and dynamics forecasting. ROMs have two main components: representation of the high-dimensional solution in terms of reduced states and dynamical evolution of these reduced states. In Section \ref{sec:POD}, we explored using POD to obtain the set of energy-dominant spatial bases and corresponding reduced states. This section focuses on the evolution of these reduced states for dynamics forecasting. Recently, there has been a growing trend towards fully data-based ROMs involving machine learning to represent the reduced states and evolve them in time \cite{Ahmed2021}. Unfortunately, these fully data-based ROMs often have limited applicability for dynamics forecasting problems, where the goal is to predict future states that are not part of the high-fidelity training data used to learn these models. This article restricts our discussion to traditional ROMs involving physical equations to evolve the reduced states. These latter ROMs are better suited for forecasting problems where the model is evaluated for future time instances that do not inform the generation of the energy-dominant basis. 

\subsection{Projection-based reduced order modeling and closure problem}

Projection-based ROMs are a model reduction approach where high-fidelity data is used to generate the spatial modes and the temporal evolution of reduced states is governed by physical equations. For incompressible flows, the reduced states can be obtained by substituting POD of velocity (\eref{PODvelThree}) and pressure (\eref{PODpresThree}) in \eref{NSE_cont} and \eref{NSE_mom} to obtain 
\begin{equation}
    \nabla \cdot (\bm{u}_c + \bm{u}_f) = 0
\end{equation}
\noindent and
\begin{equation}
    \frac{\partial (\bm{u}_c + \bm{u}_f)}{\p t} + (\bm{u}_c + \bm{u}_f) \cdot \nabla (\bm{u}_c + \bm{u}_f) = -\nabla (p_c + p_f) + \nu \nabla^2 (\bm{u}_c + \bm{u}_f).
\end{equation}
Using Galerkin projection \citep{Aubry1988}, the continuity equation is projected onto a linear subspace spanned by $\psi_k$ for $k = 1, 2, \cdot \cdot \cdot, \; r$. The resulting continuity equation is

\begin{equation}
    \Big( \psi_k, \bm{u}_m\Big) + \sum_{i=1}^r a_i \Big(\psi_k, \nabla \cdot \bm{\phi}_i \Big) + D_k^c (\bm{a}) = 0,
    \label{eq:Cont_proj}
\end{equation}

\noindent where $D_k^c (\bm{a}) = \sum_{i=r+1}^{n} a_i \Big(\psi_k, \nabla \cdot \bm{\phi}_i \Big) $ is the closure term for continuity equation.  On the other hand, the momentum equation is projected onto a linear subspace spanned by $\bm{\phi}_k$ for $k = 1, 2, \cdot \cdot \cdot, \; r$. The resulting momentum equation is

\begin{equation}
    \dot{a}_k + \sum_{i=1}^r L_{ki} a_i + \sum_{i=1}^r \sum_{j=1}^r B_{kij} a_i a_j + \sum_{i=1}^r L^P_{ki} b_i + C_k + D^m_k (\bm{a},\bm{b}) = 0, 
    \label{eq:FinalMom}
\end{equation}
\noindent where

\begin{equation}
    L_{ki} =  - \nu \Big(\bm{\phi}_k, \nabla^2 \bm{\phi}_i \Big) + \Big(\bm{\phi}_k, \bm{\phi}_i \cdot \nabla \bm{u}_m \Big) + \Big(\bm{\phi}_k, \bm{u}_m \cdot \nabla \bm{\phi}_i \Big), 
\end{equation}

\begin{equation}
    B_{kij} = \Big(\bm{\phi}_k, \bm{\phi}_i \cdot \nabla \bm{\phi}_j \Big),
\end{equation}

\begin{equation}
    L^P_{ki} = \Big(\bm{\phi}_k, \nabla \psi_i \Big),
\end{equation}

\begin{equation}
    C_k =  \Big(\bm{\phi}_k, \bm{u}_m \cdot \nabla \bm{u}_m \Big) + \Big(\bm{\phi}_k, \nabla p_m \Big) - \nu \Big(\bm{\phi}_k, \nabla^2 \bm{u}_m \Big)
\end{equation}    
\noindent and 
\begin{multline}
    D^m_k (\bm{a}, \bm{b}) =  \sum_{i=1}^r \sum_{j=r+1}^n a_i a_j \Big(\bm{\phi}_k, \bm{\phi}_i \cdot \nabla \bm{\phi}_j \Big)  +  \sum_{i=r+1}^n \sum_{j=1}^r a_i a_j \Big(\bm{\phi}_k, \bm{\phi}_i \cdot \nabla \bm{\phi}_j \Big)  \\ + \sum_{i=r+1}^n \sum_{j=r+1}^n a_i a_j \Big(\bm{\phi}_k, \bm{\phi}_i \cdot \nabla \bm{\phi}_j \Big) - \nu \sum_{i=r+1}^n a_i \Big(\bm{\phi}_k, \nabla^2 \bm{\phi}_i \Big) + \sum_{i=r+1}^{n} a_i \Big(\bm{\phi}_k, \bm{u}_m \cdot \nabla \bm{\phi}_i \Big)   \\
    + \sum_{i=r+1}^{n} a_i \Big(\bm{\phi}_k, \bm{\phi}_i \cdot \nabla \bm{u}_m \Big) + \sum_{i=r+1}^n b_i \Big( \bm{\phi}_k, \nabla \psi_i \Big)
\end{multline}
\noindent is the closure term for the momentum equation. This approach is often referred to as the variational multiscale (VMS) method for ROMs and $D^m_k$ can be thought of as the closure term that accounts for coarse and fine-scale interactions defined in VMS methods \citep{Borggard2011, Iliescu2014}. This article neglects the effect of unresolved pressure states in the closure term. With this approximation, the resulting closure modeling becomes
\begin{multline}
    D^m_k (\bm{a}, \bm{b}) \approx \bar{D}^m_k (\bm{a}) =  \sum_{i=1}^r \sum_{j=r+1}^n a_i a_j \Big(\bm{\phi}_k, \bm{\phi}_i \cdot \nabla \bm{\phi}_j \Big)  +  \sum_{i=r+1}^n \sum_{j=1}^r a_i a_j \Big(\bm{\phi}_k, \bm{\phi}_i \cdot \nabla \bm{\phi}_j \Big)  \\ + \sum_{i=r+1}^n \sum_{j=r+1}^n a_i a_j \Big(\bm{\phi}_k, \bm{\phi}_i \cdot \nabla \bm{\phi}_j \Big) - \nu \sum_{i=r+1}^n a_i \Big(\bm{\phi}_k, \nabla^2 \bm{\phi}_i \Big) + \sum_{i=r+1}^{n} a_i \Big(\bm{\phi}_k, \bm{u}_m \cdot \nabla \bm{\phi}_i \Big)   \\
    + \sum_{i=r+1}^{n} a_i \Big(\bm{\phi}_k, \bm{\phi}_i \cdot \nabla \bm{u}_m \Big).
    \label{eq:ClosureTermFinal}
\end{multline}
The incremental pressure correction scheme can implicitly account for the effect of the unresolved pressure states.

For practical near real-time prediction, it is desirable to have $r << n$, especially for non-linear equations where the cost of the non-linear terms can scale with a high exponent. For example, the cost of the online stage for the incompressible Navier-Stokes equations is $O(r^3)$, which is very expensive for large values of $r$. Unfortunately, this also implies that a cost-effective approach involves a relatively higher number of unresolved and fewer resolved states. As the dynamic evolution of unresolved states is not computed, the influence of these unresolved states on the resolved states are neglected. Therefore, if $r << n$, the accuracy of the dynamics prediction is affected, especially for strongly advective flows where the decay of Kolmogorov $n$-width is slow \cite{Greif2019}. If the influence of unresolved states is ignored by setting $a_i = 0$ and $b_i = 0$ for $i = r+1, r+2, \cdot \cdot \cdot, \;n$, the closure terms, $D^c_k$ and $D^m_k$ become zero and the equations, \eref{Cont_proj} and \eref{FinalMom} reduce to
\begin{equation}
    \Big( \psi_k, \bm{u}_m\Big) + \sum_{i=1}^r a_i \Big(\psi_k, \nabla \cdot \bm{\phi}_i \Big) = 0
\end{equation}
\noindent and 
\begin{equation}
    \dot{a}_k + \sum_{i=1}^r L_{ki} a_i + \sum_{i=1}^r \sum_{j=1}^r B_{kij} a_i a_j + \sum_{i=1}^r L^P_{ki} b_i + C_k = 0.
\end{equation}

\noindent We refer to this as the Galerkin projection-based ROMs without closure modeling. Suppose the influence of unresolved states on the dynamics of resolved states is significant. In that case, the accuracy and stability of the predicted resolved states can be severely affected in these ROMs. Therefore, accurate models that approximate $D^m_k$ without incurring a significant evaluation cost are essential. For incompressible flows, $D^c_k$ may not severely affect the dynamics of $a_i$ and $b_i$ and is set to zero in this article.


\subsection{Closure models for ROMs}

As discussed in the introduction, several closure models have been proposed over the years that account for the interaction of the unresolved states with the resolved states. These models can be classified as physics-driven or data-driven. Physics-driven models involve analysis of equations and physical insights to determine a model form and calibrate the model parameters \citep{Aubry1988, Sirisup2004, Wang2012, Osth2014}. These model forms typically have fewer parameters to ease the process of calibrating them. Most common physics-driven models involve the assumption of a model form as a linear term in \eref{FinalMom}, that is
\begin{equation}
    \bar{D}^m_k (\bm{a}) \approx D^{m-model}_k (\bm{a}) = \sum_{i=1}^r a_i \tilde{L}_{ki}, 
    \label{eq:EddyVisc}
\end{equation}
\noindent where $\tilde{L}_{ki}$ is the unknown closure term. Three model forms can approximate $\tilde{L}_{ki}$. The first model form, which we refer to as the global artificial viscosity, approximates $\tilde{L}_{ki}$ as
\begin{equation}
    \tilde{L}_{ki} = \nu^t \Big(\bm{\phi}_k, \nabla^2 \bm{\phi}_i \Big),
    \label{eq:EddyVisc1}
\end{equation}
\noindent where $\nu^t$ is the unknown global artificial viscosity. The second model form, which we refer to as the modal artificial viscosity, approximates $\tilde{L}_{ki}$ as 
\begin{equation}
    \tilde{L}_{ki} = \nu_k^t \Big(\bm{\phi}_k, \nabla^2 \bm{\phi}_i \Big),
    \label{eq:EddyVisc2}
\end{equation}
\noindent where $\nu_k^t$ is the unknown modal artificial viscosity. These two model forms are common in the literature \cite{Aubry1988, Rempfer1994, Osth2014}. In addition to these model forms, we propose a third model form, which we refer to as the tensor artificial viscosity, that approximates $\tilde{L}_{ki}$ as
\begin{equation}
    \tilde{L}_{ki} = \nu_{ik}^t \Big(\bm{\phi}_k, \nabla^2 \bm{\phi}_i \Big).
    \label{eq:EddyVisc3}
\end{equation}
where $\nu^t_{ik}$ is the $ik^{th}$ component of the unknown tensor artificial viscosity. Several strategies have been used to determine the unknown parameters in \eref{EddyVisc1} and \eref{EddyVisc2}. The mean energy balance was used in \cite{Osth2014} to obtain $\nu^t$. On the other hand, the mean of modal energy was used in \cite{Noack2005, Osth2014} to obtain $\nu^t_k$. Data-assimilation techniques were used in \citep{Cordier2013, Protas2015} to obtain the relevant unknown closure term parameters. Recently, data-driven models that use high-fidelity simulation data and machine learning techniques have been proposed. These models approximate $D_k^m$ using linear \cite{Koc2022} or quadratic operators \citep{Xie2018, Mohebujjaman2019} instead of assuming a physical model form like artificial viscosity-based closure models. As projection-based ROMs are data-intensive and rely on data to obtain the energy-dominant spatial modes, substantial data needed for determining unknown parameters in closure models is readily available.

Data-driven models \citep{Xie2018, Mohebujjaman2019, Koc2022} often used least squares minimization to obtain the unknown parameters in the model form. However, there is limited work on utilizing this calibration technique for determining the unknown parameters in physics-driven closure models \cite{Couplet2003}. Even in \cite{Couplet2003}, only the modal artificial viscosity was determined using least squares minimization, but not used as a closure model in ROMs. Therefore, one of the main contributions of this article is to provide a systematic feasibility study on using least squares minimization for learning unknown parameters in three artificial viscosity closure model forms. The least squares minimization for obtaining unknown parameters can be applied in two ways. The first and more common calibration technique is to approximate the closure term, $\bar{D}^m_k$ in \eref{ClosureTermFinal}, as accurately as possible. We refer to this technique as the closure term approximation. This technique is typically followed in the data-driven closure model literature. 
\cite{Xie2018, Mohebujjaman2019, Koc2022}.

The alternate calibration technique accurately captures the energetic interactions between resolved and unresolved states. We refer to this technique as the energy approximation. The influence of closure terms on the energy equation can be determined by examining the energy equation. This equation can be derived from the Navier-Stokes equations by taking the inner product of \eref{NSE_mom} with velocity to obtain 
\begin{equation}
    \frac{\partial }{\p t} ( \bm{u}, \bm{u} ) + \Big( \bm{u}, \bm{u} \cdot \nabla \bm{u} \Big) + \Big( \bm{u}, \nabla p \Big) + \nu \Big( \bm{u}, \nabla^2 \bm{u} \Big) = 0.
\end{equation}
Substituting POD for velocity and pressure in this equation and simplifying it, we get
\begin{multline}
     \sum_{k=1}^r \frac{\partial a_k^2}{\p t} + \sum_{i=1}^r \sum_{k=1}^r a_i a_k L_{ki} + \sum_{i=1}^r \sum_{j=1}^r \sum_{k=1}^r a_i a_j a_k B_{kij} +  \\ \sum_{i=1}^r \sum_{k=1}^r b_i a_k L^p_{ki} + \sum_{k=1}^r a_k (C_k + \bar{D}^m_k (\bm{a})) = 0.
     \label{eq:EnergyGlob}
\end{multline}
\noindent The energetic interactions of the unresolved states and their contribution towards the energy of all resolved states are accounted for in the last term: $\sum_{i=1}^r a_k D^m_k$. Each component of this term, $a_k D^m_k$, accounts for energetic interactions of unresolved states to the energy of the resolved state $a_k$. These interactions comprise dyadic energetic interactions due to linear terms and triadic energetic interactions due to non-linear terms. Similarly, the energy equation for the $k^{th}$ resolved state is
\begin{equation}
    \frac{\partial a_k^2}{\p t} + a_k \sum_{i=1}^r a_i L_{ki} + a_k \sum_{i=1}^r \sum_{j=1}^r a_i a_j B_{kij} +  \\ a_k \sum_{i=1}^r b_i L^p_{ki} + a_k C_k + a_k \bar{D}^m_k (\bm{a}) = 0.
     \label{eq:Energymodel}
\end{equation}
The modal energy equation determines the temporal evolution of energy for each resolved state. Based on these two equations, the energy approximation calibration technique for determining unknown parameters can be further classified as global and modal energy approximation techniques. The global energy approximation aims to ensure that the global energy balance (in \eref{EnergyGlob}) is accurate. Therefore, this technique estimates unknown model parameters such that $\sum_{i=1}^r a_k \bar{D}^m_k$ is accurately calculated. On the other hand, the modal energy approximation involves ensuring that the modal energy balance is accurate. Therefore, this technique estimates unknown model parameters such that $a_k \bar{D}^m_k$ is accurately calculated. Determining parameters using global energy approximation is appealing for a global artificial viscosity approximation. Meanwhile, determining parameters using modal energy approximation is a better-posed problem for modal and tensor artificial viscosity models. 

\begin{table}[t!]
    \centering
    \caption{List of the six closure models considered in this article.}    
    \begin{tabular}{|c|c|c|}
         \hline
         \textbf{Closure model} & \textbf{Model form} & \textbf{Calibration technique} \\
         \hline
         & & \\[-2.5ex]
         GV-GE & Global artificial viscosity: \eref{EddyVisc1} & Global energy: $\sum_{i=1}^r a_k \bar{D}^m_k$\\
         MV-ME & Modal artificial viscosity: \eref{EddyVisc2} & Modal energy: $a_k \bar{D}^m_k$\\
         TV-ME & Tensor artificial viscosity: \eref{EddyVisc3} & Modal energy: $ a_k \bar{D}^m_k$\\
         \hline
         & & \\[-2.5ex]         
         GV-C & Global artificial viscosity: \eref{EddyVisc1} &  \\
         MV-C & Modal artificial viscosity: \eref{EddyVisc2} & Closure term: $\bar{D}^m_k$, \eref{ClosureTermFinal}\\
         TV-C & Tensor artificial viscosity: \eref{EddyVisc3} &  \\
         \hline
    \end{tabular}
    \label{tab:ModelTable}
\end{table}

The two alternate pathways for calibrating closure models, accurate closure term and accurate energy contribution, will yield different model parameters, and it is important to study which calibration technique is better suited. Combining three closure model forms, namely global, modal, and tensor eddy artificial viscosity model forms, with these two least squares minimization-based calibration techniques, namely minimization of error in energy approximation and closure term approximation, results in six possible combinations of data-driven artificial viscosity closure models as shown in \tabref{ModelTable}.

\subsubsection{GV-GE model}

GV-GE model assumes the model form in \eref{EddyVisc1}, where the optimal value of the global artificial viscosity ($\nu^t$) ensures that the contribution from the closure term on global energy, $\sum_{k=1}^r a_k \bar{D}^m_k (\bm{a})$, is accurately estimated. Mathematically, this model is posed as
\begin{equation}
    \nu^t = \underset{\hat{\nu}}{\text{arg min}} \; \Big\vert \Big\vert \hat{\nu} \sum_{i=1}^r \sum_{k=1}^r a_i a_k \Big( \bm{\phi}_k, \nabla^2 \bm{\phi}_i \Big) - \sum_{k=1}^r a_k \bar{D}^m_k (\bm{a})\Big\vert \Big\vert^2.
\end{equation}

\noindent The same data used for obtaining POD modes is used to obtain projected reduced states, $a_i (t)$, at time $t = t_l$ where $l = 1, 2, \cdot \cdot \cdot, n$. Using this data, this problem transforms to

\begin{equation}
    \nu^t = \underset{\hat{\nu}}{\text{arg min}} \; \Big\vert \Big\vert \hat{\nu} \bm{z} - \bm{b} \Big\vert \Big\vert_2^2,
    \label{eq:nu_argmin_global}    
\end{equation}
\noindent where the $l^{th}$ component of the vector $\bm{z}$ is
\begin{equation}
    z_l = \sum_{i=1}^r \sum_{k=1}^r a_i (t_l) a_k (t_l) \Big( \bm{\phi}_k, \nabla^2 \bm{\phi}_i \Big)
\end{equation}
and the $l^{th}$ component of the vector $\bm{b}$ is
\begin{equation}
    b_l =  \sum_{k=1}^r a_k (t_l) \bar{D}^m_k (\bm{a}(t_l)).
\end{equation}

\noindent The solution of this regression problem yields the following expression for artificial viscosity, 
\begin{equation}
    \nu^t = \Big( \bm{z}^T \bm{z} \Big)^{-1} \bm{z}^T \bm{b}.
    \label{eq:EddySolGlobal}
\end{equation}
For this model, we obtain a single artificial velocity that best approximates the energy at all sampled timesteps. If the flow is not stationary, the optimal global viscosity may not be the best approximation of the energetic interactions at different time instances. Therefore, this closure model is more suited for stationary flows. Despite the model form appearing very similar to the model in \citep{Osth2014}, this model differs in the calibration technique to determine the value of $\nu^{t}$. In \citep{Osth2014}, the mean value of global energy is used to determine $\nu^{t}$, whereas raw data of global energy with least squares minimization is used in this article. 

\subsubsection{MV-ME model}

The MV-ME model assumes the model form in \eref{EddyVisc2}, where the optimal value of the modal artificial viscosity ($\nu^t_k$) ensures that the contribution of closure term on modal energy, $a_k \bar{D}^m_k (\bm{a})$, is accurately estimated. Mathematically, this model is posed as
\begin{equation}
    \nu_k^t = \underset{\hat{\nu}}{\text{arg min}} \; \Big\vert \Big\vert \hat{\nu} a_k \sum_{i=1}^r a_i \Big( \bm{\phi}_k, \nabla^2 \bm{\phi}_i \Big) - a_k \bar{D}^m_k (\bm{a}) \Big\vert \Big\vert^2.
\end{equation}
\noindent Following the procedure for other models, we use the sampled data. This problem transforms to
\begin{equation}
    \nu_k^t = \underset{\hat{\nu}}{\text{arg min}} \; \Big\vert \Big\vert \hat{\nu} \bm{z} - \bm{b} \Big\vert \Big\vert_2^2,
    \label{eq:nu_argmin_nodal}    
\end{equation}
\noindent where the $l^{th}$ component of the vector $\bm{z}$ is
\begin{equation}
    z_l = a_k (t_l) \sum_{i=1}^r a_i (t_l) \Big( \bm{\phi}_k, \nabla^2 \bm{\phi}_i \Big)
\end{equation}
and the $l^{th}$ component of the vector $\bm{b}$ is
\begin{equation}
    b_l =  a_k (t_l) \bar{D}_k^m (\bm{a} (t_l)).
    \label{eq:bRHS_tl}
\end{equation}

\noindent The solution of this regression problem yields the following expression for artificial viscosity, 
\begin{equation}
    \nu_k^t = \Big( \bm{z}^T \bm{z} \Big)^{-1} \bm{z}^T \bm{b}.
    \label{eq:EddySolModal}
\end{equation}
\noindent for the $k^{th}$ mode. This model can better approximate the dynamic interaction between unresolved and resolved states than the GV-GE model, as it is more expressive. Although the model form appears very similar to the model in \cite{Osth2014}, this model differs in the calibration technique used to determine the $\nu^{t}_k$ value. In \citep{Osth2014}, the mean value of modal energy is used to determine $\nu^{t}_k$, whereas least squares minimization with raw data of modal energy is directly used in this article. This model is similar to the one mentioned in \citep{Couplet2003} where the least squares minimization is used to obtain $\nu_k^t$. However, the applicability of this closure model for ROMs was not demonstrated in \citep{Couplet2003}.

\subsubsection{TV-ME model}

The TV-ME model assumes the model form in \eref{EddyVisc3},
where the optimal value of the tensor artificial viscosity ($\nu^t_{ki}$) is obtained to ensure that the contribution of closure term on modal energy, $a_k \bar{D}^m_k (\bm{a})$, is accurately estimated. Mathematically, this model is posed as
\begin{equation}
    \bm{\nu}_k^t = \underset{\hat{\bm{\nu}} = [\hat{\nu}_1, \; \hat{\nu}_2, \cdot \cdot \cdot, \; \hat{\nu}_r ]}{\text{arg min}} \; \Big\vert \Big\vert a_k \sum_{i=1}^r \hat{\nu}_i a_i \Big( \bm{\phi}_k, \nabla^2 \bm{\phi}_i \Big) - a_k \bar{D}^m_k (\bm{a}) \Big) \Big\vert \Big\vert^2   
\end{equation}
\noindent with $\nu^t_{ki}$ as the $i^{th}$ element of vector $\bm{\nu}^t_k$. Following the procedure for other models, we use the sampled data. This problem transforms to
\begin{equation}
    \bm{\nu}_k^t = \underset{\hat{\bm{\nu}}}{\text{arg min}} \; \Big\vert \Big\vert \hat{\bm{\nu}} \bm{Z} - \bm{b} \Big\vert \Big\vert_2^2,
    \label{eq:nu_argmin_tensor}     
\end{equation}
\noindent where the ${li}^{th}$ component of the matrix $\bm{Z}$ is
\begin{equation}
    Z_{li} = a_k (t_l) a_i (t_l) \Big( \bm{\phi}_k, \nabla^2 \bm{\phi}_i \Big)
\end{equation}
and the $l^{th}$ component of the vector $\bm{b}$ is given in \eref{bRHS_tl}. The solution of this regression problem yields the following expression for vector-valued artificial viscosity, 
\begin{equation}
    \bm{\nu}_k^t = \Big( \bm{Z}^T \bm{Z} \Big)^{-1} \bm{Z}^T \bm{b}.
    \label{eq:EddySolTensor}
\end{equation}
\noindent for the $k^{th}$ mode. This model can better approximate the dynamic interaction between unresolved and resolved states than GV-GE and MV-ME models as it is more expressive.

\subsubsection{GV-C model}

The GV-C model assumes the model form in \eref{EddyVisc1},
where the optimal value of the artificial viscosity ($\nu^t$) ensures that the closure term, $\bar{D}^m_k$, is accurately estimated. Mathematically, this model is posed as
\begin{equation}
    \nu^t = \underset{\hat{\nu}}{\text{arg min}} \; \Big\vert \Big\vert \hat{\nu} \sum_{i=1}^r a_i \Big( \bm{\phi}_k, \nabla^2 \bm{\phi}_i \Big) - \bar{D}_k^m (\bm{a}) \Big\vert \Big\vert^2,
\end{equation}
\noindent where data from all $r$ modes is used to learn the artificial viscosity. Following the procedure for other models, we use the sampled data. This problem transforms to \eref{nu_argmin_global}, where the $o^{th}$ component of the vector $\bm{z}$ is
\begin{equation}
    z_o = \sum_{i=1}^r a_i (t_l) \Big( \bm{\phi}_k, \nabla^2 \bm{\phi}_i \Big)
\end{equation}
and the $o^{th}$ component of the vector $\bm{b}$ is
\begin{equation}
    b_o =  \bar{D}^m_k (\bm{a} (t_l)) 
\end{equation}
\noindent where $o$ is a unique combination of $k$ and $l$. The solution to this regression problem, shown in \eref{EddySolGlobal}, provides the value of $\nu^t$. For this model, we identify a single artificial velocity that best approximates the closure term for all the modes at every time step. Therefore, if the flow is not stationary, the optimal global viscosity may not best approximate the closure term at different time instances.

\subsubsection{MV-C model}

The MV-C model assumes the model form in \eref{EddyVisc2}
where the optimal value of the modal artificial viscosity ($\nu_k^t$) ensures that the closure term, $\bar{D}^m_k$, is accurately estimated. Mathematically, this model is posed as
\begin{equation}
    \nu_k^t = \underset{\hat{\nu}}{\text{arg min}} \; \Big\vert \Big\vert \hat{\nu} \sum_{i=1}^r a_i \Big( \bm{\phi}_k, \nabla^2 \bm{\phi}_i \Big) - \bar{D}_k^m (\bm{a}) \Big\vert \Big\vert^2.
\end{equation}
\noindent Following the procedure for other models, we use the sampled data. This problem transforms to \eref{nu_argmin_nodal}, where the $l^{th}$ component of the vector $\bm{z}$ is
\begin{equation}
    z_l = \sum_{i=1}^r a_i (t_l) \Big( \bm{\phi}_k, \nabla^2 \bm{\phi}_i \Big)
\end{equation}
and the $l^{th}$ component of the vector $\bm{b}$ is
\begin{equation}
    b_l =  \bar{D}^m_k (\bm{a} (t_l)).
    \label{eq:bref_tl_Cterm}
\end{equation}
\noindent The solution to this regression problem, shown in \eref{EddySolModal}, provides the value of $\nu_k^t$. This model better approximates the dynamic interaction between unresolved and resolved states than the GV-C model, as it is more expressive. This model is similar to the one mentioned in \citep{Couplet2003}, although the applicability of the closure model for ROM was not demonstrated in that article.

\subsubsection{TV-C model}

The TV-C model assumes the model form in \eref{EddyVisc3}, where the optimal value of the artificial viscosity ($\nu_k^t$) ensures that the closure term, $\bar{D}^m_k$, is accurately estimated. Mathematically, this model is posed as
\begin{equation}
    \bm{\nu}_k^t = \underset{\hat{\bm{\nu}}=[\hat{\nu}_1, \; \hat{\nu}_2, \; \cdot \cdot \cdot, \; \hat{\nu}_r]}{\text{arg min}} \; \Big\vert \Big\vert \sum_{i=1}^r \hat{\nu}_i a_i \Big( \bm{\phi}_k, \nabla^2 \bm{\phi}_i \Big) - \bar{D}_k^m (\bm{a}) \Big\vert \Big\vert^2
\end{equation} 
\noindent with $\nu^t_{ki}$ as the $i^{th}$ element of vector $\bm{\nu}^t_k$. Following the procedure for other models, we use the sampled data. This problem transforms to
\eref{nu_argmin_tensor}, where the $li^{th}$ component of the matrix $\bm{Z}$ is
\begin{equation}
    Z_{li} = a_i (t_l) \Big( \bm{\phi}_k, \nabla^2 \bm{\phi}_i \Big)
\end{equation}
and the $l^{th}$ component of the vector $\bm{b}$ is given in \eref{bref_tl_Cterm}. The solution to this regression problem, shown in \eref{EddySolTensor}, provides the value of $\bm{\nu}_k^t$. This model better approximates the dynamic interaction between unresolved and resolved states than GV-C and MV-C models as it is more expressive.

\subsubsection{Differences compared to other closure modeling techniques}

As discussed earlier, several other closure models exist for ROMs. Physics-driven models are better posed as the model form is determined using the physical relationship between variables; for example, global and modal artificial viscosity models have been demonstrated to tackle the closure problem adequately \cite{Borggard2011, Osth2014, Ahmed2021}. In contrast to the conventional physics-driven models like \eref{EddyVisc1} and \eref{EddyVisc2}, the tensor artificial viscosity model form in \eref{EddyVisc3} is more expressive and can potentially give better accuracy. On the other hand, data-driven closure models are naturally suited for ROMs due to their data-intensive nature. However, common data-driven closure models that rely on least squares regression for closure modeling require regularization to remedy the ill-conditioning in the model learning problem \cite{Xie2018}. The selection of the ideal regularization strategies and constants varies with the problem definitions and introduces an additional expensive tuning process. For example, the models proposed in \citep{Xie2018, Mohebujjaman2019} have more unknown parameters; thereby, they are more expressive but also suffer from ill-conditioning and require regularization. The linear operator data-driven model in \citep{Koc2022} has fewer parameters and does not require regularization for flows considered in this article. Numerical tests not shown in this article for brevity indicated that this linear operator model yielded similar results to the TV-C model and better than models with quadratic operator model form \citep{Xie2018, Mohebujjaman2019}. The closure models discussed in this article consider several physics-driven model forms discussed above and determine unknown parameters using least squares regression, which is more commonly used for data-driven closure models.



\subsection{Incremental pressure correction schemes for projection-based reduced order modeling}
\label{sec:PresProjSection}

After including the closure model, the final ROM equations to be solved are,
\begin{equation}
    \Big( \psi_k, \bm{u}_m\Big) + \sum_{i=1}^r a_i \Big(\psi_k, \nabla \cdot \bm{\phi}_i \Big) = 0
    \label{eq:FinalFinalCont}
\end{equation}
and
\begin{equation}
    \dot{a}_k + \sum_{i=1}^r L_{ki} a_i + \sum_{i=1}^r \sum_{j=1}^r B_{kij} a_i a_j + \sum_{i=1}^r L^P_{ki} b_i + C_k + \bar{D}^{m-model}_k (\bm{a}) = 0,
    \label{eq:FinalFinalMom}    
\end{equation}
\noindent where $\bar{D}^{m-model}_k$ is determined using one of the proposed closure models. The solution of these equations also exhibits a saddle-point form like those exhibited by incompressible Navier-Stokes equations \cite{Ballarin2015}. No equation dictates the evolution of $b_i$, and hence, several alternate solution methods have been proposed to solve these problems \cite{Hughes1986, Guermond2006} in the context of FOM. Most studies on ROMs for incompressible Navier-Stokes equations assume the discrete divergence-free nature of the velocity and thereby neglect the pressure gradient term in the momentum equation \cite{Caiazzo2014}. Unfortunately, most of the practical data sources, both computational and experimental, do not satisfy the discrete divergence-free condition of velocity. Furthermore, the pressure gradient term in the momentum equation significantly influences flow behavior for several fluid flows, such as confined flows \cite{Noack2005}. In such scenarios, appropriate treatment of pressure-velocity coupling is essential.

Our solution method for these equations is inspired by the pressure-projection method used to solve incompressible Navier-Stokes equations. The most commonly used solution method is the classical pressure correction scheme in \citep{Chorin1968, Temam1968}, which uncouples pressure and velocity to avoid the solution of the saddle point problem. However, this method is first order in time and our implementation required a small time step for convergence of results. Therefore, we use incremental pressure correction that is second order in time \cite{VanKan1986}. For incompressible Navier-stokes equations, this method involves two sub-steps,
\begin{equation}
    \frac{3\Tilde{\bm{u}}^{n+1} - 4\bm{u}^n + \bm{u}^{n-1}}{2\Delta t} + \tilde{\bm{u}}^{n+1} \cdot \nabla \tilde{\bm{u}}^{n+1} + \nabla p^{n} - \nu \nabla^2 \tilde{\bm{u}}^{n+1} = 0,
    \label{eq:NSE_mom_increpres_1}    
\end{equation}
\noindent
\begin{equation}
    \frac{3\bm{u}^{n+1} - 3\Tilde{\bm{u}}^{n+1}}{2\Delta t} + \nabla p^{n+1} - \nabla p^n = 0,
    \label{eq:NSE_mom_increpres_2}    
\end{equation}
\noindent and
\begin{equation}
    \nabla \cdot \bm{u}^{n+1} = 0.
    \label{eq:NSE_mom_increpres_3}       
\end{equation}

A similar solution method can be used for Galerkin projection-based ROMs by substituting \eref{PODvelandpres} in \eref{NSE_mom_increpres_1}, \eref{NSE_mom_increpres_2} and \eref{NSE_mom_increpres_3}. The resulting momentum equation, \eref{NSE_mom_increpres_1}, is projected onto the linear subspace spanned by the $\bm{\phi}_i$ for $i = 1, 2, \cdot \cdot \cdot, \; r$. We first take the divergence of \eref{NSE_mom_increpres_2} and then project the resulting equation along with the continuity equation \eref{NSE_mom_increpres_3} onto the linear subspace spanned by the $\psi_i$ for $i = 1, 2, \cdot \cdot \cdot, \; r$. The resulting equations are,

\begin{multline}
    \frac{3\tilde{a}^{n+1}_k - 4 a^{n}_k + a^{n-1}_k}{2 \Delta t} + \sum_{i=1}^r \tilde{a}^{n+1}_i L_{ki} + \sum_{i=1}^r \sum_{j=1}^r \tilde{a}^{n+1}_i \tilde{a}^{n+1}_j B_{kij} + \sum_{i=1}^r b^{n}_i L^P_{ki}  \\ + C_k + \bar{D}^{m-model}_k (\tilde{\bm{a}}^{n+1}) = 0, \quad \text{for} \quad k = 1, 2, \cdot \cdot \cdot, r,
    \label{eq:Proj_EqMom_1}
\end{multline}
\noindent
\begin{equation}    
    \sum_{i=1}^n (a^{n+1}_i  - \tilde{a}^{n+1}_i) (\psi_k, \nabla \cdot \bm{\phi}_i) - \frac{2 \Delta t}{3} \sum_{i=1}^r (b^{n+1}_i - b_i^n) (\nabla \psi_k, \nabla \psi_i) = 0, \quad \text{for} \quad k = 1, 2, \cdot \cdot \cdot, r
    \label{eq:Proj_EqMom_2}
\end{equation}
\noindent and
\begin{equation}
     \Big( \psi_k, \nabla \cdot \bm{u}_m\Big) + \sum_{i=1}^n a^{n+1}_i (\psi_k, \nabla \cdot \bm{\phi}_i) = 0,
    \label{eq:Proj_EqMom_3}
\end{equation}
where $a_k^n = a_k (t_n)$ and $b_k^n = b_k (t_n)$. As discussed in Section \ref{sec:NSE}, \textit{inf-sup} condition restricts the choice of function spaces for pressure and velocity. For example, mixed elements such as Taylor-hood elements \cite{Taylor1973} are often used to overcome this restriction. On the other hand, if equal order interpolation functions are chosen for velocity and pressure, additional stabilization techniques, such as the pressure-stabilizing Petrov-Galerkin (PSPG) method \cite{Hughes1986}, may be needed. In the context of ROMs, the spatial basis for both pressure and velocity is determined from data, and there is limited literature on identifying adequate function spaces for velocity and pressure POD basis. The literature in this area highlights using stabilization techniques \cite{Bergmann2009, Caiazzo2014} and supremizers \cite{Rozza2007, Ballarin2015}. In the context of FOMs, such as those using finite element methods, pressure-projection schemes have been shown to overcome the restriction posed by \textit{inf-sup} condition on selecting velocity and pressure spaces \citep{Guermond1998}. Therefore, these schemes are also expected to work well for ROMs and this article demonstrates this through numerical simulations. As pressure correction scheme \citep{Chorin1968, Temam1968} is shown to be of similar form to PSPG scheme in \cite{Hughes1986}, pressure correction schemes could implicitly act as a model for unresolved pressure states \cite{Rannacher1992} and thereby improve pressure stability. Detailed numerical analysis of the stability and convergence of this method in the context of ROMs will be considered a future study. 

Despite good predictions of $a_i$ using the incremental pressure correction scheme for ROMs, the accurate temporal evolution of $b_i$ was not observed. This observation could be attributed to the role of pressure to enforce the incompressibility condition. Therefore, the evolution of $b_i$ is obtained in the post-processing stage by utilizing the $a_i$ and solving the pressure-Poisson equation as also done in \cite{Noack2005, Akhtar2009}. These equations are obtained by taking the pressure-Poisson equation,
\begin{equation}
    \nabla^2 p = - \nabla \cdot (\bm{u} \cdot \nabla \bm{u}),
\end{equation}
\noindent substituting \eref{PODvelandpres} in it and then projecting it on a linear subspace formed by $\psi_i$ for $i = 1, 2, \cdot \cdot \cdot \; r$, giving us,
\begin{multline}
    \Big( \psi_k, \nabla^2 p_m \Big) + \sum_{i=1}^r b_i \Big(\psi_k, \nabla^2 \psi_i \Big) + \sum_{i=1}^r \sum_{j=1}^r a_i a_j \Big( \psi_k, \nabla \cdot (\bm{\phi}_i \cdot \nabla \bm{\phi}_j) \Big) + \sum_{i=1}^r a_i \Big( \psi_k, \nabla \cdot (\bm{\phi}_i \cdot \nabla \bm{u}_m) \Big) \\ + \sum_{i=1}^r a_i \Big( \psi_k, \nabla \cdot (\bm{u}_m \cdot \nabla \bm{\phi}_i) \Big) + \Big( \psi_k, \nabla \cdot (\bm{u}_m \cdot \nabla \bm{u}_m) \Big), \quad \text{for} \quad k = 1, 2, \cdot \cdot \cdot, \; r.
    \label{eq:PresPoissROM}
\end{multline}
\noindent Note that the values of $b_i$ obtained through this post-processing step are solely used to compute the pressure distribution and not used in the evolution of velocity states.  The use of pressure-Poisson for post-processing to obtain pressure is also common when a discretely divergence free condition is used to simulate incompressible flows, as pressure is not computed during the solution stage. On the other hand, other common solution methods for resolved pressure states \cite{Caiazzo2014} typically involve the solution of the pressure-Poisson equation during the solution step. 

\section{Numerical experiments}

\label{sec:Results}

This section aims to validate the performance of the ROMs with proposed closure models and compare the results to ROMs without any closure model. First, we describe the details of the FOM simulation setup for the validation case: 2-D flow over a cylinder \citep{Schafer1996}. This flow has been frequently considered for validating ROM solution methods and closure models \citep{Xie2018, Mohebujjaman2019}. Second, we give details for the construction of the operators and time evolution of reduced states. Third, we compare the performance of proposed closure models at three Reynolds numbers: $Re = 200$, $Re = 500$ and $Re = 1,000$.

\subsection{Full order model details}
\label{sec:Results_FOM}

\begin{figure}[t]
    \centering
    \includegraphics[width=0.8\textwidth]{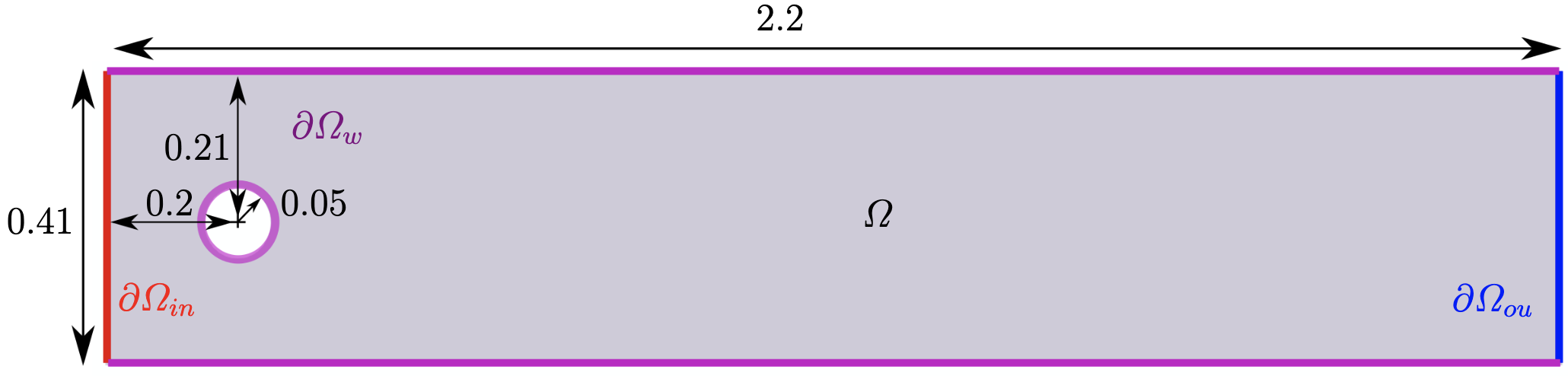}
    \caption{Domain for the flow over a 2-D cylinder.}
    \label{fig:CylinderDomain}
\end{figure}

 FOM simulations use a finite element code developed using the DOLFINx library \cite{Baratta2023}. The finite element code closely follows openly available tutorials in \citep{Langtangen2016, DokkenLink}. The commonly used Taylor-Hood elements (P2-Q1) are chosen as the spatial discretization for the FOM simulation. The problem can be posed as: 

\bigskip
\noindent Find $\bm{u} \in \mathcal{V}^h \subset \mathcal{V} = ( H_0^1 (\Omega))^2$ and $p \in \mathcal{Q}^h \subset \mathcal{Q} = \{ q \in L^2 (\Omega) : \int_{\Omega} q dx = 0 \} $, such that

\begin{equation}
    \Big( \bm{v}, \frac{\partial \bm{u}}{\p t} \Big) + \Big( \bm{v}, \bm{u} \cdot \nabla \bm{u} \Big) = - \Big( \bm{v}, \nabla p \Big) + \Big( \bm{v}, \nu \nabla^2 \bm{u} \Big), \quad \forall \quad \bm{v} \in \mathcal{V}^h
\end{equation}
and
\begin{equation}
    \Big( q, \nabla \cdot \bm{u} \Big) = 0 \quad \forall \quad q \in \mathcal{Q}^h
\end{equation}
subject to boundary conditions
\begin{equation}
    \bm{u} \vert_{\partial \Omega_D} = g \quad \text{and} \quad p \vert_{\partial \Omega_N} = h,
\end{equation}
where $\partial \Omega_D$ is the Dirichlet boundary and $\partial \Omega_N$ is the Neumann boundary. The Dirichlet boundary for the 2-D flow over a cylinder that encompasses of wall boundary ($\Omega_w$), inlet boundary ($\Omega_{in}$), and outlet boundary ($\Omega_{ou}$), that is $\partial \Omega_D = \partial \Omega_w \bigoplus \partial \Omega_{in} \bigoplus \partial \Omega_{ou}$). The domain and locations of the boundaries are shown in \figref{CylinderDomain}. The wall boundary condition is
\begin{equation}
    \bm{u} \vert_{\partial \Omega_w} = 0, 
\end{equation}
the inlet boundary condition is
\begin{equation}
    \bm{u} \vert_{\partial \Omega_{in}} = \Big[ 4 U_0 \frac{y (0.41 - y)}{0.41^2}, \; 0 \Big], \quad \text{where} \quad U_0 = 1.5 \; m/s,
\end{equation}
and the outlet boundary condition is
\begin{equation}
    p \vert_{\partial \Omega_{ou}} = 0.
\end{equation}

\noindent The simulation setup is similar to the well-known benchmark \cite{Schafer1996}. Unlike the benchmark, which uses a sinusoidally varying velocity, the flow considered in this article has a constant inflow velocity. These equations are solved using the Crank-Nicholson and semi-implicit Adams-Bashforth time integration scheme with the above boundary conditions. The mesh for $Re = 200$ was composed of quadrilateral elements formed using approximately $14,000$ nodes. Meanwhile, the $Re = 500$ and $1,000$ meshes comprised around $22,000$ nodes. This mesh resolution possibly underresolves the flow at the given Reynolds numbers. However, this mesh works well to demonstrate the applicability of the closure model and ROM solution method for incompressible fluid flows. The FOM simulation is started from a zero velocity and pressure initial condition. The simulation is integrated in time until $T_f = 10 s$ with a timestep of $1/1,600$. For $Re = 500$ and $1,000$, a lower timestep was also tested to ensure that the lower Courant–Friedrichs–Lewy (CFL) number remains similar between simulations of different Reynolds numbers. However, the behavior of FOM simulations and ROMs remained insensitive to this selection.

\subsection{Reduced order modeling details}

The offline step of ROMs involves the computation of operators as these are expensive to compute, for example, the non-linear term of Navier-Stokes equations in our case. ROMs can be categorized as discrete ROMs or continuous ROMs following the approach taken to construct operators in \eref{Proj_EqMom_1}, \eref{Proj_EqMom_2} and \eref{Proj_EqMom_3}. The key differences between these ROMs are discussed in \cite{Kalashnikova2014, Parish2023}. In this article, we work with continuous ROMs as it is mathematically more consistent with FOM than discrete ROMs \cite{Ingimarson2022, Parish2023}. Continuous ROMs leverage interpolation and numerical integration routines used in the corresponding FOM solver, favoring the consistency between ROMs and FOMs. The first step for assembling these ROMs involves interpolating the POD basis, $\bm{\phi}$ and $\psi$, to a mesh with a suitable polynomial degree. The polynomial degree for the spatial discretization can be arbitrarily chosen but could introduce consistency issues if not matched with the spatial discretization in the FOM \cite{Ingimarson2022}. In this article, a more consistent approach is undertaken by selecting function spaces for $\bm{\phi}$ and $\psi$ to be the same as those for $\bm{u}$ and $p$, that is, $\mathcal{V}^h$ and $\mathcal{Q}^h$ respectively.  Additionally, we use the same quadrature rule as FOMs for operator integrals. 

The online step of ROMs involves the time integration of the dynamical equations of reduced states. This article uses a second-order implicit time stepping scheme as discussed in Section \ref{sec:PresProjSection}. As this scheme is not the same as the time discretization used for the FOM, some inconsistency in time discretization may be introduced. This decision was still made to demonstrate the applicability of pressure correction schemes for the temporal evolution of reduced states in ROMs. The equations \eref{Proj_EqMom_1}, \eref{Proj_EqMom_2} and \eref{Proj_EqMom_3} are solved using SNES non-linear solver from PETSc \cite{Dalcin2011}. The time step for ROMs was chosen to be the same as the one for FOMs after a systematic timestep convergence study, which is not included in this article for brevity. The initial transient of the FOM simulation, that is, $t \in [0,\; 5] s$, is ignored. The data generated by FOM is uniformly sampled in a time interval of $T_{pred} = t \in [5, \;10] s$ and used for testing the ROMs. The POD basis, equation operators and closure models are obtained using uniformly sampled data in $t \in [5, \; 6] s$. Data at every $20^{th}$ timestep is selected in this time interval, resulting in 80 data points. Other time intervals and data selection frequencies were also considered and yielded similar conclusions. The goal of these computational tests is to demonstrate the ability of ROM for forecasting problems. For all Reynolds numbers, we are interested in answering the question: Do ROMs obtained using data from $t \in [5, \; 6] s$ give good predictions at a future time: $t \in [5, \; 10] s$ (renamed as $t \in [0, \; 5] s$)? 

\subsection{Cylinder flow at $Re = 200$}

We compare the energy ($E$), drag coefficient ($C_D$), and lift coefficient ($C_L$) for different ROMs with corresponding values of FOM. These are defined as:
\begin{equation}
    E (\bm{u},t) = \Big( \bm{u}(\bm{x},t), \bm{u} (\bm{x},t) \Big),
\end{equation}
\begin{equation}
    C_D (\bm{u},p,t) = \frac{2}{L U^2_0} \int_{\partial \Omega_S} \nu n \cdot \nabla u_{t_s} (t) n_y - p(t) n_x d s,
\end{equation}
and
\begin{equation}
    C_L (\bm{u},p,t) = - \frac{2}{L U^2_0} \int_{\partial \Omega_S} \nu n \cdot \nabla u_{t_s} (t) n_x + p(t) n_y d s,
\end{equation}
where $\partial \Omega_S$ is the surface of the cylinder and $u_{t_s} (= \bm{u} \cdot (n_y, - n_x))$ is the tangential component of the velocity with $n_x$ and $n_y$ as the components of the normal to the cylinder surface. In particular, we compare error in energy ($e_E$), error in $C_D$ ($e_{C_L}$) and error in $C_L$ ($e_{C_L}$) which are defined as,
\begin{equation}
     e_E (t) = E (\bm{u}^{FOM}, t) - E (\bm{u}^{ROM}, t),
\end{equation}
\begin{equation}
   e_{C_L} (t) = C_L (\bm{u}^{FOM}, p^{FOM}, t) - C_L (\bm{u}^{ROM}, p^{ROM}, t),
\end{equation}
and
\begin{equation}
    e_{C_D} (t) = C_D (\bm{u}^{FOM}, p^{FOM}, t) - C_D (\bm{u}^{ROM}, p^{ROM}, t).
\end{equation}
Note that errors in $C_L$ and $C_D$ have a mean bias introduced due to the difference in pressure computation between ROMs and FOM simulations. ROM determines pressure by solving pressure-Poisson equation \eref{PresPoissROM}, whereas no such equations are solved in FOM simulations.

\begin{figure}
    \centering
    \subfigure[\label{fig:ROM_200_Energy_None_ModeComp}]{\includegraphics[width=0.49\textwidth,trim={0cm 0 1cm 0},clip]{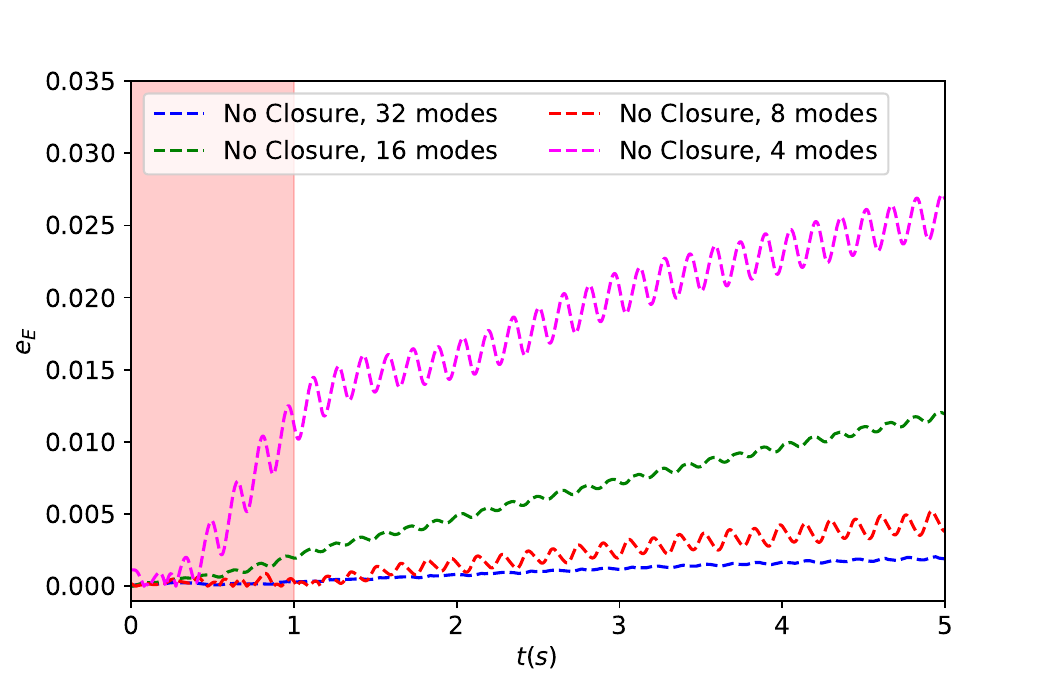}}
    \subfigure[\label{fig:ROM_200_CL_None_ModeComp}]{\includegraphics[width=0.49\textwidth,trim={0cm 0 1cm 0},clip]{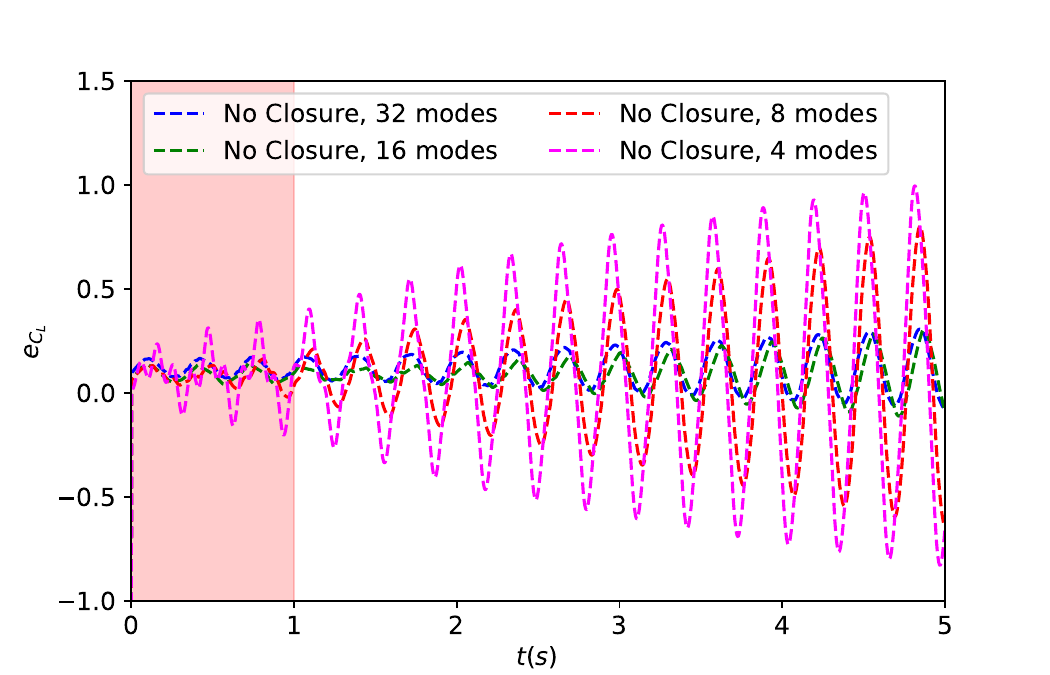}}
    \subfigure[\label{fig:ROM_200_CD_None_ModeComp}]{\includegraphics[width=0.49\textwidth,trim={0cm 0 1cm 0},clip]{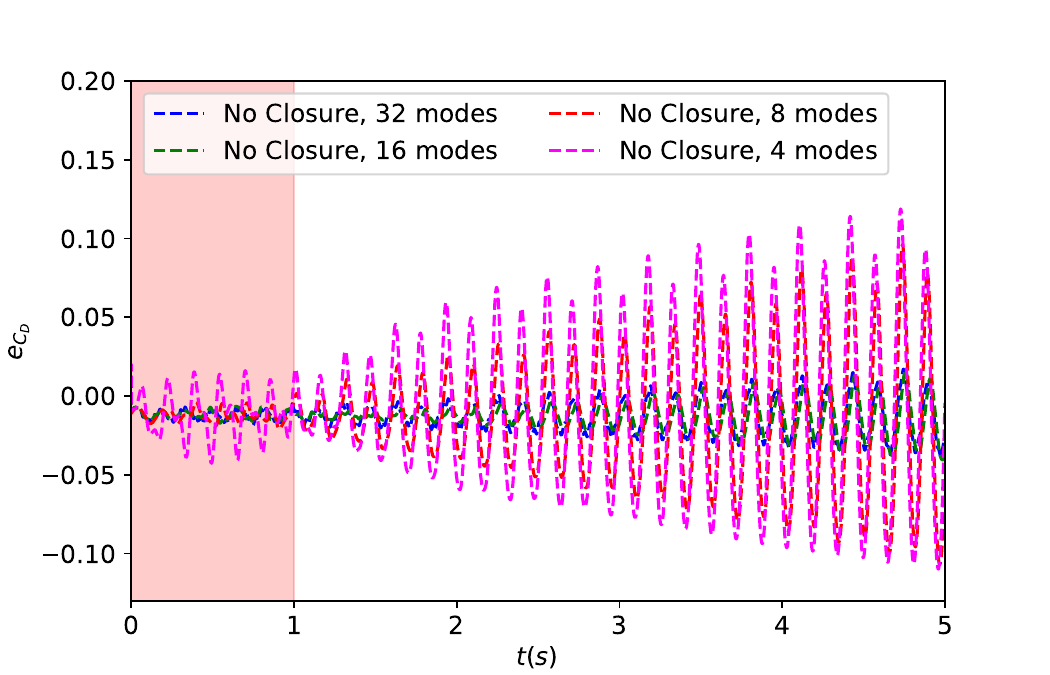}}    
    \vspace{-3mm}
    \caption{Error in (a) energy, (b) $C_L$ and (c) $C_D$ for 2-D cylinder flow at $Re = 200$ for Galerkin ROM without closure model. The shaded area is the time interval from which data is extracted to determine the POD basis, equation operators and closure models.}
    \label{fig:ROM_200_None_ModeComp}
\end{figure}

We first demonstrate the performance of the ROM without using a closure model. The evolution of energy, $C_L$ and $C_D$ for ROMs without a closure model for different numbers of modes is shown in \figref{ROM_200_None_ModeComp}. For all selection of modes, we observe an increase in the error with an increase in time. Using eight modes for the ROM gives better energy prediction than using 16 modes. This result implies that increasing the number of modes for representing the solution does not improve the errors in energy approximation in time. On the other hand, results for $C_D$ and $C_L$ give a more consistent picture as the error for eight modes is much higher than the error for 16 modes. The error in these aerodynamic coefficients reduces significantly with an increase in the number of modes. These numerical results demonstrate the applicability of an incremental pressure correction scheme for treating pressure-velocity coupling in the context of incompressible Navier-Stokes equations. 

These results indicate that ROMs would require many modes to achieve accurate predictions without using any closure model. As the online cost of ROMs scales as $r^3$, where $r$ is the number of modes, using many modes to achieve more accurate results becomes computationally expensive and is not ideal. These observations and drawbacks of Galerkin ROMs are well-known in the literature. This discussion motivates the need for closure models that ensure higher accuracy without increasing the online cost of ROMs.

\begin{figure}
    \centering    
    
\includegraphics[width=\textwidth,trim={0 0 0 41cm},clip]{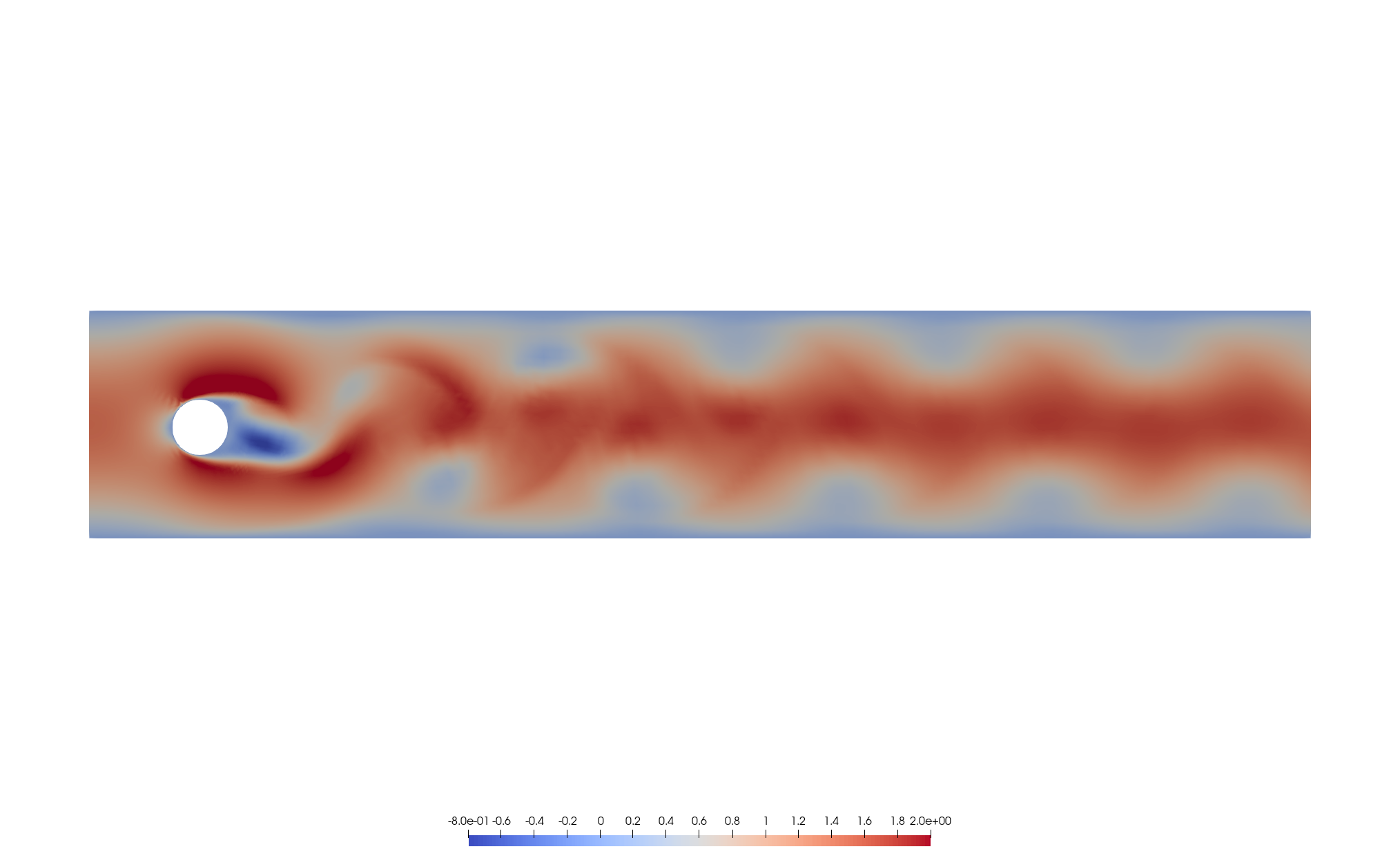}
    \subfigure[\label{fig:Re200_Contour_uROM_4modes_FOM}]{\includegraphics[width=0.49\textwidth,trim={0 0 5cm 27cm},clip]{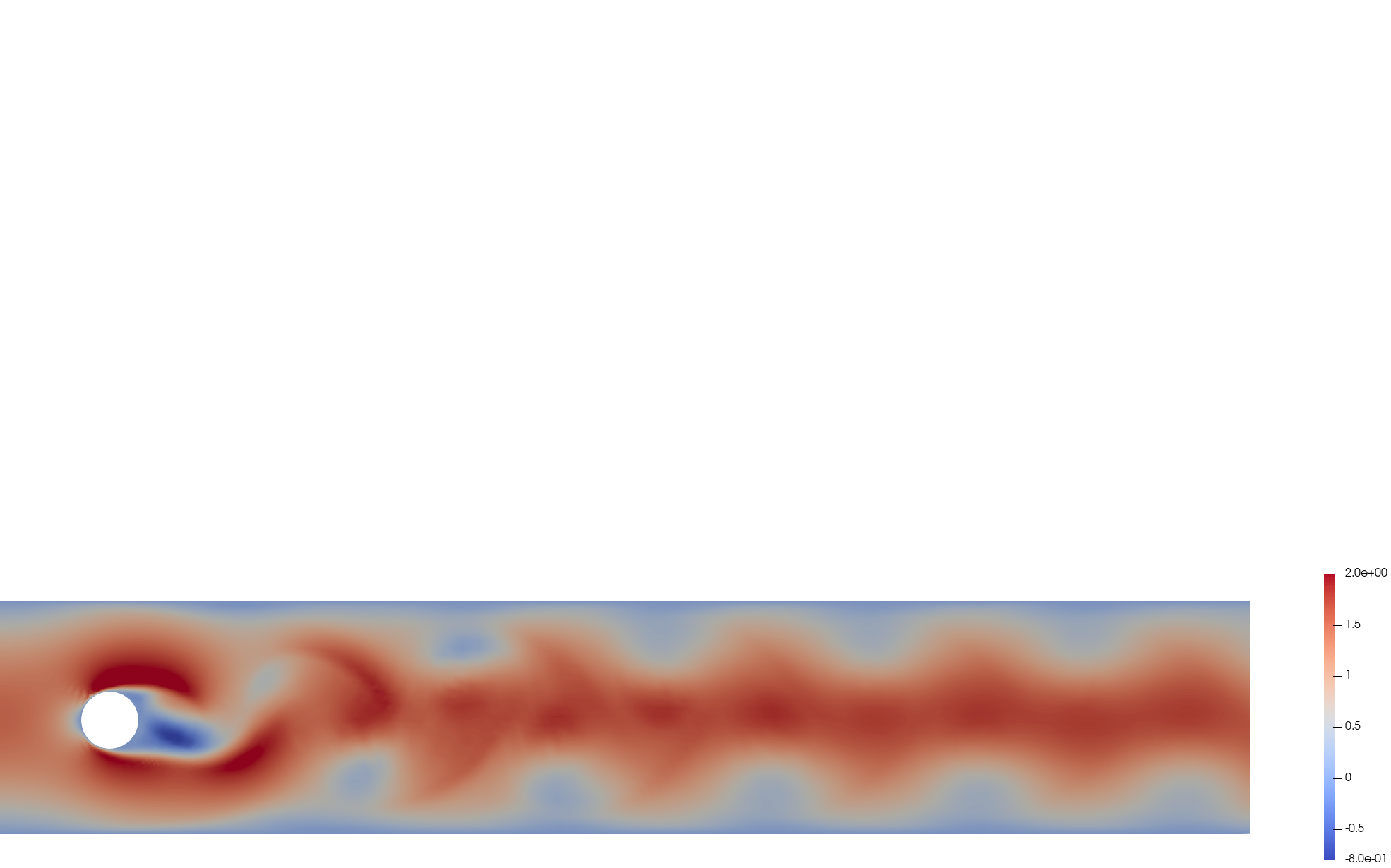}}
    \subfigure[\label{fig:Re200_Contour_uROM_4modes_NM}]{\includegraphics[width=0.49\textwidth,trim={0 0 5cm 27cm},clip]{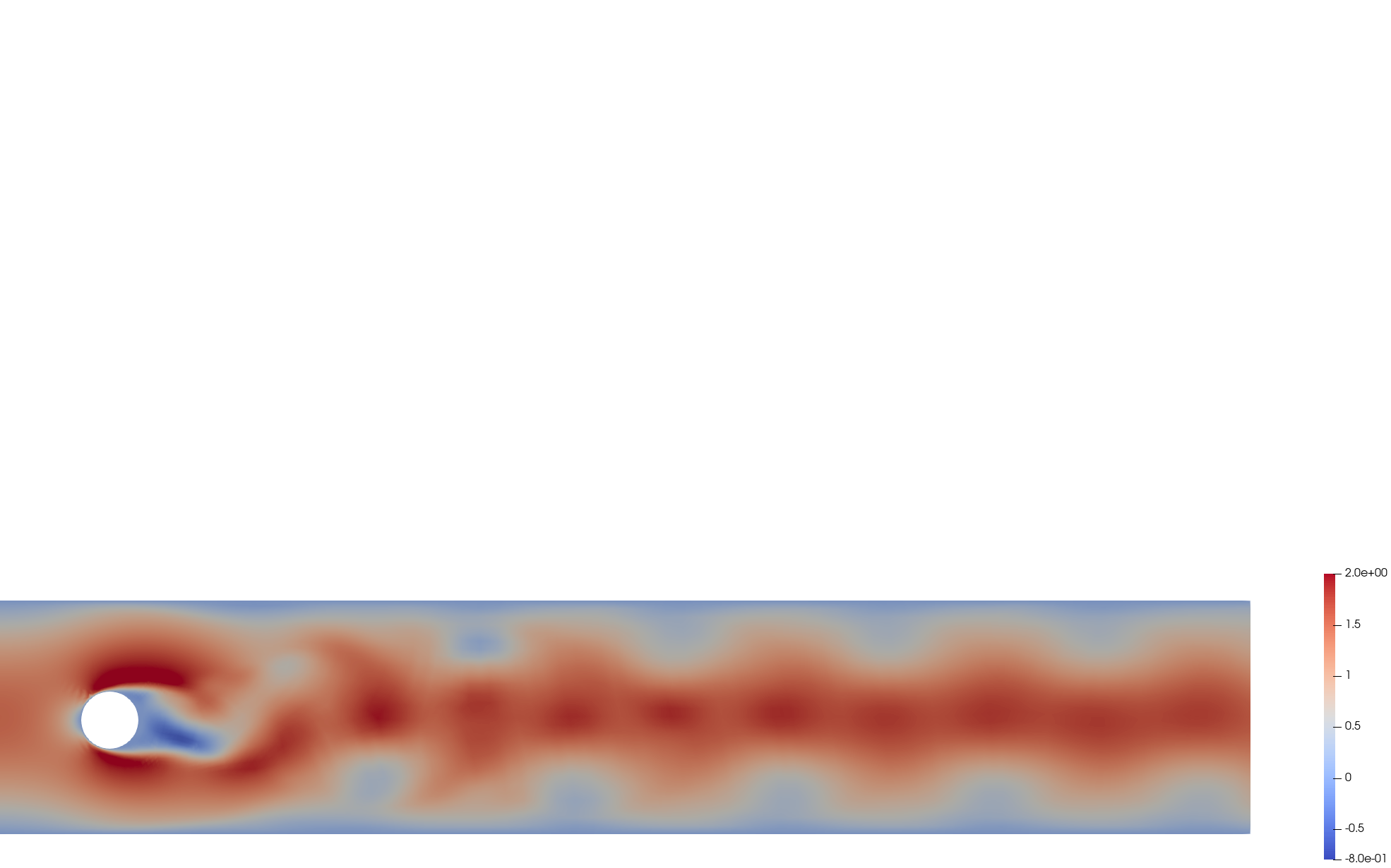}}
    \subfigure[\label{fig:Re200_Contour_uROM_4modes_M3}]{\includegraphics[width=0.49\textwidth,trim={0 0 5cm 27cm},clip]{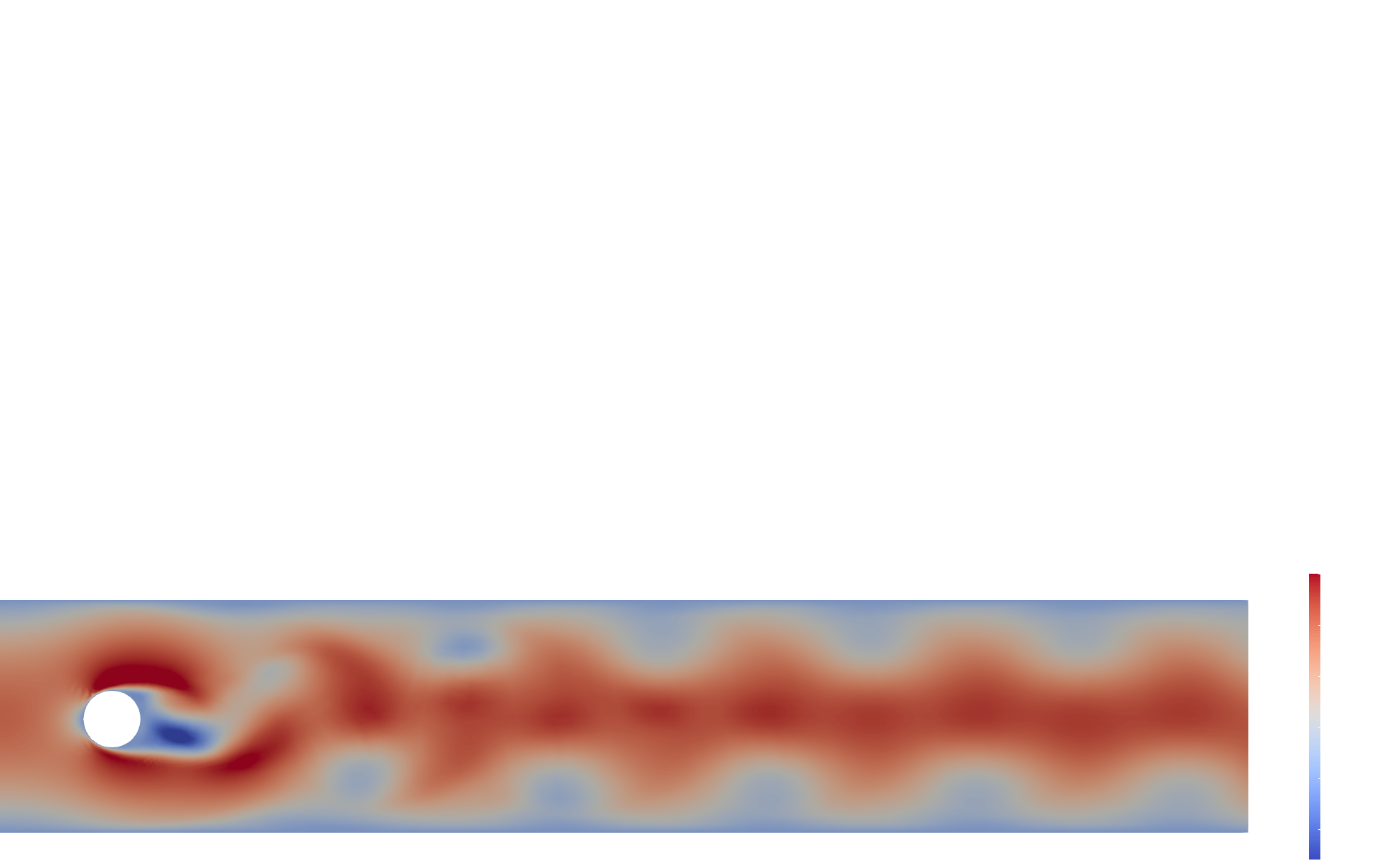}}    
    \subfigure[\label{fig:Re200_Contour_uROM_4modes_M6}]{\includegraphics[width=0.49\textwidth,trim={0 0 5cm 27cm},clip]{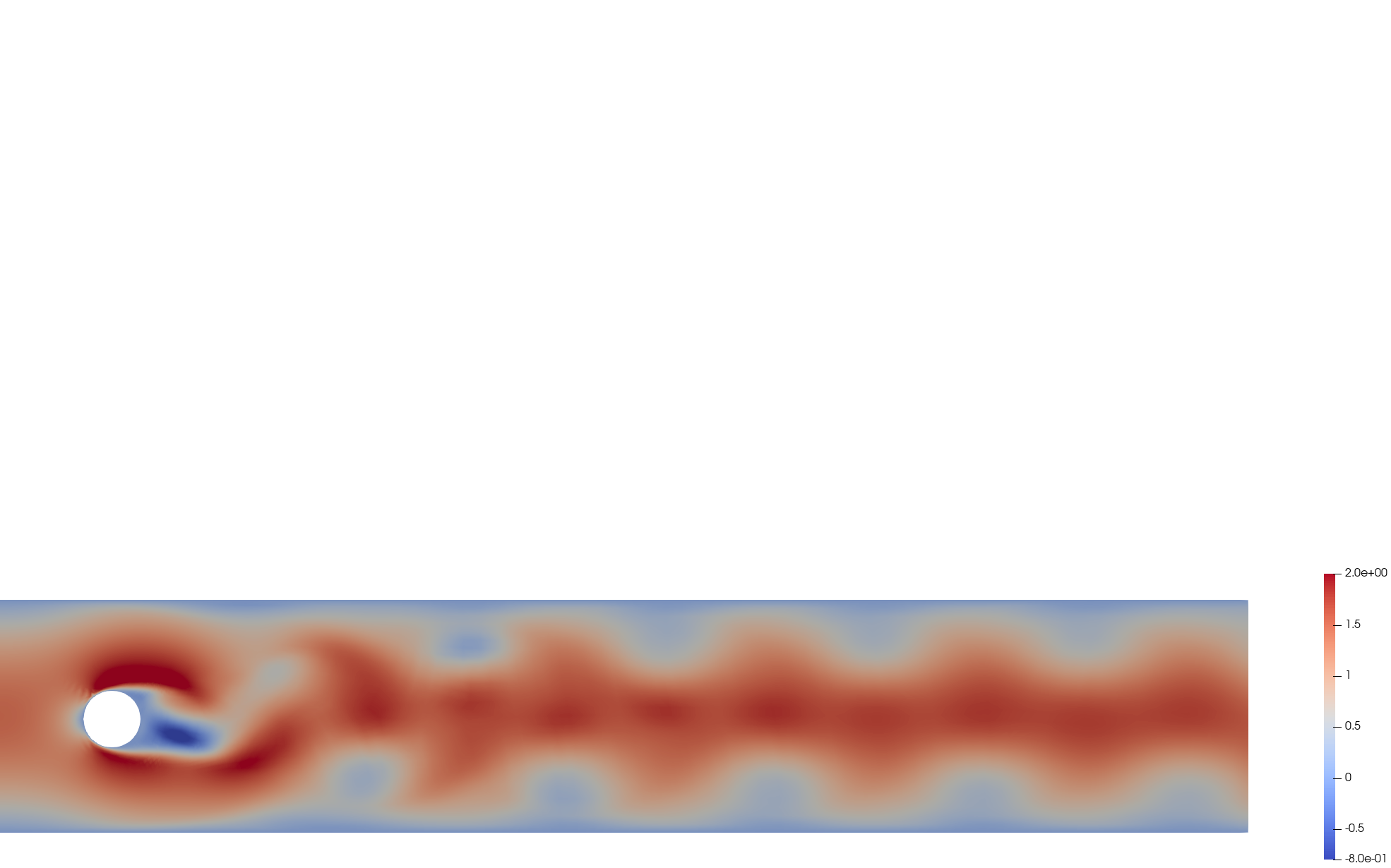}}       
    \vspace{-3mm}
    \caption{Streamwise velocity component ($u_1$) field at t = 5s for (a) FOM, (b) Galerkin ROM with no closure model, (c) Galerkin ROM with the TV-ME closure model and (d) Galerkin ROM with the TV-C closure model when $r = 4$ modes are used.}
    \label{fig:Re200_Contour_uROM_4modes}
\end{figure}

\begin{figure}
    \centering

    \includegraphics[width=\textwidth,trim={0 0 0 41cm},clip]{Re200_Contour_uFOM_ColorMap.png}
    \subfigure[\label{fig:Re200_Contour_pROM_4modes_FOM}]{\includegraphics[width=0.49\textwidth,trim={0 0 5cm 27cm},clip]{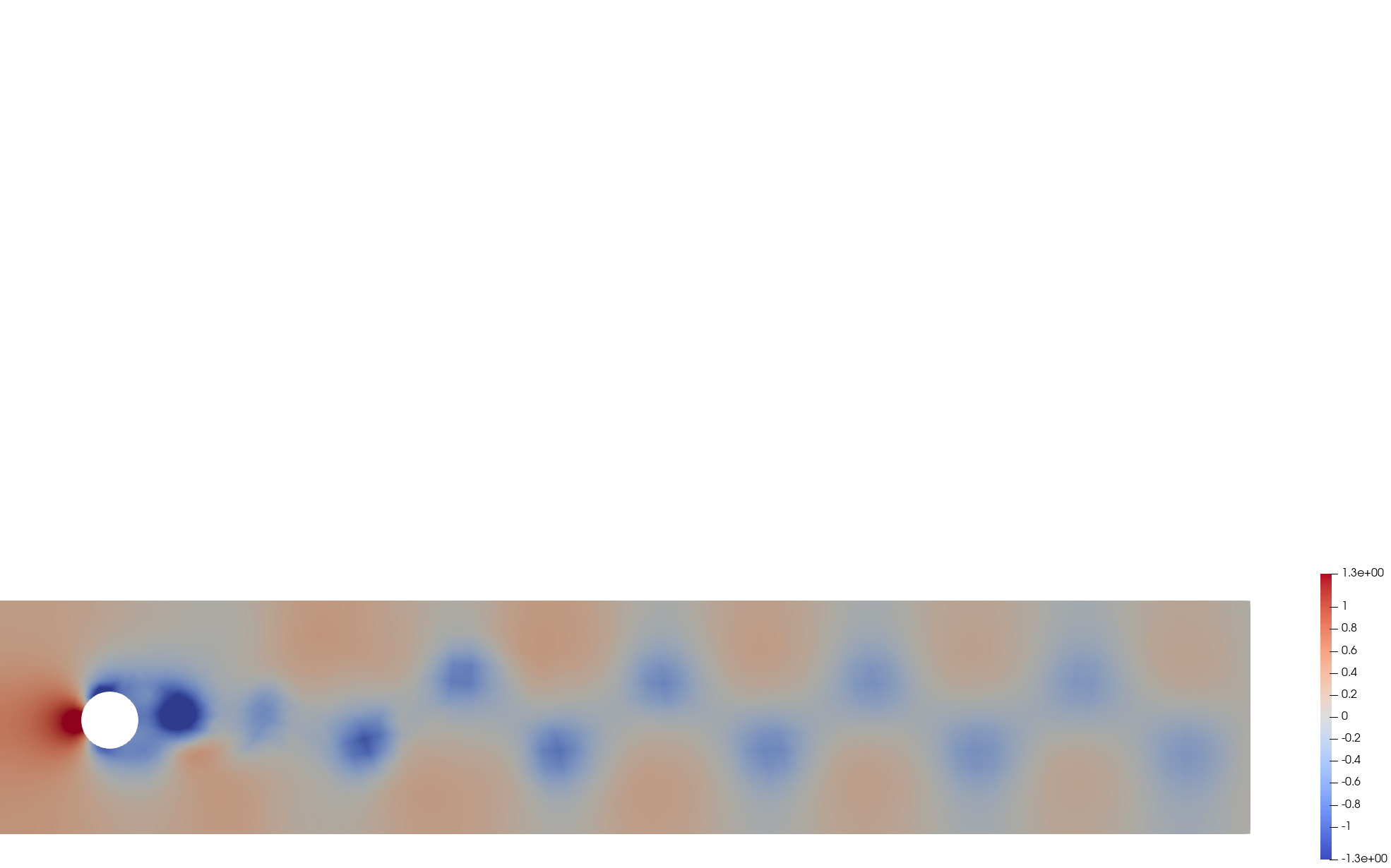}}
    \subfigure[\label{fig:Re200_Contour_pROM_4modes_NM}]{\includegraphics[width=0.49\textwidth,trim={0 0 5cm 27cm},clip]{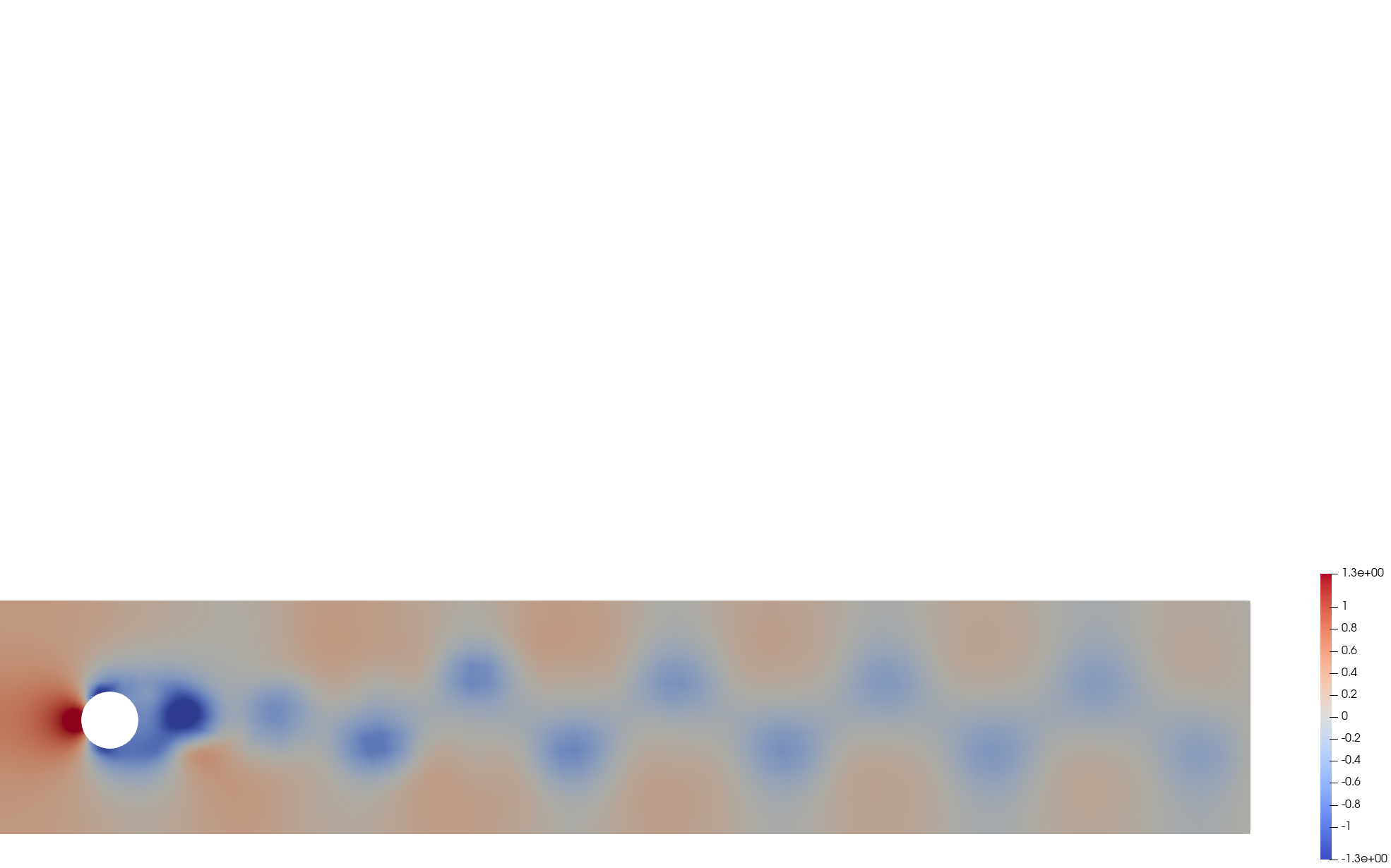}}
    \subfigure[\label{fig:Re200_Contour_pROM_4modes_M3}]{\includegraphics[width=0.49\textwidth,trim={0 0 5cm 27cm},clip]{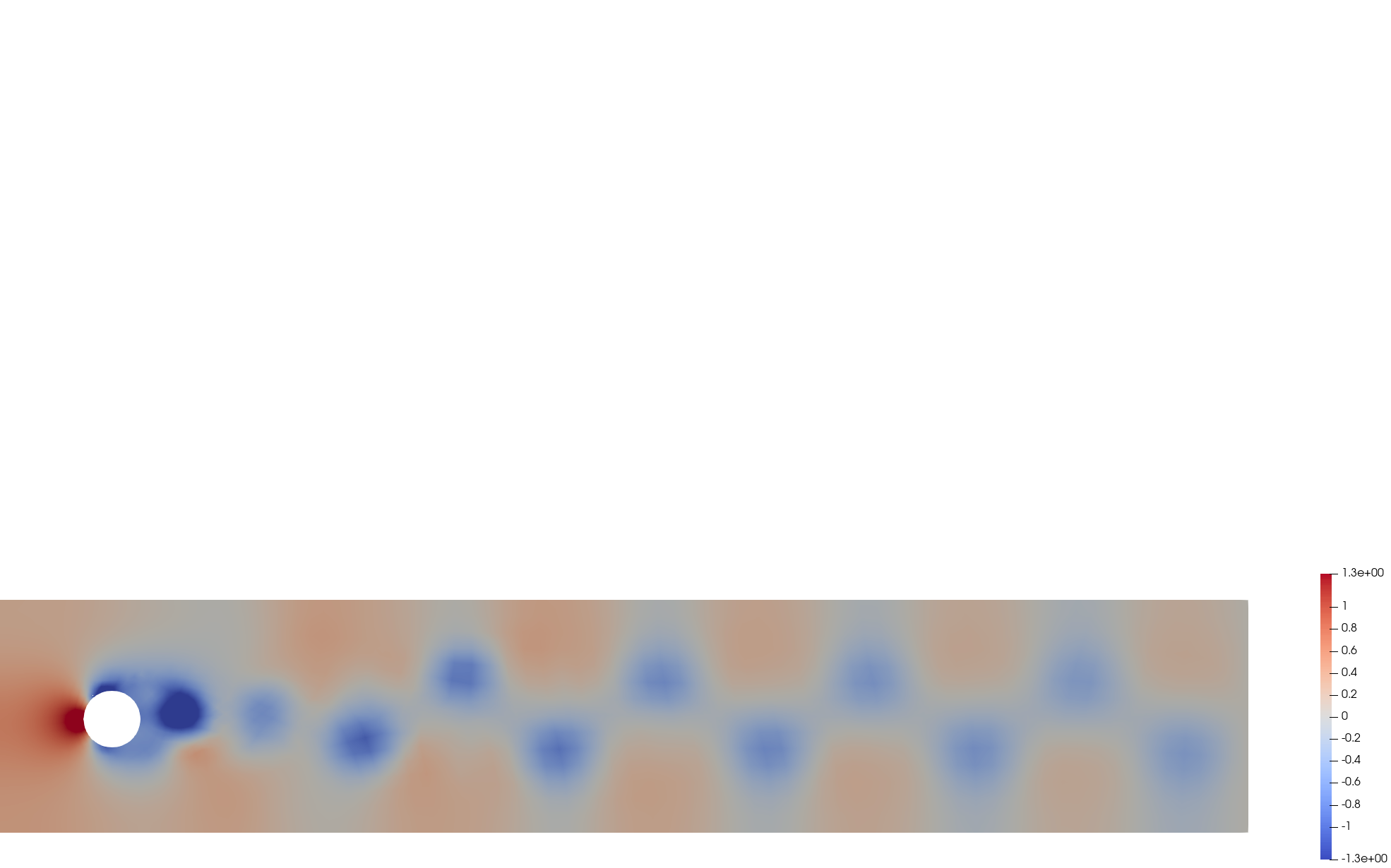}}    
    \subfigure[\label{fig:Re200_Contour_pROM_4modes_M6}]{\includegraphics[width=0.49\textwidth,trim={0 0 5cm 27cm},clip]{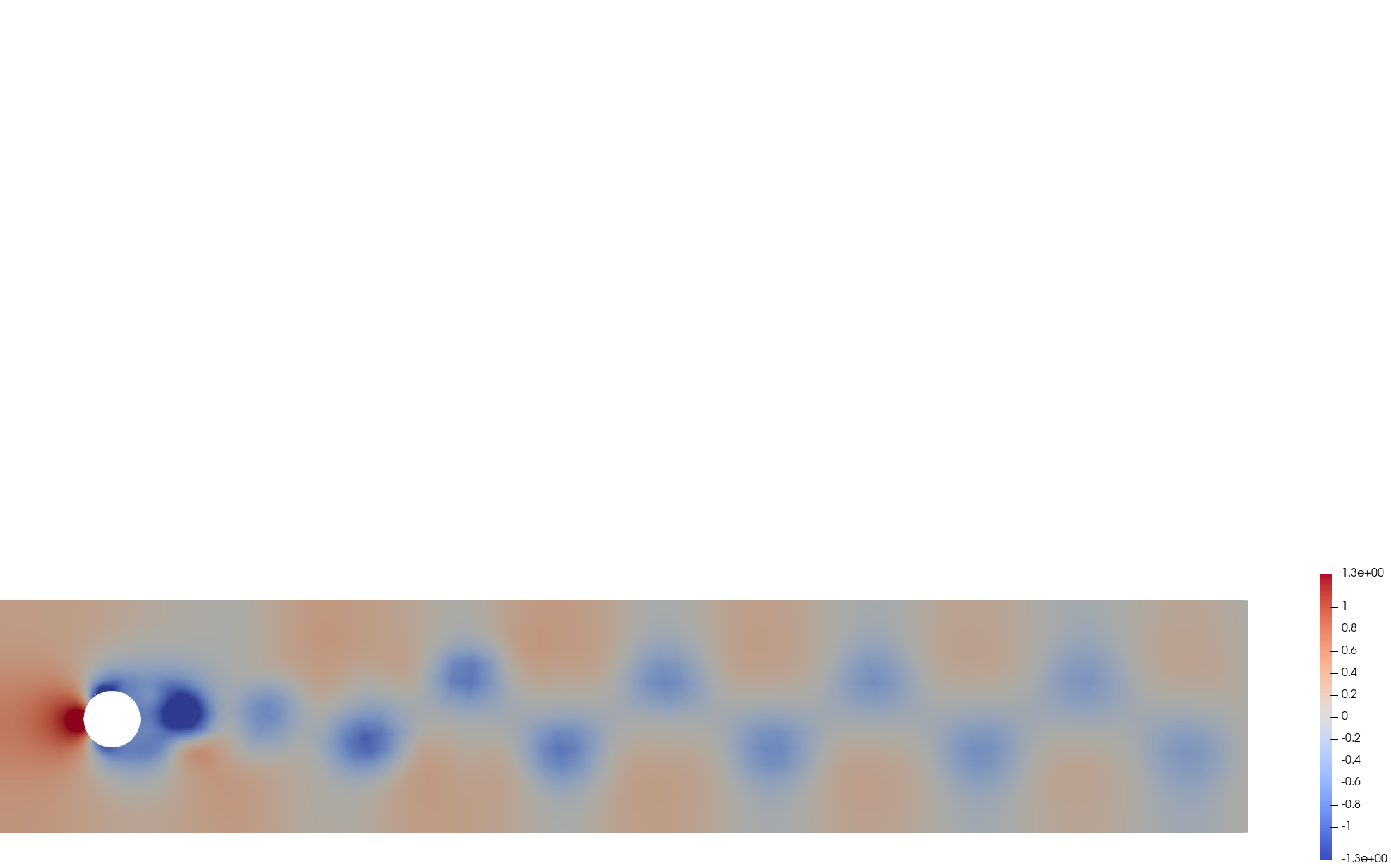}}       \vspace{-3mm}
    \caption{Pressure field at t = 5s for (a) FOM, (b) Galerkin ROM with no closure model, (c) Galerkin ROM with the TV-ME closure model and (d) Galerkin ROM with the TV-C closure model when $r = 4$ modes are used.}
    \label{fig:Re200_Contour_pROM_4modes}
\end{figure}

The streamwise velocity component and pressure fields at $t = 5s$ are shown in \figref{Re200_Contour_uROM_4modes} and \figref{Re200_Contour_pROM_4modes} respectively. The pressure distribution for several ROMs tested in this study is similar to those for FOM. These observations highlight the robustness of pressure prediction even without closure models. As discussed in Section \ref{sec:PresProjSection}, the numerical solution scheme may implicitly act as a model for unresolved pressure states, thereby providing good pressure predictions. ROMs without a closure model lead to a slightly different prediction of the streamwise velocity in the wake region closer to the cylinder. ROMs with TV-ME and TV-C closure models in this region yield a closer wake structure to FOM. In the far-wake region, the flow structures predicted by FOMs and ROMs slightly differ with TV-ME and TV-C closure models; however, these are still in better agreement compared to ROMs without a closure model used. These results demonstrate the applicability of the incremental pressure correction scheme-based ROMs to predict similar flow structures as FOMs. 

\begin{figure}
    \centering
    \subfigure[\label{fig:ROM_200_Energy_4mode_ModelComp}]{\includegraphics[width=0.49\textwidth,trim={0cm 0 1cm 0},clip]{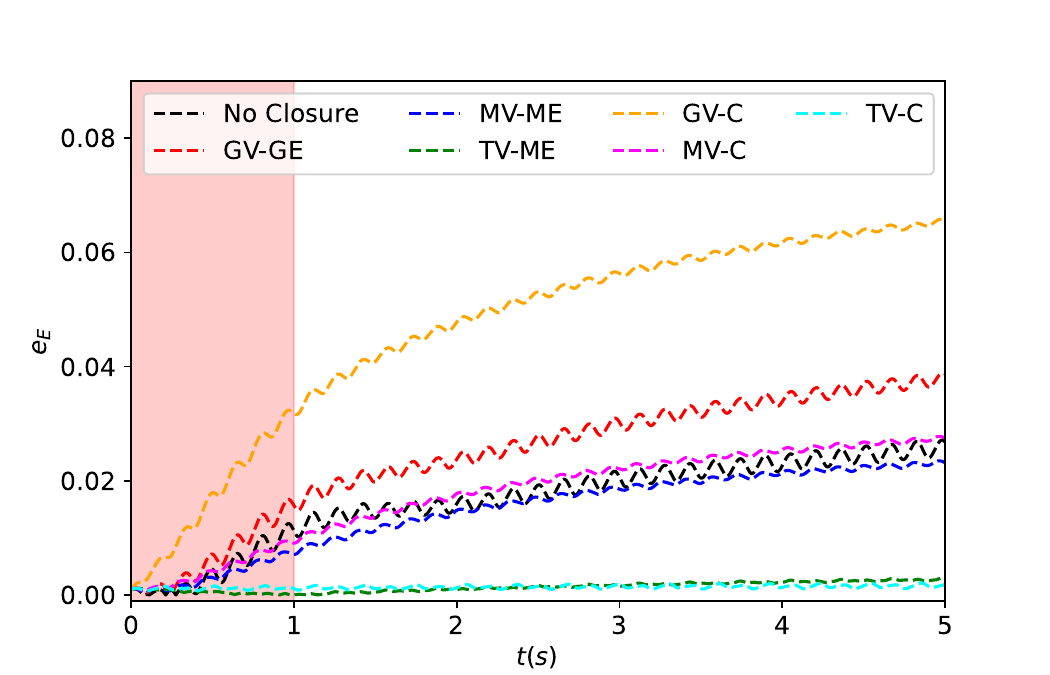}}
    \subfigure[\label{fig:ROM_200_Energy_6mode_ModelComp}]{\includegraphics[width=0.49\textwidth,trim={0cm 0 1cm 0},clip]{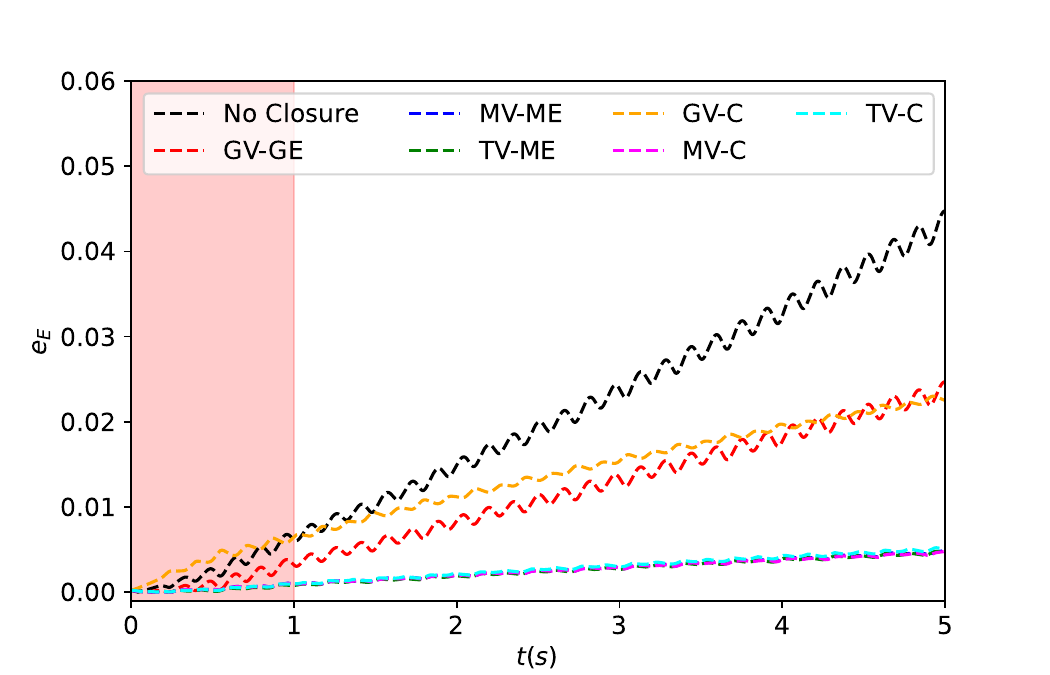}}
    \vspace{-3mm}
    \caption{Error in energy (a) four modes and (b) six modes for 2-D cylinder flow at $Re = 200$. The shaded area is the time interval from which data is extracted to determine the POD basis, equation operators and closure models.}
    \label{fig:ROM_200_Energy_ModelComp}
\end{figure}

The temporal evolution of error in energy for four and six mode ROMs with different closure models is shown in \figref{ROM_200_Energy_ModelComp}. A significant error growth is observed when no closure model is used for four mode ROMs. ROMs with GV-GE and GV-C closure models provide even higher errors than those without a closure model. Similarly, ROMs with MV-ME and MV-C closure models give errors similar to ROMs without a closure model. ROMs with TV-ME and TV-C closure models provide significantly lower errors compared to ROMs with other and without closure models. ROMs with any of the proposed closure models give better energy prediction and lower error than ROMs without a closure model for six mode ROMs. ROMs with GV-GE and GV-C closure models exhibit higher errors than ROMs with other closure models. ROMs with MV-ME, MV-C, TV-ME and TV-C closure models show very low and almost overlapping errors, thereby demonstrating the promise of these closure models over ROMs without closure models.

\begin{figure} [htp]
    \centering
    \subfigure[\label{fig:ROM_200_CL_4mode_ModelComp}]{\includegraphics[width=0.49\textwidth,trim={0cm 0 1cm 0},clip]{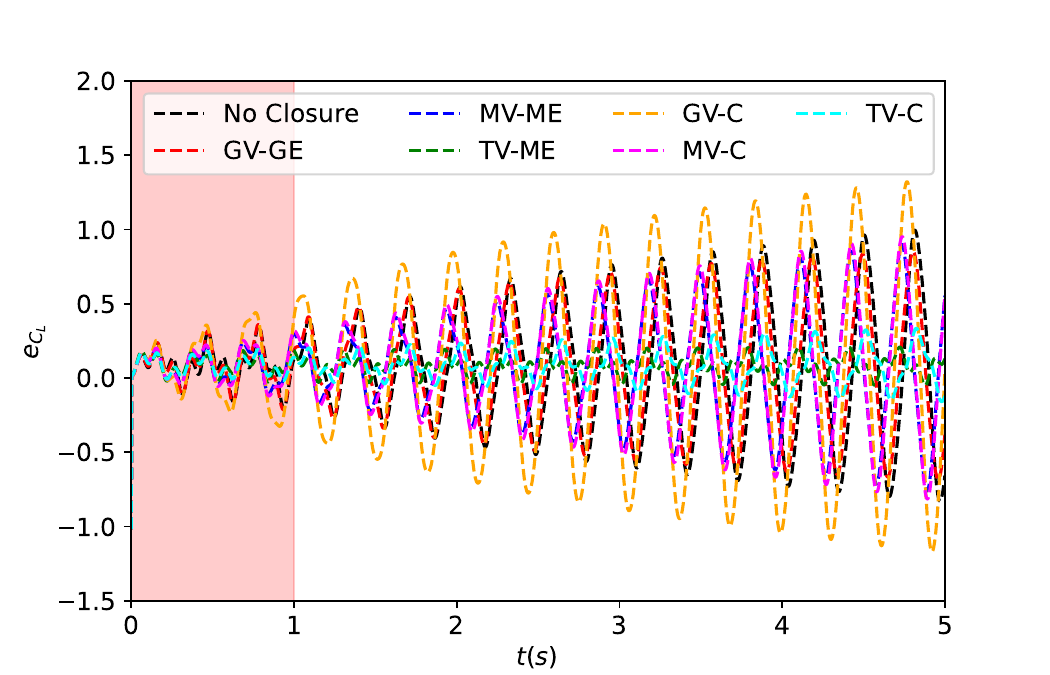}}    
    \subfigure[\label{fig:ROM_200_CL_6mode_ModelComp}]{\includegraphics[width=0.49\textwidth,trim={0cm 0 1cm 0},clip]{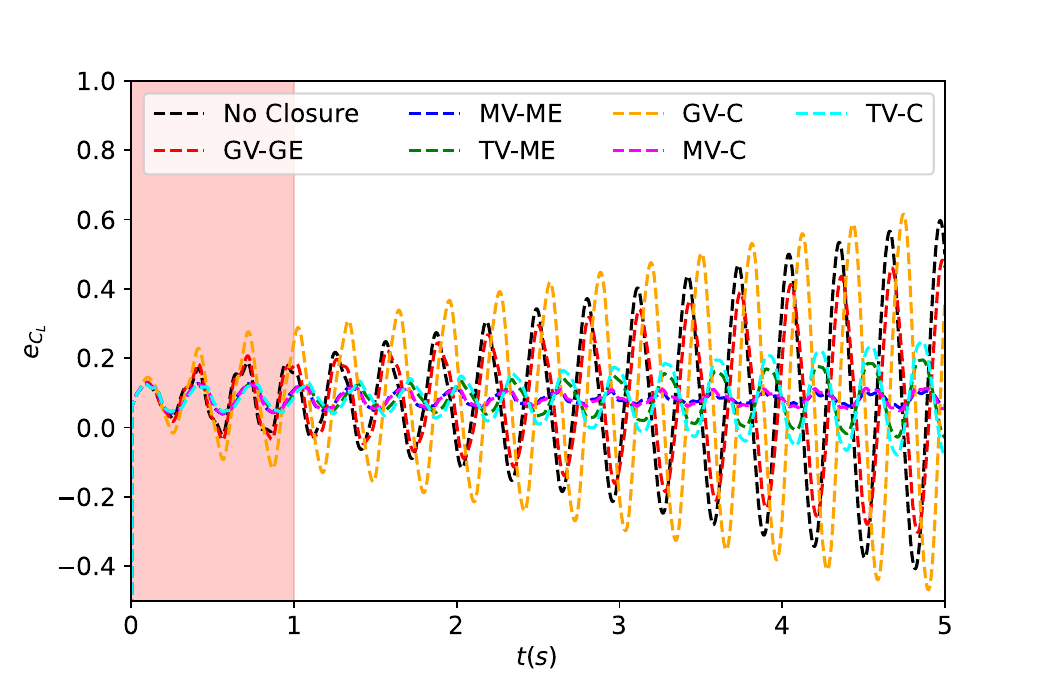}}
    \vspace{-3mm}
    \caption{Error in $C_L$ (a) four modes and (b) six modes for 2-D cylinder flow at $Re = 200$. The shaded area is the time interval from which data is extracted to determine the POD basis, equation operators and closure models.}
    \label{fig:ROM_200_CL_ModelComp}
\end{figure}

\begin{figure} [htp]
    \centering
    \subfigure[\label{fig:ROM_200_CD_4mode_ModelComp}]{\includegraphics[height=0.32\textwidth,trim={0cm 0 1cm 0},clip]{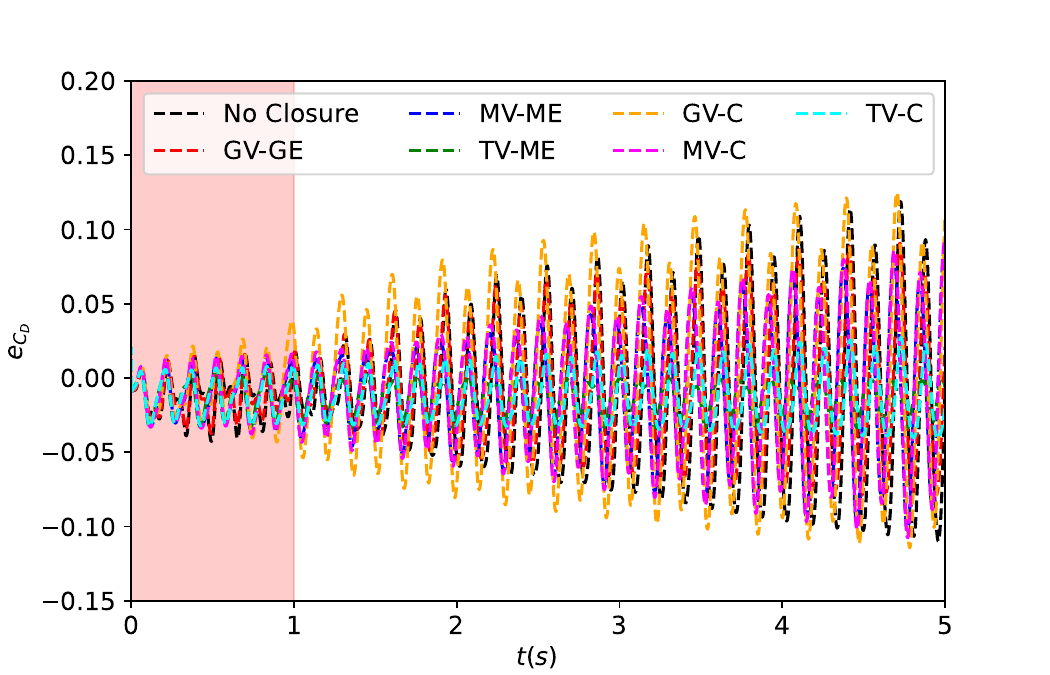}}    
    \subfigure[\label{fig:ROM_200_CD_6mode_ModelComp}]{\includegraphics[height=0.32\textwidth,trim={0cm 0cm 1cm 0},clip]{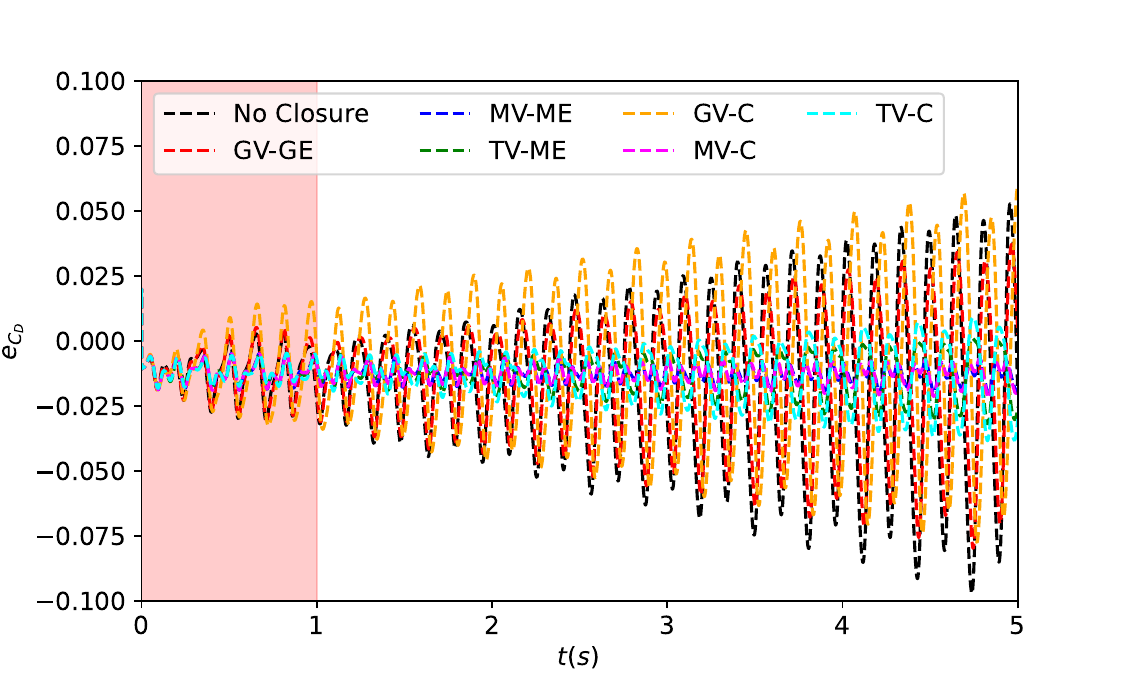}}    
    \vspace{-3mm}
    \caption{Error in $C_D$ (a) four modes and (b) six modes for 2-D cylinder flow at $Re = 200$. The shaded area is the time interval from which data is extracted to determine the POD basis, equation operators and closure models.}
    \label{fig:ROM_200_CD_ModelComp}
\end{figure}

The temporal evolution of errors in $C_L$ and $C_D$ for ROMs with different closure models is shown in \figref{ROM_200_CL_ModelComp} and \figref{ROM_200_CD_ModelComp} respectively. The errors in $C_L$ and $C_D$ exhibit a similar trend to the error in energy. For four mode ROMs, closure models TV-ME and TV-C provide low errors in $C_L$ and $C_D$ compared to other closure models and ROMs without closure models. For six mode ROMs, MV-ME and MV-C closure models exhibit better performance with the lowest error in time, which is even smaller than the error for ROMs with TV-ME and TV-C closure models. The error in these aerodynamic coefficients remains high when GV-GE and GV-C closure models are used despite the slightly lower error in energy approximation. These results indicate that ROM without a closure model is insufficient for dynamics prediction as the error rapidly increases outside the interval for which the data is used to obtain the energetic spatial basis. However, this drawback of Galerkin ROMs can be rectified by selecting an appropriate closure model. As the error does not grow significantly in time when MV-ME, MV-C, TV-ME and TV-C closure models are used, Galerkin ROM with these closure models appear to be well suited for time dynamics prediction for this Reynolds number. 

A more complete comparison of the model requires a comparison of these errors for other selections of the number of modes. Some efficient metrics for comparing the model performance of the different number of modes are the relative integrated error in energy ($\eta_E$), relative integrated error in $C_L$ ($\eta_{C_L}$) and relative integrated error in $C_D$ ($\eta_{C_D}$) which are defined as follows:
\begin{equation}
    \eta_E = \frac{\int_{0}^{5} E (\bm{u}^{ROM}, t) - E(\bm{u}^{FOM}, t) \; dt}{\int_{0}^{5} E(\bm{u}^{FOM}, t) \; dt}, 
\end{equation}
\begin{equation}
    \eta_{C_L} = \frac{\int_{0}^{5} C_L (\bm{u}^{ROM}, p^{ROM}, t) - C_L (\bm{u}^{FOM}, p^{FOM}, t) \; dt}{\int_{0}^{5} C_L (\bm{u}^{FOM}, p^{FOM}, t) \; dt}
\end{equation}
and
\begin{equation}
    \eta_{C_D} = \frac{\int_{0}^{5} C_D (\bm{u}^{ROM}, p^{ROM}, t) - C_D (\bm{u}^{FOM}, p^{FOM}, t) \; dt}{\int_{0}^{5} C_D (\bm{u}^{FOM}, p^{FOM}, t) \; dt}.
\end{equation}

\begin{figure}
    \centering
    \subfigure[\label{fig:Int_ROM_200_Energy_None_ModeComp}]{\includegraphics[width=0.33\textwidth,trim={0cm 0 1.5cm 0},clip]{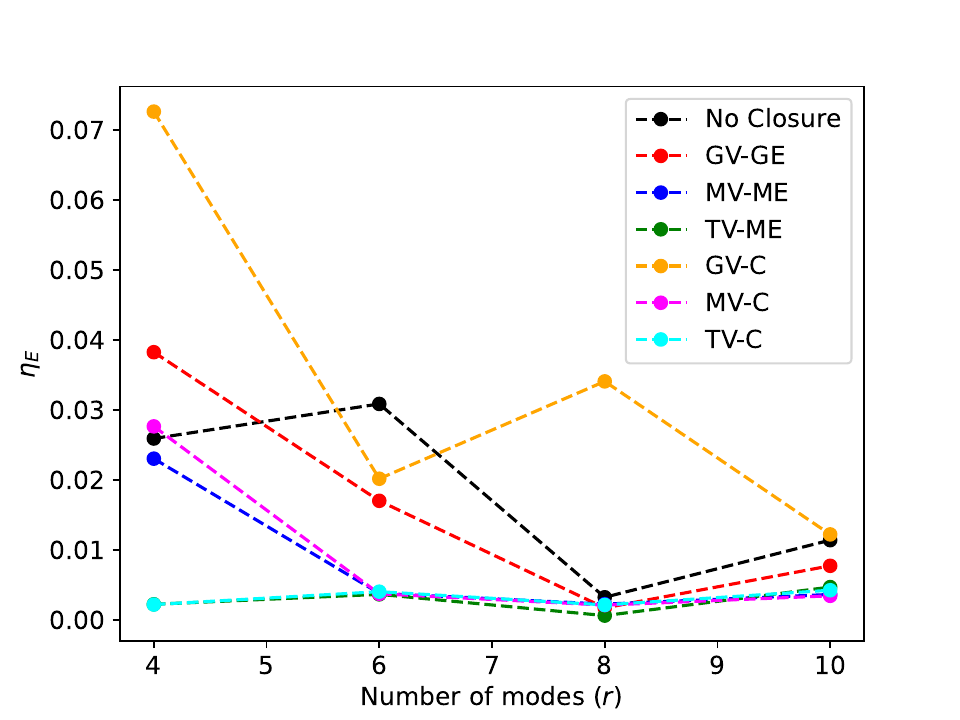}}\subfigure[\label{fig:Int_ROM_200_CL_None_ModeComp}]{\includegraphics[width=0.33\textwidth,trim={0cm 0 1.5cm 0},clip]{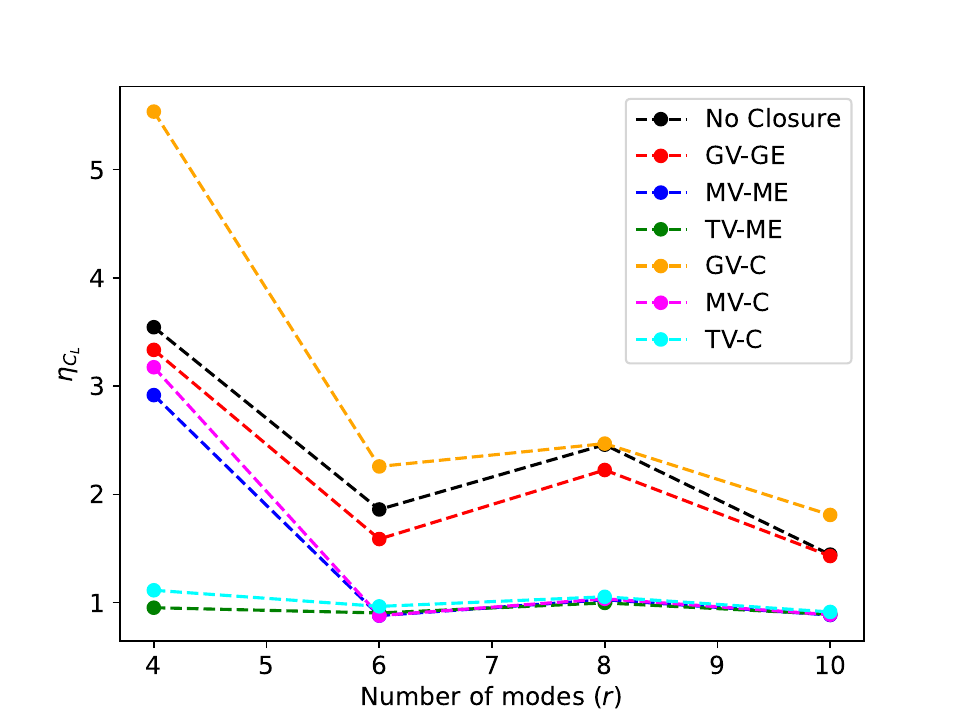}}\subfigure[\label{fig:Int_ROM_200_CD_None_ModeComp}]{\includegraphics[width=0.33\textwidth,trim={0cm 0 1.5cm 0},clip]{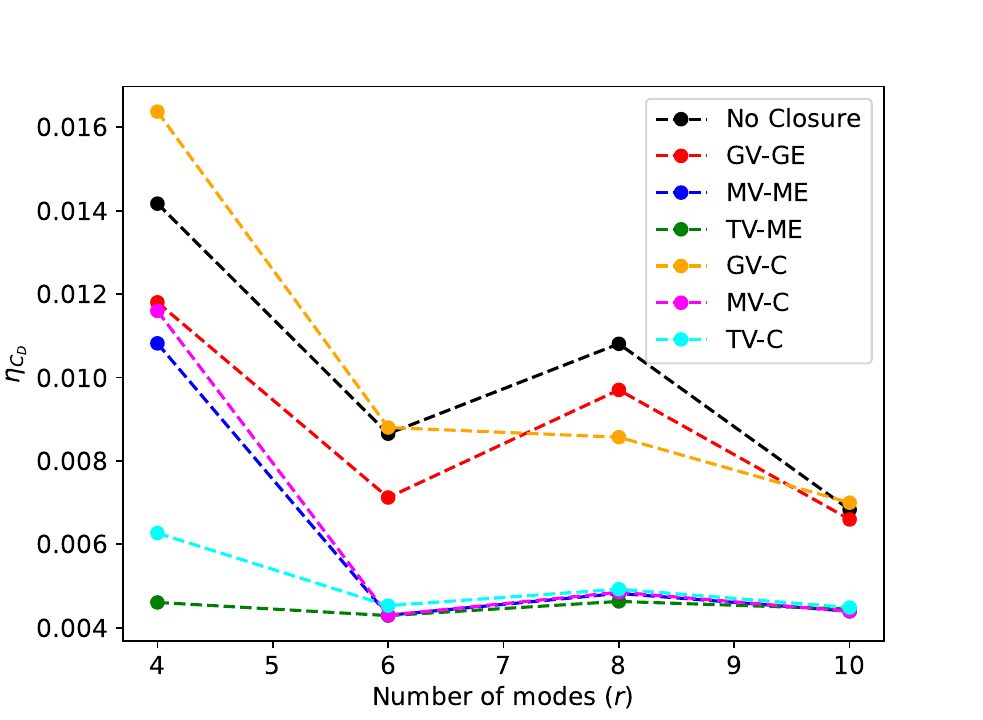}}
    \vspace{-3mm}
    \caption{Integrated error in (a) energy, (b) $C_L$ and (c) $C_D$ for 2-D cylinder flow at $Re = 200$.}
    \label{fig:Int_ROM_200_None_ModeComp}
\end{figure}

The variation of the integrated error in energy, $C_L$ and $C_D$ with the number of modes for several closure models is shown in \figref{Int_ROM_200_None_ModeComp}.  A non-monotonic change in the integrated error is observed with increased modes for ROMs with and without closure models. The results indicate a higher error for ROMs with the GV-C closure model for several modes than when ROMs without a closure model are used, indicating that this closure model does not perform well. For a lower number of modes, ROMs with GV-GE, MV-ME and MV-C closure models appear to have a higher error in energy but lower errors in $C_L$ and $C_D$ compared to ROMs without a closure model. These results indicate that these models perform similarly to the ROMs without closure model for fewer modes. This behavior is drastically different when a higher number of modes are used. The results indicate that ROMs with GV-GE, MV-ME, and MV-C closure models give lower errors in energy, $C_L$, and $C_D$ for a higher number of modes. The error for ROMs with the GV-GE closure model is slightly lower than those without a closure model. In contrast, ROMs with MV-ME and MV-C closure models give similar results with significantly low errors in $C_L$ and $C_D$ compared to ROMs without closure models.
At a higher number of modes, ROMs with MV-ME and MV-C closure models give similar predictions to those from ROMs with TV-ME and TV-C closure models. The results also indicate that ROMs with MV-ME and MV-C closure models do not yield the best results at fewer modes. ROMs with TV-ME and TV-C closure models give similar results and the lowest errors for all modes. The predictions using these models exhibit a much lower error in energy, $C_L$ and $C_D$ compared to ROMs without closure models. These results indicate that ROMs with TV-ME and TV-C closure models consistently yield the best results for this Reynolds number. 

\subsection{Cylinder flow at $Re = 500$}

\begin{figure}
    \centering
    \subfigure[\label{fig:ROM_500_Energy_4mode_ModelComp}]{\includegraphics[width=0.49\textwidth,trim={0cm 0 1cm 0},clip]{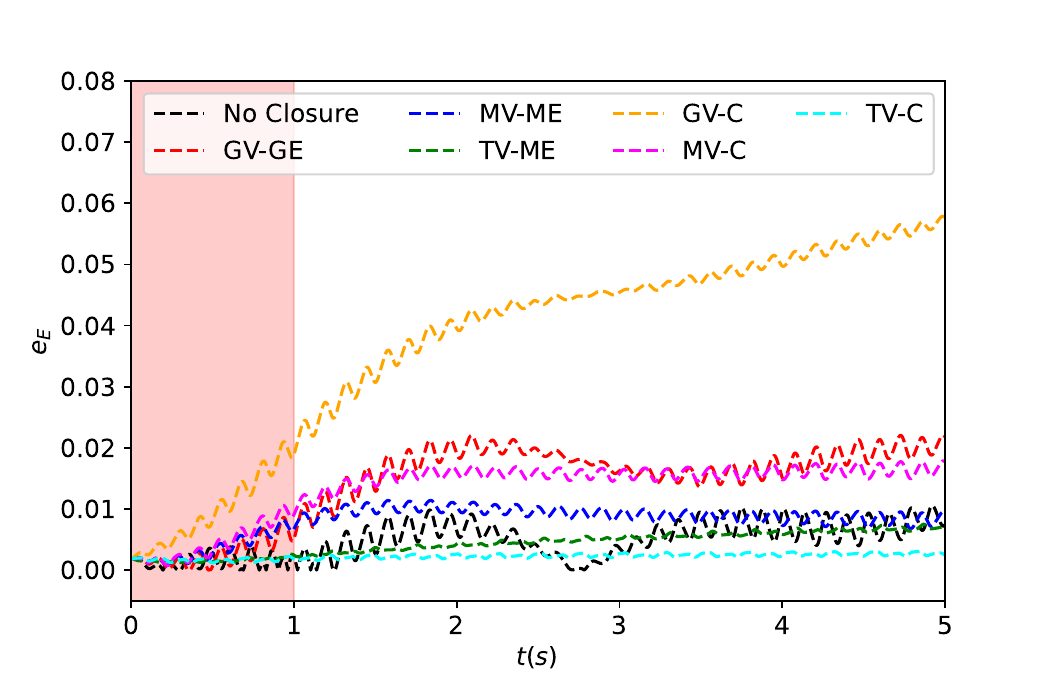}}
    \subfigure[\label{fig:ROM_500_Energy_6mode_ModelComp}]{\includegraphics[width=0.49\textwidth,trim={0cm 0 1cm 0},clip]{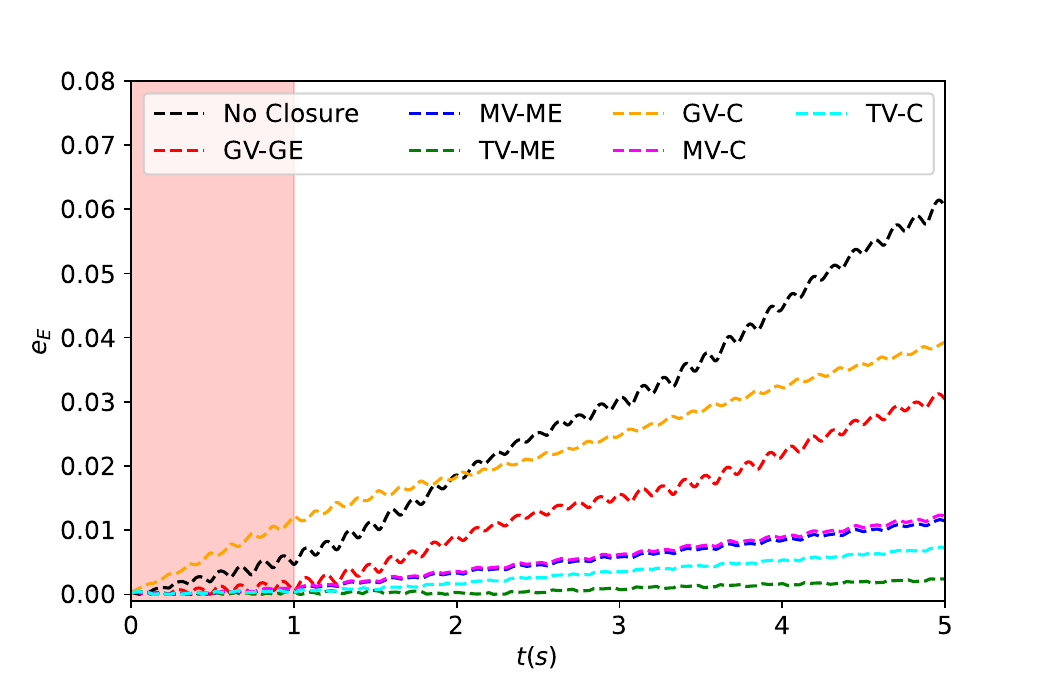}}
    \vspace{-3mm}
    \caption{Error in energy for (a) four modes and (b) six modes for 2-D cylinder flow at $Re = 500$. The shaded area is the time interval from which data is extracted to determine the POD basis, equation operators and closure models.}
    \label{fig:ROM_500_Energy_ModelComp}
\end{figure}

The temporal evolution of error in energy for four and six mode ROMs with different closure models is shown in \figref{ROM_500_Energy_ModelComp}. As the error in energy for the four mode ROM without a closure model is surprisingly low, ROMs with GV-GE, MV-ME, GV-C and MV-C closure models give comparatively worse errors. This lower energy error for the four mode ROM without a closure model is an anomaly, as we will observe by assessing other mode ROMs and looking at different quantities of interest. For the four mode scenario, the ROM with the TV-ME closure model gives results similar to the ROM without a closure model. In contrast, the ROM with the TV-C closure model gives lower errors in energy than ROMs with other models and without a closure model. For the case of six mode ROMs, the ROM without a closure model exhibits a high error that rapidly rises outside the data generation window. ROMs with GV-GE and GV-C closure models display lower errors than those without a closure model. However, this error increases rapidly outside the data generation window. ROMs with MV-ME and MV-C closure models exhibit very low and similar errors. Lastly, ROMs with TV-ME and TV-C closure models exhibit even lower errors, with the former ROM displaying the lowest error among all ROMs. 

\begin{figure}
    \centering
    \subfigure[\label{fig:ROM_500_CL_4mode_ModelComp}]{\includegraphics[width=0.49\textwidth,trim={0cm 0 1cm 0},clip]{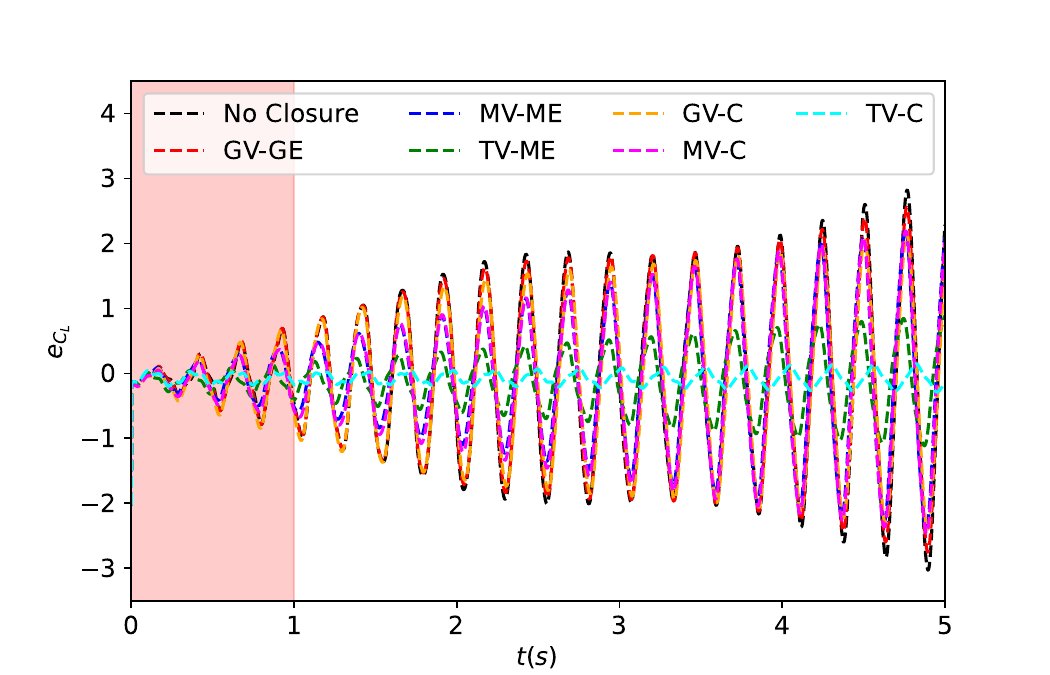}}    
    \subfigure[\label{fig:ROM_500_CL_6mode_ModelComp}]{\includegraphics[width=0.49\textwidth,trim={0cm 0 1cm 0},clip]{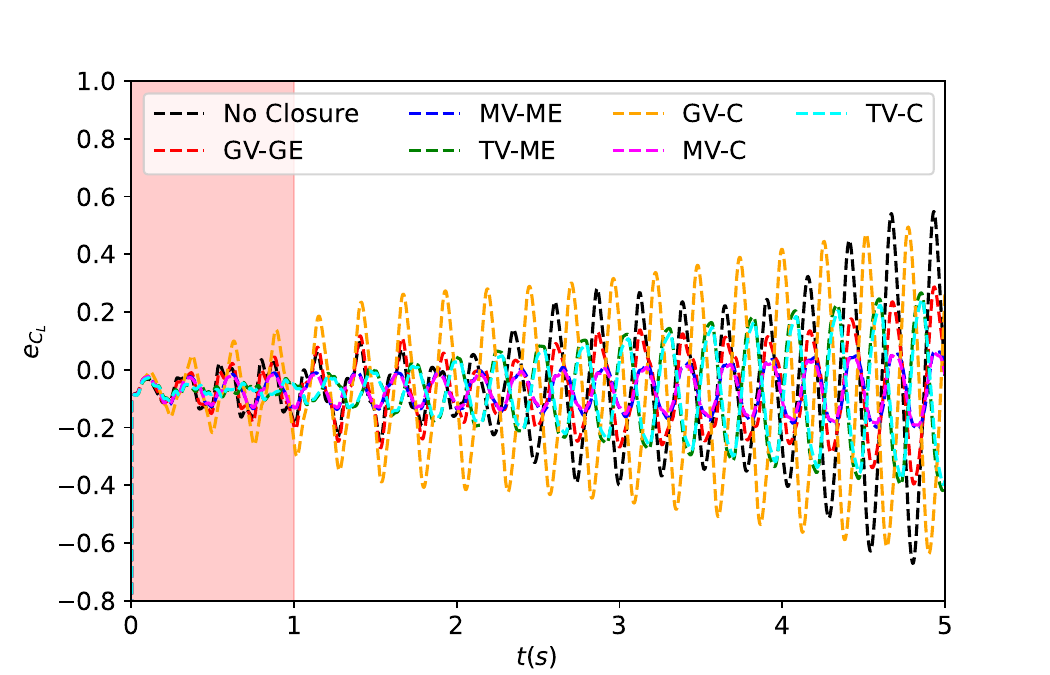}}
    \vspace{-3mm}
    \caption{Error in $C_L$ for (a) four modes and (b) six modes for 2-D cylinder flow at $Re = 500$. The shaded area is the time interval from which data is extracted to determine the POD basis, equation operators and closure models.}
    \label{fig:ROM_500_CL_ModelComp}
\end{figure}

\begin{figure}
    \centering
    \subfigure[\label{fig:ROM_500_CD_4mode_ModelComp}]{\includegraphics[width=0.49\textwidth,trim={0cm 0 1cm 0},clip]{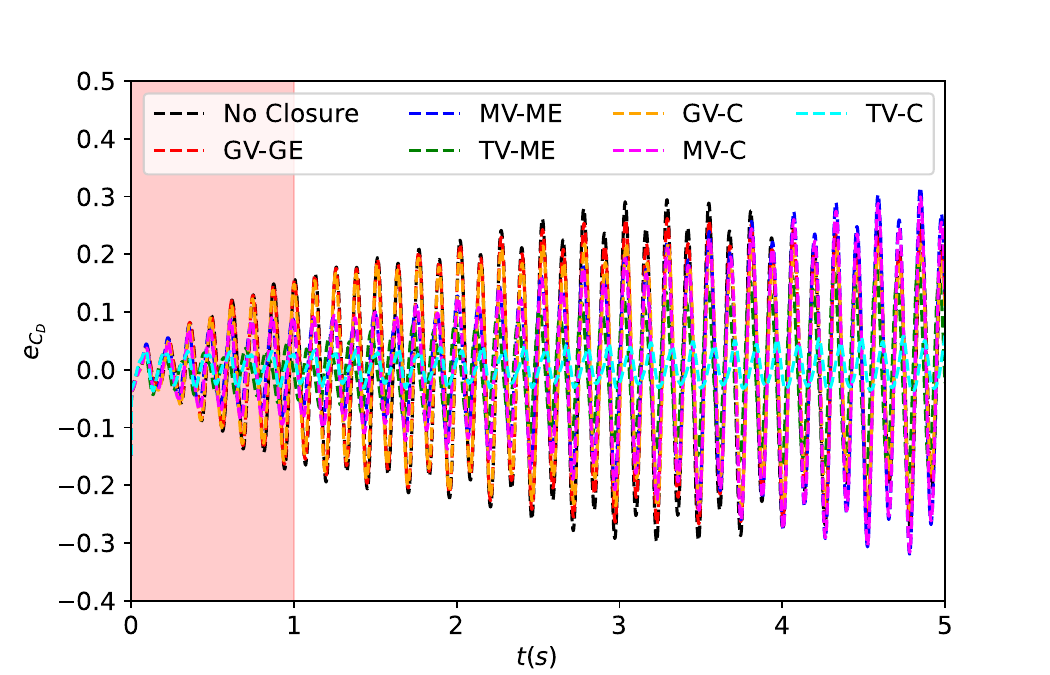}}    
    \subfigure[\label{fig:ROM_500_CD_6mode_ModelComp}]{\includegraphics[width=0.49\textwidth,trim={0cm 0 1cm 0},clip]{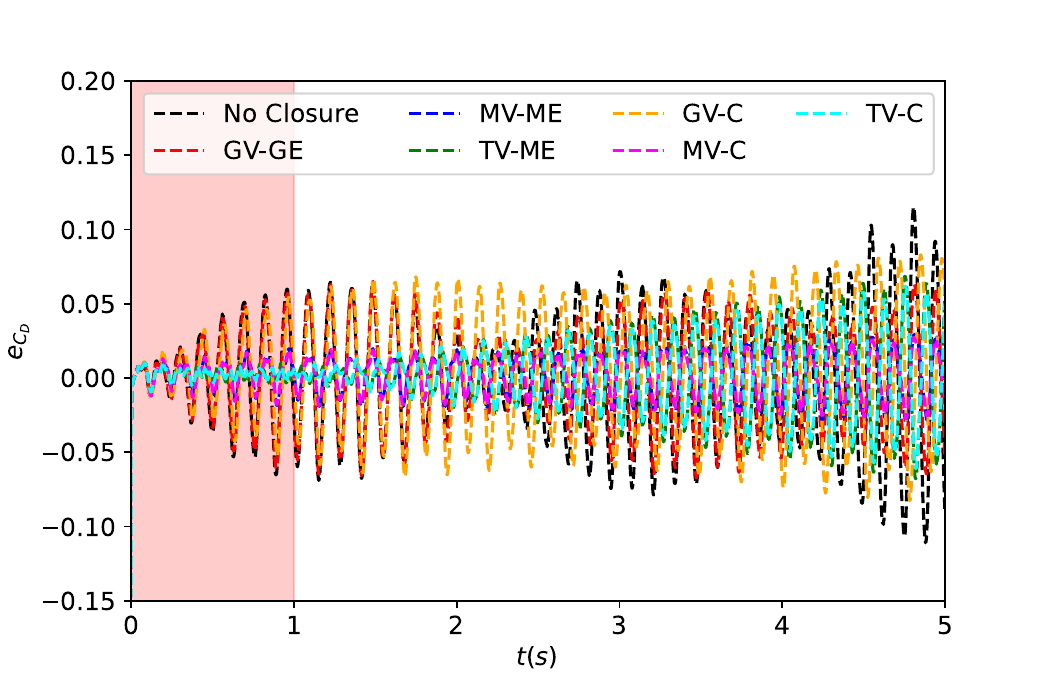}}
    \vspace{-3mm}
    \caption{Error in $C_D$ for (a) four modes and (b) six modes for 2-D cylinder flow at $Re = 500$. The shaded area is the time interval from which data is extracted to determine the POD basis, equation operators and closure models.}
    \label{fig:ROM_500_CD_ModelComp}
\end{figure}

The temporal evolution of error in $C_L$ and $C_D$ for $4$ and $6$ mode ROMs with different closure models is shown in \figref{ROM_500_CL_ModelComp} and \figref{ROM_500_CD_ModelComp} respectively. When GV-GE, GV-C, MV-ME and MV-C closure models are used with ROMs consisting of 4 modes, errors in $C_L$ and $C_D$ are very similar to ROMs without closure models, especially outside the initial data generation window. On the other hand, ROMs with TV-ME and TV-C closure models exhibit a much lower error for these aerodynamic coefficients, with the latter model providing the lowest errors. The results for six mode ROMs indicate that ROMs with MV-ME and MV-C closure models give the lowest error in $C_L$ and $C_D$ for $t > 2s$. ROMs with GV-GE, TV-ME, and TV-C closure models give much lower errors than those without closure models, but this error is slightly higher than those for ROMs with MV-ME and MV-C closure models. 

\begin{figure}
    \centering\subfigure[\label{fig:Int_ROM_500_Energy_None_ModeComp}]{\includegraphics[width=0.33\textwidth,trim={0cm 0 1.5cm 0},clip]{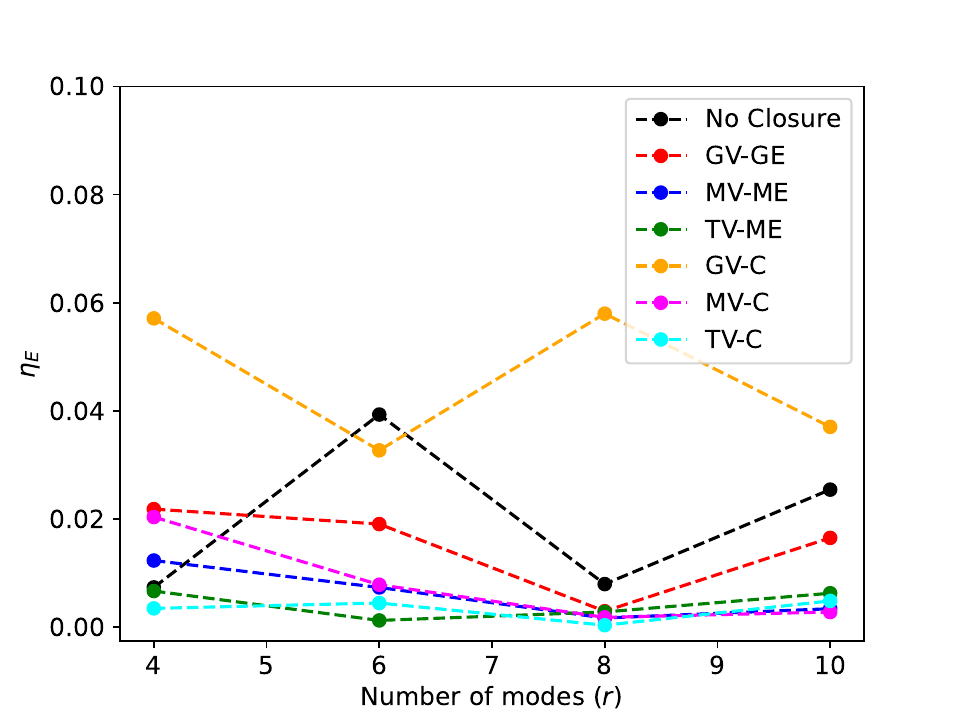}} \subfigure[\label{fig:Int_ROM_500_CL_None_ModeComp}]{\includegraphics[width=0.33\textwidth,trim={0cm 0 1.5cm 0},clip]{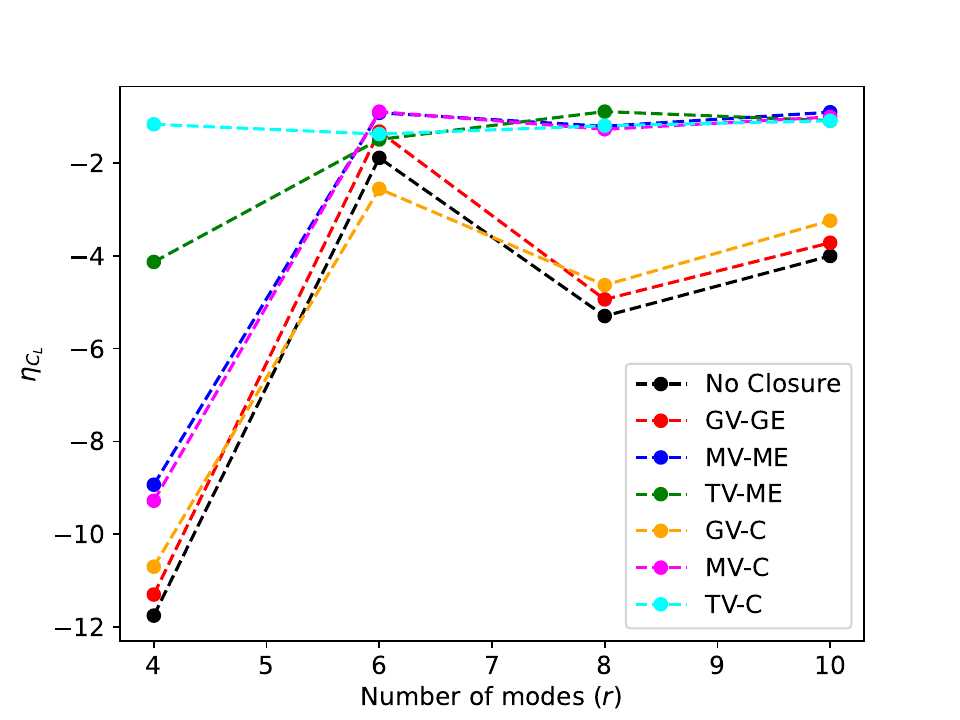}}\subfigure[\label{fig:Int_ROM_500_CD_None_ModeComp}]{\includegraphics[width=0.33\textwidth,trim={0cm 0 1.5cm 0},clip]{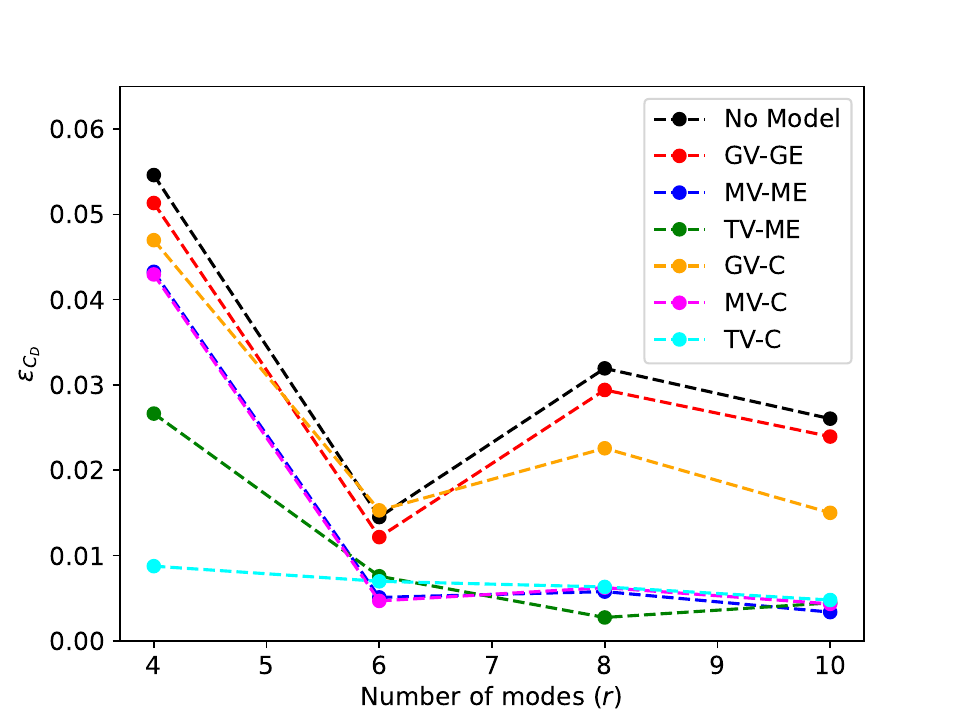}}  
    \vspace{-3mm}
    \caption{Integrated error in (a) energy, (b) $C_L$ and (c) $C_D$ for 2-D cylinder flow at $Re = 500$.}
    \label{fig:Int_ROM_500_None_ModeComp}
\end{figure}

The integrated error in energy, $C_L$ and $C_D$ with different closure models is shown in \figref{Int_ROM_500_None_ModeComp}. The integrated error in energy for ROMs with the GV-C closure model is mostly higher than the ROMs without closure models, indicating this closure model downgrades the performance of ROM compared to the baseline ROMs without any closure model. ROMs with the GV-C closure model exhibit a slightly lower error in $C_L$ and $C_D$ than ROMs without a closure model. ROMs with the GV-GE closure model give lower errors in $C_L$, $C_D$ and energy than ROMs without the closure model for all modes except when six mode ROMs are used. As highlighted earlier, the lower error for four mode ROM without a closure model is not expected and is peculiar for this test case. ROMs with MV-ME and MV-C closure models yield high errors for four mode ROMs. However, these errors decrease significantly for six to 10 mode ROMs. ROMs with TV-ME and TV-C closure models consistently give the lowest errors for all mode ROMs considered in this study. 

\subsection{Cylinder flow at $Re = 1,000$}

\begin{figure}
    \centering
    \subfigure[\label{fig:ROM_1000_Energy_4mode_ModelComp}]{\includegraphics[width=0.49\textwidth,trim={0cm 0 1cm 0},clip]{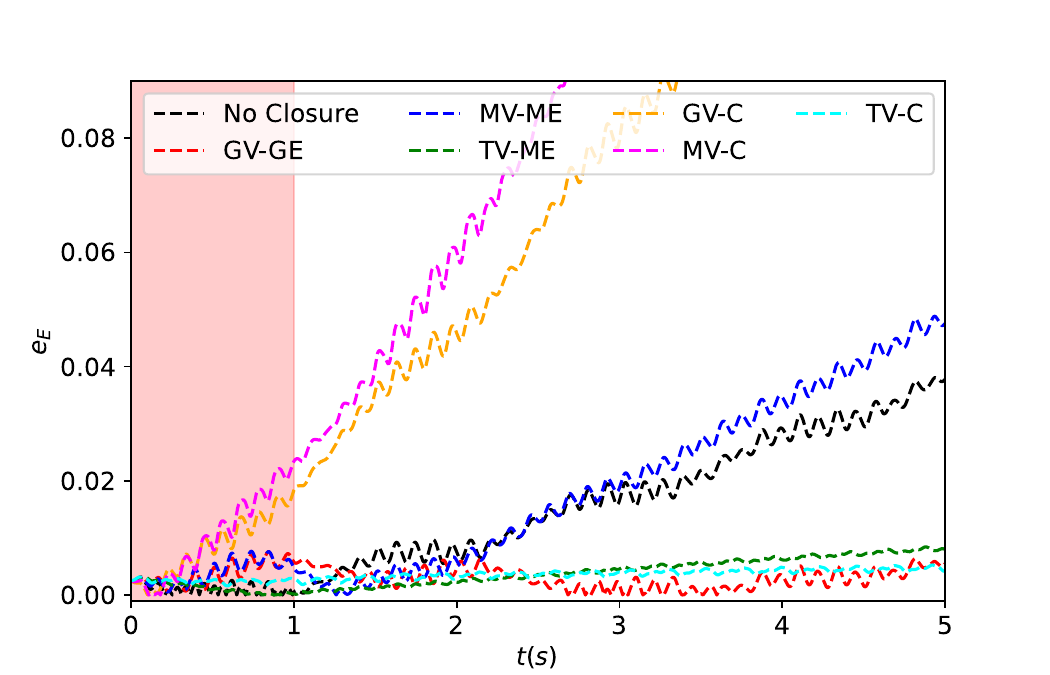}}
    \subfigure[\label{fig:ROM_1000_Energy_6mode_ModelComp}]{\includegraphics[width=0.49\textwidth,trim={0cm 0 1cm 0},clip]{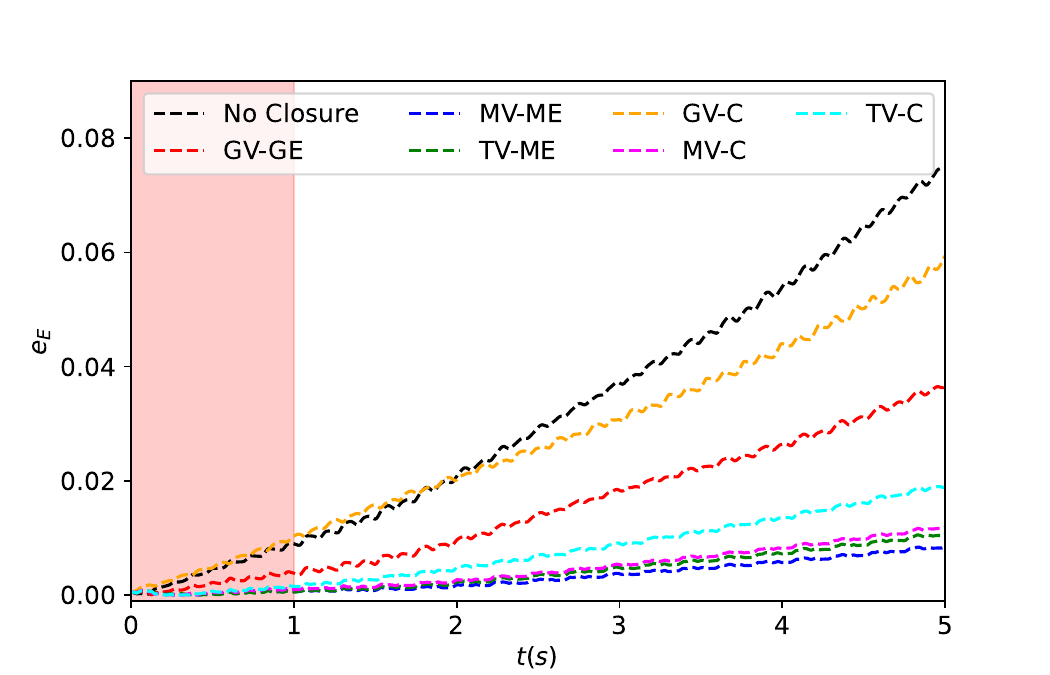}}
    \vspace{-3mm}
    \caption{Error in energy for (a) four modes and (b) six modes for 2-D cylinder flow at $Re = 1,000$. The shaded area is the time interval from which data is extracted to determine the POD basis, equation operators and closure models.}
    \label{fig:ROM_1000_Energy_ModelComp}
\end{figure}

The temporal evolution of error in energy for four and six mode ROMs with different closure models is shown in \figref{ROM_1000_Energy_ModelComp}. For four mode ROMs, the error in energy for ROMs with GV-C, MV-ME and MV-C closure models is higher than the ROM without any closure model. On the other hand, ROMs with GV-GE, TV-ME and TV-C closure models yield much lower errors in energy. For six mode ROMs, ROMs with any closure model deliver lower errors compared to the ROM without a closure model. ROMs with MV-ME, MV-C and TV-ME yield the lowest errors in energy that do not rise considerably even outside the initial time used for computing ROM operators. The errors for the ROM with the TV-C closure model increase after $t > 1$s. However, these errors are still very low, even for the prediction time window. On the other hand, the errors for ROMs with GV-GE and GV-C closure models rise considerably, with the latter giving much higher errors. 

\begin{figure}
    \centering
    \subfigure[\label{fig:ROM_1000_CL_4mode_ModelComp}]{\includegraphics[width=0.49\textwidth,trim={0cm 0 1cm 0},clip]{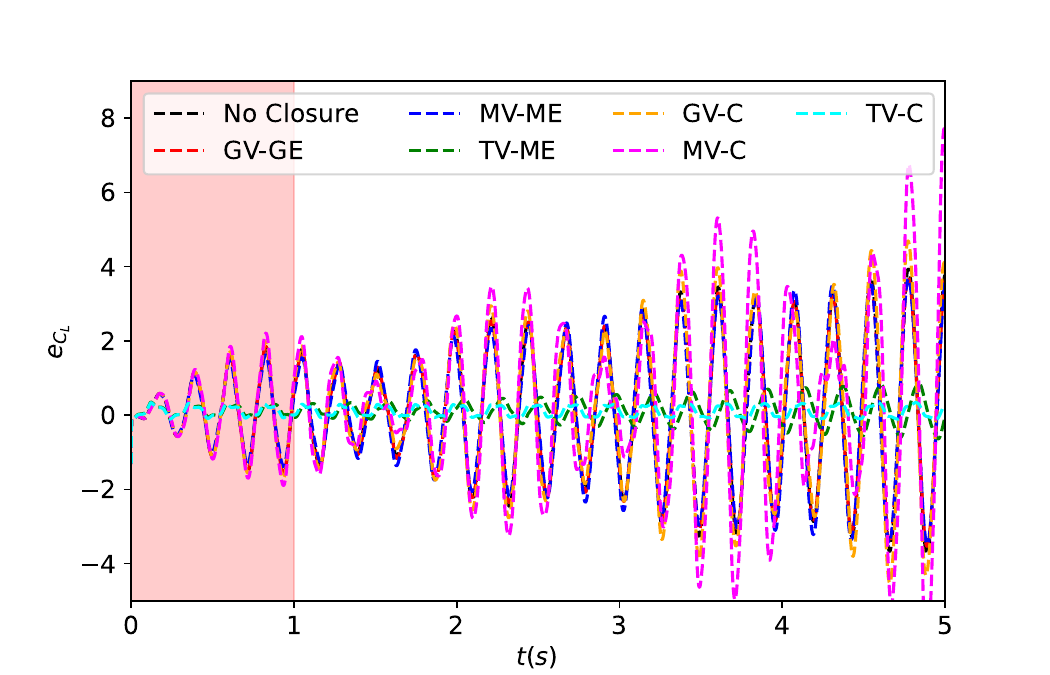}}    
    \subfigure[\label{fig:ROM_1000_CL_6mode_ModelComp}]{\includegraphics[width=0.49\textwidth,trim={0cm 0 1cm 0},clip]{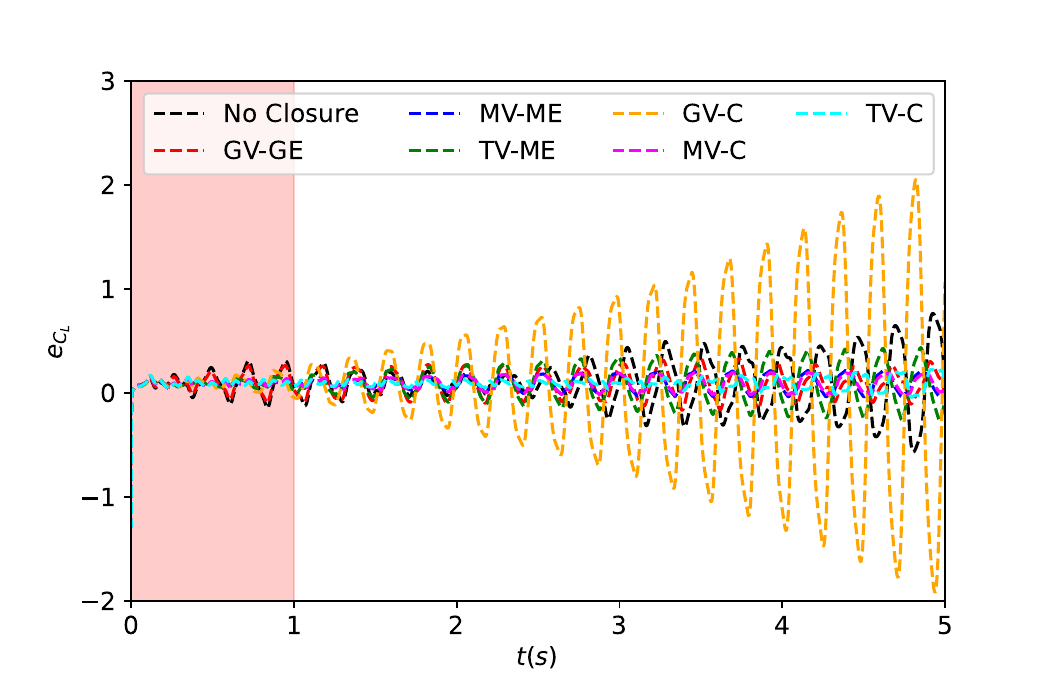}}
    \vspace{-3mm}
    \caption{Error in $C_L$ for (a) four modes and (b) six modes for 2-D cylinder flow at $Re = 1,000$. The shaded area is the time interval from which data is extracted to determine the POD basis, equation operators and closure models.}
    \label{fig:ROM_1000_CL_ModelComp}
\end{figure}
\begin{figure}
    \centering
    \subfigure[\label{fig:ROM_1000_CD_4mode_ModelComp}]{\includegraphics[width=0.49\textwidth,trim={0cm 0 1cm 0},clip]{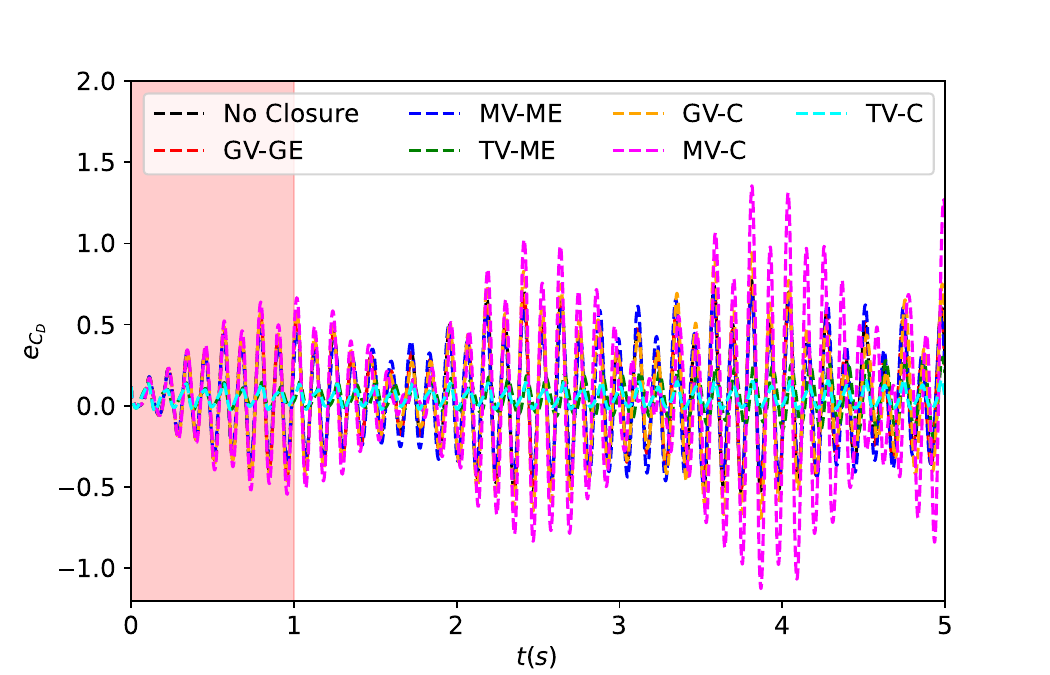}}    
    \subfigure[\label{fig:ROM_1000_CD_6mode_ModelComp}]{\includegraphics[width=0.49\textwidth,trim={0cm 0 1cm 0},clip]{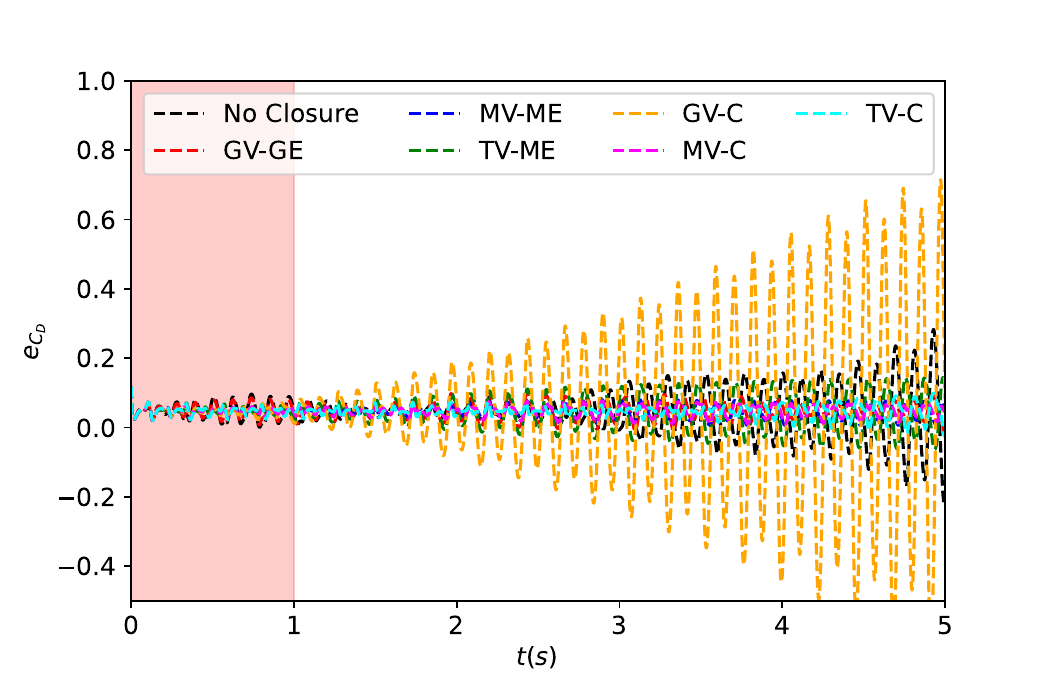}}    
    \vspace{-3mm}
    \caption{Error in $C_D$ for (a) four modes and (b) six modes for 2-D cylinder flow at $Re = 1,000$. The shaded area is the time interval from which data is extracted to determine the POD basis, equation operators and closure models.}
    \label{fig:ROM_1000_CD_ModelComp}
\end{figure}

The temporal evolution of error in $C_L$ and $C_D$ for four and six mode ROMs with different closure models is shown in \figref{ROM_1000_CL_ModelComp} and \figref{ROM_1000_CD_ModelComp} respectively. For four mode ROMs, the error in $C_L$ and $C_D$ is much lower when TV-ME and TV-C closure models are used than ROMs with other closure models and without a closure model. Similarly, for six mode ROMs, ROMs with MV-ME, MV-C and TV-C closure models yield the lowest errors, whereas ROMs with GV-GE and TV-ME closure models exhibit slightly higher errors. The errors for ROMs with these closure models are still much lower than those without a closure model, indicating the feasibility of most closure models, except for the ROM with the GV-C closure model. 

\begin{figure}
    \centering
    \subfigure[\label{fig:Int_ROM_1000_Energy_None_ModeComp}]{\includegraphics[width=0.33\textwidth,trim={0cm 0 1.5cm 0},clip]{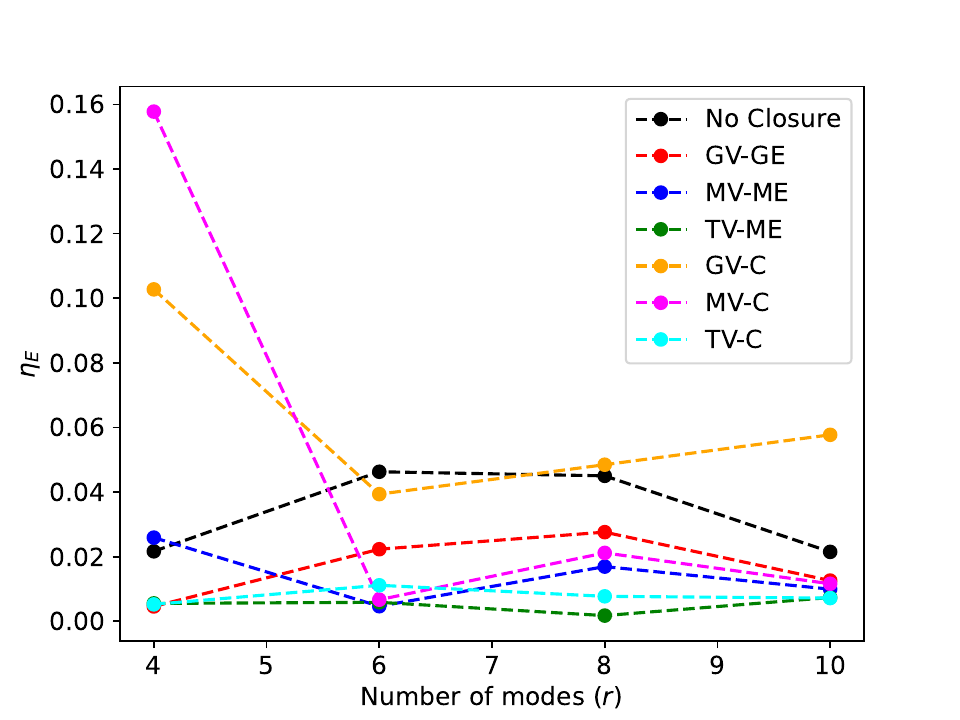}}\subfigure[\label{fig:Int_ROM_1000_CL_None_ModeComp}]{\includegraphics[width=0.33\textwidth,trim={0cm 0 1.5cm 0},clip]{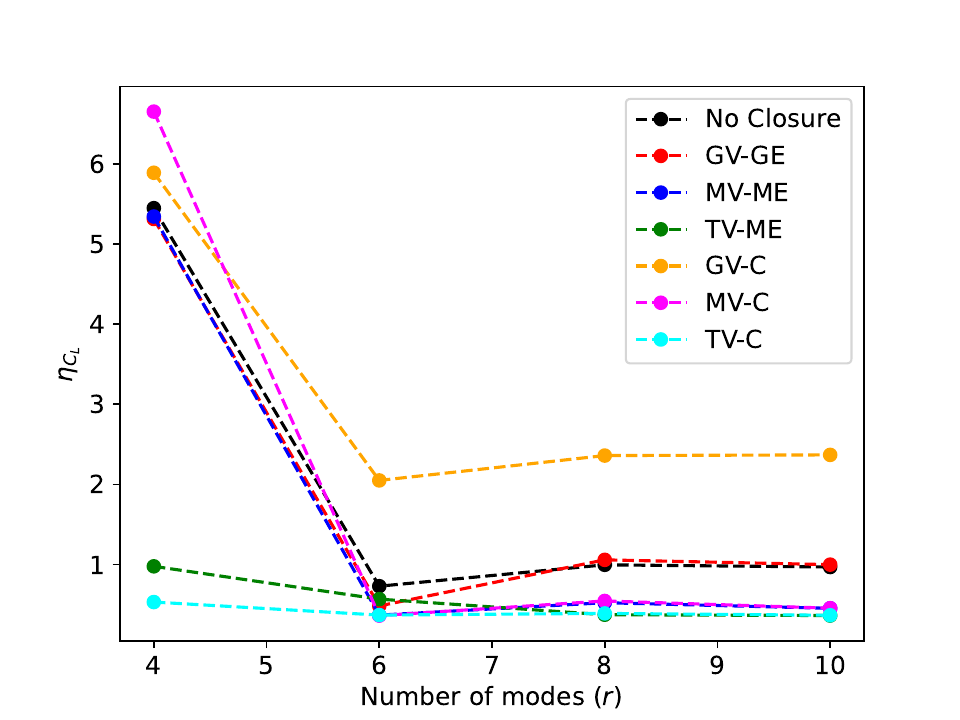}}\subfigure[\label{fig:Int_ROM_1000_CD_None_ModeComp}]{\includegraphics[width=0.33\textwidth,trim={0cm 0 1.5cm 0},clip]{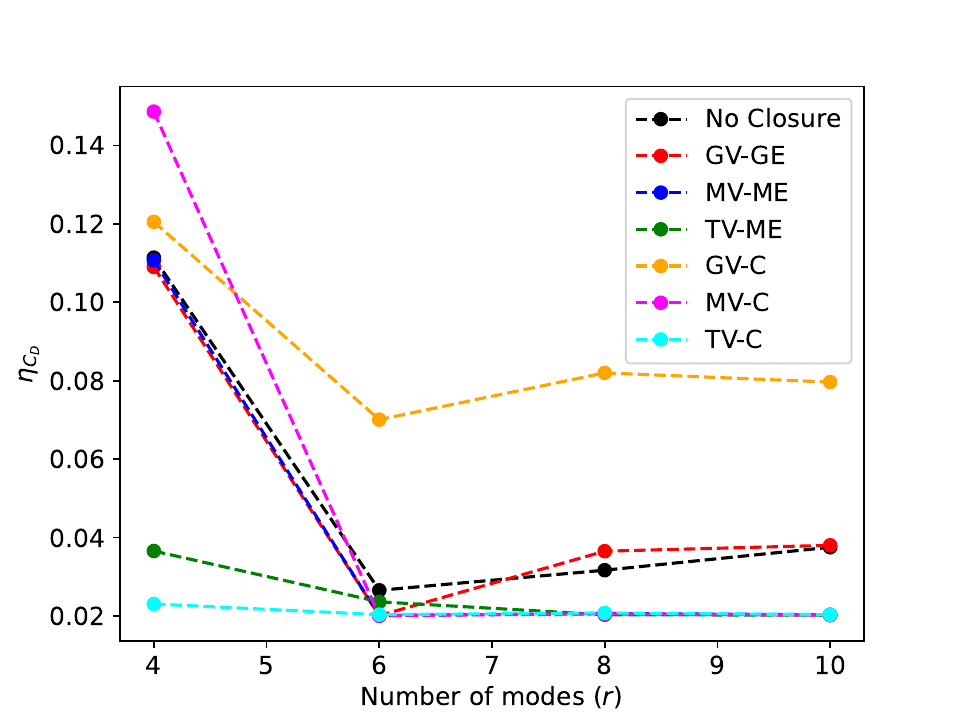}}  
    \vspace{-3mm}
    \caption{Integrated error in (a) energy, (b) $C_L$ and (c) $C_D$ for 2-D cylinder flow at $Re = 1,000$.}
    \label{fig:Int_ROM_1000_None_ModeComp}
\end{figure}

The integrated error in energy, $C_L$ and $C_D$ with different closure models is shown in \figref{Int_ROM_1000_None_ModeComp}. Similar to the results for the other two Reynolds numbers, the ROMs with the GV-C closure model exhibit higher errors than ROMs without closure models for all mode ROMs considered in this article. ROMs with the GV-GE closure model show a significantly lower energy error than ROMs without a closure model. However, the differences in errors in aerodynamic coefficients between ROMs with these two closure models are minor. ROMs with TV-ME and TV-C closure models yield the lowest errors in all quantities for all modes, where the latter gives slightly better $C_L$ and $C_D$ for four modes and the former yields more accurate energy for six and eight mode ROMs. ROMs with MV-ME and MV-C closure models also yield similar low error at a higher number of modes, that is, for six to 10 mode ROMs. However, this error rises sharply for four mode ROMs. These results indicate that ROMs with TV-ME and TV-C closure models yield consistently low errors for all modes; therefore, these are the preferred closure models.

\subsection{Discussion}

The following conclusions can be drawn from the results obtained from the flow over a 2-D cylinder at three Reynolds numbers:
\begin{itemize}
    \item The incremental pressure correction scheme provides accurate estimates of the several quantities of interest for different mode selections and gives significantly lower errors when coupled with an appropriate closure model. 
    \item A constant artificial viscosity model form may not be expressive enough to resolve the closure problem and provide significantly lower errors than Galerkin ROMs without a closure model. 
    \item The parameters of modal and tensor artificial viscosity closure models obtained using the two calibration techniques, based on energy and closure term approximation, do not lead to an appreciable difference in the closure model performance.
    \item Both modal and tensor artificial viscosity model forms are sufficient for a higher number of modes to account for the interaction of unresolved states on resolved states. However, at a lower number of modes, ROMs with tensor artificial viscosity models consistently yield better results than ROMs with modal artificial viscosity closure models. 
\end{itemize}
These observations inform decision-making for selecting the appropriate solution method and closure models for ROMs for incompressible flows. The inapplicability of global artificial viscosity closure models is demonstrated through results indicating that GV-GE and GV-C closure models do not consistently yield lower errors than ROMs without a closure model. This conclusion resonates with the observation in \citep{Osth2014}, where a modal artificial viscosity was used to overcome this drawback. A comparison of the two calibration techniques indicated that energy transfer between unresolved and resolved states is sufficient to account for the non-linear dynamics, even for the highest Reynolds numbers under consideration.  This result echoes the observation in \citep{Couplet2003} where the mean modal artificial viscosities computed using two calibration techniques are similar. Lastly, the modal artificial viscosity closure form is insufficient to capture the dynamics between the more energetic resolved and unresolved states, which appear to be better handled by a tensor artificial viscosity closure form. These conclusions guide the selection of appropriate model forms for even higher Reynolds number flows where the closure problem becomes even more prominent and ROMs with more modes become less favorable due to higher evaluation costs during the online stage. 


\section{Conclusions}

ROMs are essential for multi-query applications, real-time control and dynamics forecasting. These models must achieve high accuracy while exhibiting a lower online computation cost. The cost of the online stage scales as $r^3$ where $r$ is the number of resolved states for incompressible fluid flows. Therefore, it is desirable to use a lower number of resolved states. However, with a decrease in the number of resolved states, the need for accounting for the interaction between unresolved and resolved states becomes more important. The first contribution of this article is to propose closure models for the interaction between the unresolved and resolved states. We consider six different closure models determined using three model forms, based on global, modal and tensor artificial viscosity, where the unknown coefficients are determined by two calibration techniques: least squares minimization of error in energy approximation and closure term approximation. These closure models can be considered hybrid data-physics models as physical arguments inspire the model form, but the model parameters are directly learned through raw data. The proposed models did not require regularization for the cases considered in this study, unlike other common data-driven closure models \citep{Xie2018, Mohebujjaman2019}. The flow over a 2D cylinder is considered at three Reynolds numbers and errors in energy, $C_L$ and $C_D$ are compared to validate these closure models. The results indicate that modal and tensor artificial viscosity model forms yield more accurate ROMs than those without any closure model. ROMs with tensor artificial viscosity model form deliver even more consistent results as the results did not deteriorate even for four mode ROMs, indicating that this model form is the most consistent amongst those considered in this article. Lastly, the two calibration techniques used for determining the unknown parameters in the closure model forms, least squares minimization for energy error or closure term error, provide closure models with slight differences in results. This observation indicates that both calibration techniques are equally well suited for determining modal and tensor artificial viscosity closure model parameters.

ROMs for incompressible fluid flows exhibit coupling between the reduced pressure states and reduced velocity states, where the primary role of the reduced pressure states is to enforce the incompressibility constraint. The unique structure of these equations results in a saddle-point problem that requires special solution techniques to obtain the dynamic evolution of velocity and pressure states. The second contribution of this article is to demonstrate the applicability of a solution technique: incremental pressure correction for projection-based ROMs. This method is adapted from the incremental pressure correction scheme readily used in full order modeling of incompressible fluid flows \citep{VanKan1986}. The ROMs obtained through this solution method give low errors for incompressible flows under consideration, thereby demonstrating the applicability of this method for ROM of incompressible fluid flows.

As more complex and challenging fluid flow applications that require ROMs emerge, the solution method and the closure models proposed in this article will require further validation to ensure the accuracy remains. Furthermore, a detailed numerical analysis of the stability and convergence of the incremental pressure correction scheme in the context of ROMs will be performed. Alternative solution methods for saddle-point problems, such as velocity correction schemes, exist in the full order modeling literature. The applicability of schemes for ROMs must be validated. The closure models considered in this article are linear as they depend linearly on reduced states. These models could not adequately account for the interaction between resolved and unresolved states when less than four modes were used for ROMs. Non-linear model forms have the potential to be more accurate for such a lower number of modes and will be explored in future work. Despite the potential to be more accurate than linear models, these models would involve a higher online stage cost. Therefore, a comprehensive study is required to investigate the trade-off between cost and accuracy. As part of future work, we also plan to explore novel applications of ROMs in material transport regulation of neurite networks \cite{Li2021, Li2022, Li2022b, Li2023} and heat exchanger design for additive manufacturing \cite{Liang2022, Liang2022b} where complex and challenging fluid flows need to be addressed.

\section{Acknowledgements}

The authors would like to acknowledge the support from the National Science Foundation (NSF) grant, CMMI-1953323, for the funds used towards this project. The research in this paper was sponsored by the Army Research Laboratory and was accomplished under Cooperative Agreement Number W911NF-20-2-0175. The views and conclusions contained in this document are those of the authors and should not be interpreted as representing the official policies, either expressed or implied, of the Army Research Laboratory or the U.S. Government. The U.S. Government is authorized to reproduce and distribute reprints for Government purposes notwithstanding any copyright notation herein.

\bibliographystyle{unsrt}
\bibliography{main_bbl.bib}

\begin{thebibliography}{10}

\bibitem{Sirovich1987}
L.~Sirovich.
\newblock {Turbulence and the dynamics of coherent structures part III: Dynamics and scaling }.
\newblock {\em Quarterly of Applied Mathematics}, 45(3):583--590, 1987.

\bibitem{Aubry1988}
N.~Aubry, P.~Holmes, J.~L. Lumley, and E.~Stone.
\newblock The dynamics of coherent structures in the wall region of a turbulent boundary layer.
\newblock {\em Journal of Fluid Mechanics}, 192:115–173, 1988.

\bibitem{Lumley1967}
J.~L. Lumley.
\newblock The structure of inhomogeneous turbulent flows.
\newblock In {\em Atmospheric Turbulence and Radio Wave Propagation}, pages 166--178. Moscow: Nauka, 1967.

\bibitem{Lumley1981}
J.~L. Lumley.
\newblock Coherent structures in turbulence.
\newblock In Richard~E. Meyer, editor, {\em Transition and Turbulence}, pages 215--242. Academic Press, 1981.

\bibitem{Hesthaven2016}
J.~S. Hesthaven, G.~Rozza, and B.~Stamm.
\newblock {\em Certified Reduced Basis Methods for Parametrized Partial Differential Equations}.
\newblock Springer, 2016.

\bibitem{Sirisup2004}
S.~Sirisup and G.~E. Karniadakis.
\newblock {A spectral viscosity method for correcting the long-term behavior of POD models}.
\newblock {\em Journal of Computational Physics}, 194(1):92--116, 2004.

\bibitem{Wang2012}
Z.~Wang, I.~Akhtar, J.~Borggaard, and T.~Iliescu.
\newblock Proper orthogonal decomposition closure models for turbulent flows: A numerical comparison.
\newblock {\em Computer Methods in Applied Mechanics and Engineering}, 237-240:10--26, 2012.

\bibitem{Bergmann2009}
M.~Bergmann, C.-H. Bruneau, and A.~Iollo.
\newblock {Enablers for robust POD models}.
\newblock {\em Journal of Computational Physics}, 228(2):516--538, 2009.

\bibitem{Carlberg2011}
K.~Carlberg, C.~Bou-Mosleh, and C.~Farhat.
\newblock {Efficient non-linear model reduction via a least-squares Petrov–Galerkin projection and compressive tensor approximations}.
\newblock {\em International Journal for Numerical Methods in Engineering}, 86(2):155--181, 2011.

\bibitem{Ahmed2021}
S.~E. Ahmed, S.~Pawar, O.~San, A.~Rasheed, T.~Iliescu, and B.~R. Noack.
\newblock {On closures for reduced order models - A spectrum of first-principle to machine-learned avenues}.
\newblock {\em Physics of Fluids}, 33(9), 2021.

\bibitem{Lee2017}
K.~Carlberg, M.~Barone, and H.~Antil.
\newblock {Galerkin v. least-squares Petrov–Galerkin projection in nonlinear model reduction}.
\newblock {\em Journal of Computational Physics}, 330:693--734, 2017.

\bibitem{Parish2020}
E.~J. Parish, C.~R. Wentland, and K.~Duraisamy.
\newblock {The adjoint Petrov–Galerkin method for non-linear model reduction}.
\newblock {\em Computer Methods in Applied Mechanics and Engineering}, 365:112991, 2020.

\bibitem{Parish2023}
E.~Parish, M.~Yano, I.~Tezaur, and T.~Iliescu.
\newblock Residual-based stabilized reduced-order models of the transient convection-diffusion-reaction equation obtained through discrete and continuous projection.
\newblock {\em arXiv: 2302.09355}, 2023.

\bibitem{Iliescu2014}
T.~Iliescu and Z.~Wang.
\newblock {Variational multiscale proper orthogonal decomposition: Navier-Stokes equations}.
\newblock {\em Numerical Methods for Partial Differential Equations}, 30(2):641--663, 2014.

\bibitem{Rempfer1994}
D.~Rempfer and H.~F. Fasel.
\newblock Dynamics of three-dimensional coherent structures in a flat-plate boundary layer.
\newblock {\em Journal of Fluid Mechanics}, 275:257–283, 1994.

\bibitem{Couplet2003}
M.~Couplet, P.~Sagaut, and C.~Basdevant.
\newblock {Intermodal energy transfers in a proper orthogonal decomposition – Galerkin representation of a turbulent separated flow}.
\newblock {\em Journal of Fluid Mechanics}, 491:275–284, 2003.

\bibitem{Borggard2011}
J.~Borggaard, T.~Iliescu, and Z.~Wang.
\newblock Artificial viscosity proper orthogonal decomposition.
\newblock {\em Mathematical and Computer Modelling}, 53(1):269--279, 2011.

\bibitem{Osth2014}
J.~Östh, B.~R. Noack, S.~Krajnović, D.~Barros, and J.~Borée.
\newblock {On the need for a nonlinear subscale turbulence term in POD models as exemplified for a high-Reynolds-number flow over an Ahmed body}.
\newblock {\em Journal of Fluid Mechanics}, 747:518–544, 2014.

\bibitem{Xie2017}
X.~Xie, D.~Wells, Z.~Wang, and T.~Iliescu.
\newblock Approximate deconvolution reduced order modeling.
\newblock {\em Computer Methods in Applied Mechanics and Engineering}, 313:512--534, 2017.

\bibitem{Reyes2020}
R.~Reyes and R.~Codina.
\newblock Projection-based reduced order models for flow problems: A variational multiscale approach.
\newblock {\em Computer Methods in Applied Mechanics and Engineering}, 363:112844, 2020.

\bibitem{Xie2018b}
X.~Xie, D.~Wells, Z.~Wang, and T.~Iliescu.
\newblock Numerical analysis of the leray reduced order model.
\newblock {\em Journal of Computational and Applied Mathematics}, 328:12--29, 2018.

\bibitem{Cordier2013}
L.~Cordier, B.~R. Noack, G.~Tissot, G.~Lehnasch, J.~Delville, M.~Balajewicz, G.~Daviller, and R.~K. Niven.
\newblock {Identification strategies for model-based control topics in flow control.}
\newblock {\em Experiments in Fluids}, 54(8), 2013.

\bibitem{Protas2015}
B.~Protas, B.~R. Noack, and J.~Östh.
\newblock {Optimal nonlinear eddy viscosity in Galerkin models of turbulent flows}.
\newblock {\em Journal of Fluid Mechanics}, 766:337–367, 2015.

\bibitem{Ahmed2020}
S.~E. Ahmed, K.~Bhar, O.~San, and A.~Rasheed.
\newblock Forward sensitivity approach for estimating eddy viscosity closures in nonlinear model reduction.
\newblock {\em Physics Review E}, 102, 2020.

\bibitem{Xie2018}
X.~Xie, M.~Mohebujjaman, L.~G. Rebholz, and T.~Iliescu.
\newblock Data-driven filtered reduced order modeling of fluid flows.
\newblock {\em SIAM Journal on Scientific Computing}, 40(3):B834--B857, 2018.

\bibitem{Duraisamy2019}
K.~Duraisamy, G.~Iaccarino, and H.~Xiao.
\newblock Turbulence modeling in the age of data.
\newblock {\em Annual Review of Fluid Mechanics}, 51(1):357--377, 2019.

\bibitem{Prakash2022}
A.~Prakash, K.~E. Jansen, and J.~A. Evans.
\newblock Invariant data-driven subgrid stress modeling in the strain-rate eigenframe for large eddy simulation.
\newblock {\em Computer Methods in Applied Mechanics and Engineering}, 399:115457, 2022.

\bibitem{Prakash2023a}
A.~Prakash, K.~E. Jansen, and J.~A. Evans.
\newblock Invariant data-driven subgrid stress modeling on anisotropic grids for large eddy simulation.
\newblock {\em arXiv: 2212.00332}, 2023.

\bibitem{Prakash2023c}
A.~Prakash.
\newblock {\em Scale-Resolving Simulations and Data-Driven Subgrid Modeling for Complex Turbulent Boundary Layer Flows}.
\newblock PhD thesis, University of Colorado Boulder, 2023.

\bibitem{Koc2022}
B.~Koc, C.~Mou, H.~Liu, Z.~Wang, G.~Rozza, and T.~Iliescu.
\newblock Verifiability of the data-driven variational multiscale reduced order model.
\newblock {\em Journal of Scientific Computing}, 93(2):54, 2022.

\bibitem{Mohebujjaman2019}
M.~Mohebujjaman, L.G. Rebholz, and T.~Iliescu.
\newblock Physically constrained data-driven correction for reduced-order modeling of fluid flows.
\newblock {\em International Journal for Numerical Methods in Fluids}, 89(3):103--122, 2019.

\bibitem{San2018}
O.~San and R.~Maulik.
\newblock {Neural network closures for nonlinear model order reduction}.
\newblock {\em Advances in Computational Mathematics}, 44(6):1717--1750, 2018.

\bibitem{Cordier2010}
L.~Cordier, B.~Abou El~Majd, and J.~Favier.
\newblock {Calibration of POD reduced-order models using Tikhonov regularization}.
\newblock {\em International Journal for Numerical Methods in Fluids}, 63(2):269--296, 2010.

\bibitem{Noack2005}
B.~R. Noack, P.~Papas, and P.~A. Monkewitz.
\newblock {The need for a pressure-term representation in empirical Galerkin models of incompressible shear flows}.
\newblock {\em Journal of Fluid Mechanics}, 523:339–365, 2005.

\bibitem{Caiazzo2014}
A.~Caiazzo, T.~Iliescu, V.~John, and S.~Schyschlowa.
\newblock A numerical investigation of velocity–pressure reduced order models for incompressible flows.
\newblock {\em Journal of Computational Physics}, 259:598--616, 2014.

\bibitem{Ballarin2015}
F.~Ballarin, A.~Manzoni, A.~Quarteroni, and G.~Rozza.
\newblock {Supremizer stabilization of POD–Galerkin approximation of parametrized steady incompressible Navier–Stokes equations}.
\newblock {\em International Journal for Numerical Methods in Engineering}, 102(5):1136--1161, 2015.

\bibitem{Decaria2020}
V.~DeCaria, T.~Iliescu, W.~Layton, M.~McLaughlin, and M.~Schneier.
\newblock An artificial compression reduced order model.
\newblock {\em SIAM Journal on Numerical Analysis}, 58(1):565--589, 2020.

\bibitem{Robino2020}
Samuele Rubino.
\newblock {Numerical analysis of a projection-based stabilized POD-ROM for incompressible flows}.
\newblock {\em SIAM Journal on Numerical Analysis}, 58(4):2019--2058, 2020.

\bibitem{Guermond2006}
J.~L. Guermond, P.~Minev, and J.~Shen.
\newblock An overview of projection methods for incompressible flows.
\newblock {\em Computer Methods in Applied Mechanics and Engineering}, 195(44):6011--6045, 2006.

\bibitem{Prakash2023b}
A.~Prakash, K.~E. Jansen, and J.~A. Evans.
\newblock {Extension of the Smagorinsky subgrid stress model to anisotropic filters}.
\newblock In {\em AIAA SCITECH 2023 Forum}, 2023.

\bibitem{Abba2023}
A.~Abb{\`a} and L.~Valdettaro.
\newblock {Tensorial turbulent viscosity model for LES: properties and applications}.
\newblock In {\em From Kinetic Theory to Turbulence Modeling}, pages 9--20, Singapore, 2023. Springer Nature Singapore.

\bibitem{Babuska1973}
I.~Babu{\v{s}}ka.
\newblock {The finite element method with Lagrangian multipliers}.
\newblock {\em Numerische Mathematik}, 20(3):179--192, 1973.

\bibitem{Brezzi1974}
F.~Brezzi.
\newblock {On the existence, uniqueness and approximation of saddle-point problems arising from Lagrangian multipliers}.
\newblock {\em R.A.I.R.O. Analyse Numérique}, 8:129--151, 1974.

\bibitem{Rowley2004}
C.~W. Rowley, T.~Colonius, and R.~M. Murray.
\newblock {Model reduction for compressible flows using POD and Galerkin projection}.
\newblock {\em Physica D: Nonlinear Phenomena}, 189(1):115--129, 2004.

\bibitem{Kalashnikova2010}
I.~Kalashnikova and M.~F. Barone.
\newblock {On the stability and convergence of a Galerkin reduced order model (ROM) of compressible flow with solid wall and far-field boundary treatment}.
\newblock {\em International Journal for Numerical Methods in Engineering}, 83(10):1345--1375, 2010.

\bibitem{Greif2019}
C.~Greif and K.~Urban.
\newblock {Decay of the Kolmogorov N-width for wave problems}.
\newblock {\em Applied Mathematics Letters}, 96:216--222, 2019.

\bibitem{Hughes1986}
T.~J.R. Hughes, L.~P. Franca, and M.~Balestra.
\newblock {A new finite element formulation for computational fluid dynamics: V. Circumventing the Babuška-Brezzi condition: a stable Petrov-Galerkin formulation of the Stokes problem accommodating equal-order interpolations}.
\newblock {\em Computer Methods in Applied Mechanics and Engineering}, 59(1):85--99, 1986.

\bibitem{Chorin1968}
A.~J. Chorin.
\newblock {Numerical Solution of the Navier-Stokes Equations}.
\newblock {\em Mathematics of Computation}, 22(104):745--762, 1968.

\bibitem{Temam1968}
R~T{\'{e}}mam.
\newblock {Sur l'approximation de la solution des {\'{e}}quations de Navier-Stokes par la m{\'{e}}thode des pas fractionnaires (I)}.
\newblock {\em Archive for Rational Mechanics and Analysis}, 32(2):135--153, 1969.

\bibitem{VanKan1986}
J.~V.~Kan.
\newblock A second-order accurate pressure-correction scheme for viscous incompressible flow.
\newblock {\em SIAM Journal on Scientific and Statistical Computing}, 7(3):870--891, 1986.

\bibitem{Taylor1973}
C.~Taylor and P.~Hood.
\newblock {A numerical solution of the Navier-Stokes equations using the finite element technique}.
\newblock {\em Computers \& Fluids}, 1(1):73--100, 1973.

\bibitem{Rozza2007}
G.~Rozza and K.~Veroy.
\newblock {On the stability of the reduced basis method for Stokes equations in parametrized domains}.
\newblock {\em Computer Methods in Applied Mechanics and Engineering}, 196(7):1244--1260, 2007.

\bibitem{Guermond1998}
J.~L. Guermond and L.~Quartapelle.
\newblock {On stability and convergence of projection methods based on pressure Poisson equation}.
\newblock {\em International Journal for Numerical Methods in Fluids}, 26(9):1039--1053, 1998.

\bibitem{Rannacher1992}
R.~Rannacher.
\newblock {On Chorin's projection method for the incompressible Navier-Stokes equations}.
\newblock In {\em The Navier-Stokes Equations II --- Theory and Numerical Methods}, pages 167--183. Springer Berlin Heidelberg, 1992.

\bibitem{Akhtar2009}
I.~Akhtar, A.~H. Nayfeh, and C.~J. Ribbens.
\newblock {On the stability and extension of reduced-order Galerkin models in incompressible flows}.
\newblock {\em Theoretical and Computational Fluid Dynamics}, 23(3):213--237, 2009.

\bibitem{Schafer1996}
M.~Sch{\"a}fer, S.~Turek, F.~Durst, E.~Krause, and R.~Rannacher.
\newblock Benchmark computations of laminar flow around a cylinder.
\newblock In {\em Flow Simulation with High-Performance Computers II: DFG Priority Research Programme Results 1993--1995}, pages 547--566. Vieweg+Teubner Verlag, 1996.

\bibitem{Baratta2023}
I.~A. Baratta, J.~P. Dean, J.~S. Dokken, M.~Habera, J.~S. Hale, C.~N. Richardson, M.~E. Rognes, M.~W. Scroggs, N.~Sime, and G.~N. Wells.
\newblock {DOLFINx: The next generation FEniCS problem solving environment}, December 2023.

\bibitem{Langtangen2016}
H.~P. Langtangen and A.~Logg.
\newblock {\em Solving PDEs in Python: The FEniCS Tutorial I}.
\newblock Springer, 2016.

\bibitem{DokkenLink}
{J. S. Dokken}.
\newblock {The FEniCSx tutorial}.
\newblock \url{https://jsdokken.com/dolfinx-tutorial/}.
\newblock Accessed: 2024-01-07.

\bibitem{Kalashnikova2014}
I.~Kalashnikova, M.~F. Barone, S.~Arunajatesan, and B.~G. V.~B. Waanders.
\newblock Construction of energy-stable projection-based reduced order models.
\newblock {\em Applied Mathematics and Computation}, 249:569--596, 2014.

\bibitem{Ingimarson2022}
S.~Ingimarson, L.~G. Rebholz, and T.~Iliescu.
\newblock Full and reduced order model consistency of the nonlinearity discretization in incompressible flows.
\newblock {\em Computer Methods in Applied Mechanics and Engineering}, 401:115620, 2022.

\bibitem{Dalcin2011}
L.~D. Dalcin, R.~R. Paz, P.~A. Kler, and A.~Cosimo.
\newblock {Parallel distributed computing using Python}.
\newblock {\em Advances in Water Resources}, 34(9):1124 -- 1139, 2011.

\bibitem{Li2021}
Angran Li, Amir {Barati Farimani}, and Yongjie~Jessica Zhang.
\newblock {Deep learning of material transport in complex neurite networks}.
\newblock {\em Scientific Reports}, 11(1):11280, 2021.

\bibitem{Li2022}
Angran Li and Yongjie~Jessica Zhang.
\newblock {Modeling material transport regulation and traffic jam in neurons using PDE-constrained optimization}.
\newblock {\em Scientific Reports}, 12(1):3902, 2022.

\bibitem{Li2022b}
Angran Li and Yongjie~Jessica Zhang.
\newblock {Modeling intracellular transport and traffic jam in 3D neurons using PDE-constrained optimization}.
\newblock {\em Journal of Mechanics}, 38:44--59, 03 2022.

\bibitem{Li2023}
Angran Li and Yongjie~Jessica Zhang.
\newblock Isogeometric analysis-based physics-informed graph neural network for studying traffic jam in neurons.
\newblock {\em Computer Methods in Applied Mechanics and Engineering}, 403:115757, 2023.

\bibitem{Liang2022}
Xuan Liang, Angran Li, Anthony~D Rollett, and Yongjie~Jessica Zhang.
\newblock {An isogeometric analysis-based topology optimization framework for 2D cross-flow heat exchangers with manufacturability constraints}.
\newblock {\em Engineering with Computers}, 38(6):4829--4852, 2022.

\bibitem{Liang2022b}
Xuan Liang, Lisha White, Jonathan Cagan, Anthony~D. Rollett, and Yongjie~Jessica Zhang.
\newblock {Unit-Based Design of Cross-Flow Heat Exchangers for LPBF Additive Manufacturing}.
\newblock {\em Journal of Mechanical Design}, 145(1):012002, 10 2022.

\end{thebibliography}

\end{document}